\documentclass[twocolumn]{aastex63} 

\usepackage[normalem]{ulem}
\usepackage[version=4]{mhchem}
\usepackage{longtable}
\graphicspath{{./}{figures/}}

\shorttitle{Multi-wavelength water survey}

\begin{document}

\title{The kinematics and excitation of infrared water vapor emission from planet-forming disks: \\ results from spectrally-resolved surveys and guidelines for JWST spectra}

\correspondingauthor{Andrea Banzatti}
\email{banzatti@txstate.edu}

\author{Andrea Banzatti}
\affil{Department of Physics, Texas State University, 749 N Comanche Street, San Marcos, TX 78666, USA}

\author{Klaus M. Pontoppidan}
\affiliation{Space Telescope Science Institute 3700 San Martin Drive Baltimore, MD 21218, USA}

\author{Jos\'e P\'erez Ch\'avez}
\affil{Department of Physics, Texas State University, 749 N Comanche Street, San Marcos, TX 78666, USA}

\author{Colette Salyk}
\affil{Department of Physics and Astronomy, Vassar College, 124 Raymond Avenue, Poughkeepsie, NY 12604, USA}

\author{Lindsey Diehl}
\affil{Department of Physics, Texas State University, 749 N Comanche Street, San Marcos, TX 78666, USA}

\author{Simon Bruderer}
\affiliation{Max-Planck-Institut für extraterrestrische Physik, Gießenbachstraße 1, 85748 Garching bei München}

\author{Gregory J. Herczeg}
\affil{Kavli Institute for Astronomy and Astrophysics, Peking University, Beijing 100871, China}

\author{Andres Carmona}
\affiliation{Univ. Grenoble Alpes, CNRS, IPAG, F-38000 Grenoble, France}

\author{Ilaria Pascucci}
\affil{Department of Planetary Sciences, University of Arizona, 1629 East University Boulevard, Tucson, AZ 85721, USA}

\author{Sean Brittain}
\author{Stanley Jensen}
\affiliation{Department of Physics \& Astronomy, 118 Kinard Laboratory, Clemson University, Clemson, SC 29634, USA}

\author{Sierra Grant}
\affiliation{Max-Planck-Institut für extraterrestrische Physik, Gießenbachstraße 1, 85748 Garching bei München}

\author{Ewine F. van Dishoeck}
\affiliation{Max-Planck-Institut für extraterrestrische Physik, Gießenbachstraße 1, 85748 Garching bei München}
\affiliation{Leiden Observatory, Leiden University, P.O. Box 9513, 2300 RA Leiden, the Netherlands}

\author{Inga Kamp}
\affiliation{Kapteyn Astronomical Institute, University of Groningen, PO Box 800, 9700 AV Groningen,The Netherlands}

\author{Arthur D. Bosman}
\affil{Department of Astronomy, University of Michigan, 1085 S. University Ave, Ann Arbor, MI 48109}

\author{Karin I. \"Oberg}
\affiliation{Center for Astrophysics, Harvard \& Smithsonian, 60 Garden St., Cambridge, MA 02138, USA}

\author{Geoff A. Blake}
\affiliation{Division of Geological \& Planetary Sciences, MC 150-21, California Institute of Technology, Pasadena, CA 91125, USA}

\author{Michael R. Meyer}
\affil{Department of Astronomy, University of Michigan, 1085 S. University Ave, Ann Arbor, MI 48109}

\author{Eric Gaidos}
\affiliation{Department of Earth Sciences, University of Hawaii at Manoa, 1680 East-West Rd, Honolulu, HI 96822, USA}

\author{Adwin Boogert}
\author{John T. Rayner}
\affiliation{Institute for Astronomy, University of Hawaii, 2680 Woodlawn Drive, Honolulu, HI, USA 96822}

\author{Caleb Wheeler}
\affiliation{Simons Observatory - Center for Computational Astrophysics, Flatiron Institute, 162 5th Ave, NY 10010}

\begin{abstract}
This work presents ground-based spectrally-resolved water emission at R = 30000--100000 over infrared wavelengths covered by JWST (2.9--12.8~$\mu$m).
Two new surveys with iSHELL and VISIR are combined with previous spectra from CRIRES and TEXES to cover parts of multiple ro-vibrational and rotational bands observable within telluric transmission bands, for a total of $\approx160$ spectra and 85 disks (30 of which are JWST targets in Cycle 1). 
The general expectation of a range of regions and excitation conditions traced by infrared water spectra is for the first time supported by the combined kinematics and excitation as spectrally resolved at multiple wavelengths.
The main findings from this analysis are: 
1) water lines are progressively narrower from the ro-vibrational bands at 2--9~$\mu$m to the rotational lines at 12~$\mu$m, and partly match a broad (BC) and narrow (NC) emission components, respectively, as extracted from ro-vibrational CO spectra;
2) rotation diagrams of resolved water lines from upper level energies of 4000--9500~K show vertical spread and curvatures indicative of optically thick emission ($\approx 10^{18}$~cm$^{-2}$) from a range of excitation temperatures ($\approx 800$--1100~K);
3) the new 5~$\mu$m spectra demonstrate that slab model fits to the rotational lines at $> 10$~$\mu$m strongly over-predict the ro-vibrational emission bands at $< 9$~$\mu$m, implying non-LTE vibrational excitation.
We discuss these findings in the context of emission from a disk surface and a molecular inner disk wind, and provide a list of guidelines to support the analysis of spectrally-unresolved JWST spectra.
\end{abstract}


\section{Introduction} \label{sec: intro}

The inner regions of young protoplanetary disks within $\approx$~5--10~au from the central stars present the right conditions for molecular chemistry to thrive in a warm disk layer irradiated by stellar and accretion-shock radiations \citep[e.g.][]{glassgold09,najita11,du14,walsh15,bosman17,woitke18,anderson21}. This inner molecular layer is observed through a forest of emission lines at infrared wavelengths from a number of molecules, especially from CO, \ce{H2O}, \ce{OH}, \ce{HCN}, \ce{C2H2}, and \ce{CO2} \citep[e.g.][]{najita03,cn08,pont10,salyk08,salyk11,fedele11,mandell12,pascucci13,banz17,banz22}. These molecular spectra have been found to reflect the irradiation, physical, and chemical structure of inner disks through their excitation and kinematics. Line fluxes from different molecules present trends with stellar temperature \citep{pont10,pascucci13}, with stellar and accretion luminosity \citep{salyk11_spitz,brittain16,banz17}, with the formation of an inner disk dust cavity \citep{salyk11_spitz,banz17}, and with the dust disk mass and radius as observed at millimeter wavelengths \citep{najita13,banz20}. The line kinematics, when spectrally-resolved in high-resolution spectra, reveal inner regions of gas depletion \citep{brittain03,salyk09,bp15}, a different gas excitation in dust-free vs dust-rich regions \citep{brittain07,vdplas15,heinbert16,banz18}, variability in line kinematics \citep[e.g.][]{brittain13,banz22}, and kinematic profiles tracing both a disk surface in Keplerian rotation and a slow wind launched from inner disks \citep{pont11,bast11,brown13,banz22}. Infrared molecular spectra therefore provide fundamental diagnostics of the physics and chemistry of inner disks at the time of disk dissipation and planet formation.

\begin{figure*}
\centering
\includegraphics[width=1\textwidth]{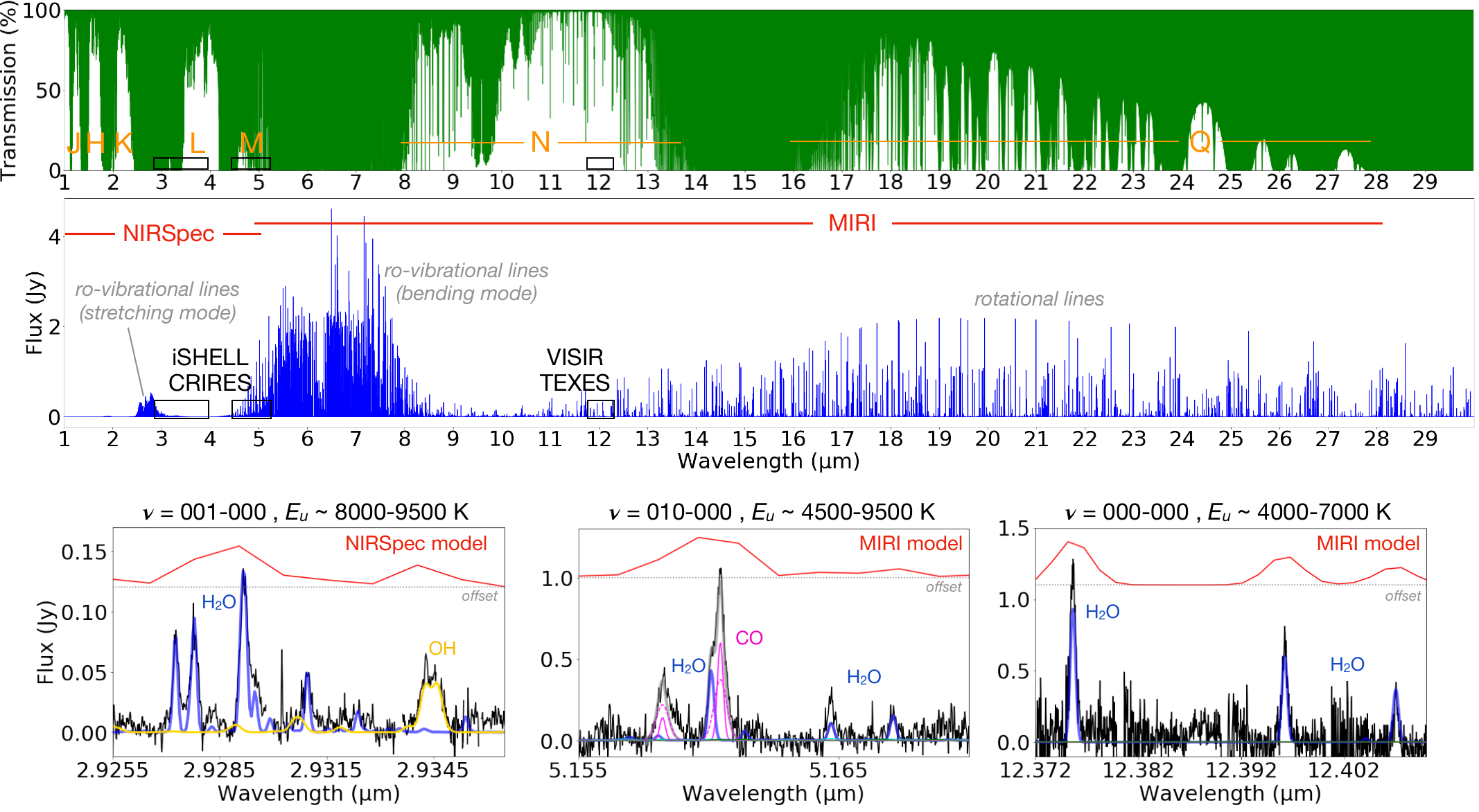} 
\caption{Overview of near- and mid-infrared water emission bands as observed from space and the ground (see Section \ref{sec: water_overview}). \textit{Top}: Earth's atmospheric transmission (in green), with observing bands labelled in orange. \textit{Middle}: water emission model for illustration of its multi-band structure (not a fit to the data; model fits to the data are instead shown in Sections \ref{sec: data} and \ref{sec: slab_fits}). The coverage of different spectrographs is shown with black boxes to illustrate which parts of the spectrum can be observed: high-resolution (R $>$~30000) spectrographs from the ground can spectrally-resolve relatively weak high-energy lines in small spectral windows, while JWST from space can observe a much wider spectral range but only at moderate resolution (R $\sim$~1500--3700). \textit{Bottom}: zoomed-in spectral regions to illustrate the difference in resolution between JWST and ground-based data \citep[the CRIRES, iSHELL, and TEXES spectra are for DR Tau and adopted from this work and][]{banz17,salyk19}. 
}
\label{fig: spec_overview}
\end{figure*}

Among the multiple molecules observed at infrared wavelengths, water presents unique opportunities and challenges. Water can be observed from molecular clouds, through the phases of star and planet formation, all the way to disks and exoplanets \citep[e.g.][]{vandishoeck14}. Water is also expected to play a major role in planet formation from the dynamics \citep[e.g.][]{cieslacuzzi06}, to the accretion of solids \citep[e.g.][]{ros13}, to habitability \citep[e.g.][]{krijt22}, and it is a major driver of exoplanet science today. However, the molecular structure of water, an asymmetric top molecule with three vibrational ($v_1 ~ v_2 ~ v_3$)\footnote{for symmetric stretching, bending, and asymmetric stretching modes respectively.} and three rotational ($J ~ K_a ~ K_c$) quantum numbers, produces a complex rotational and ro-vibrational spectrum that spreads across infrared wavelengths, posing multiple technical challenges in observations (see Section \ref{sec: water_overview}) and in the analysis and interpretation of spectra \citep[e.g.][]{meijerink09,kamp13}.

Observing water and its evolution in disks at the time of planet formation is one of the major drivers of scientific investigations in the field of planet formation with the \textit{James Webb Space Telescope} (JWST), which is uniquely suited for water observations thanks to the wide spectral coverage, high sensitivity, and the location outside of Earth's atmosphere. Despite the amount of water spectra from disks collected to date \citep[see previous summary in][]{banz17}, a number of fundamental questions remain:

1) which inner disk region(s) do the infrared water lines trace, and are there multiple water reservoirs (with different temperature and density);

2) what is the water abundance in inner disks and what determines it (chemistry vs dynamics);

3) what is the relative role of different excitation processes, and is water emission in local thermodynamic equilibrium (LTE);

4) is water present in a molecular inner disk wind; and

5) how can we correctly interpret the complex water spectra observed across infrared wavelengths within a unified picture of inner disks?

In this work we offer a contribution to this field at a time when the first water spectra are being observed with JWST \citep{yang22}. We present and analyze spectrally-resolved water emission lines from ro-vibrational and rotational bands at multiple wavelengths between 2.9 and 12.9~$\mu$m tracing upper level energies between 4000 and 9500~K, and compare them to the velocity and excitation of ro-vibrational CO spectra observed at 4.5--5.25~$\mu$m tracing the same range in upper level energies. One goal of this work is to present guidelines to support the analysis of both individual and large samples of spectrally-unresolved water spectra from JWST.

\subsection{Observing water in disks: the need for a synergy between space and ground data} \label{sec: water_overview}

Water spectra in protoplanetary disks have so far been observed with instruments that fall under two main categories: space-based low-resolution spectrographs (e.g. IRS on \textit{Spitzer} with maximum resolving power R~$\approx 700$), and ground-based high-resolution echelle spectrographs (e.g. CRIRES on the VLT with R~$\approx 95000$). Each type of instrument presents specific advantages and disadvantages, which we illustrate in Figure \ref{fig: spec_overview}. 
Space-based spectrographs provide wide spectral coverage (e.g. 10--37~$\mu$m with Spitzer-IRS, and now 4.9--28~$\mu$m with JWST-MIRI) but lose kinematic information on the observed emission, and blend together lines from multiple emission/absorption components and different chemical species.
Conversely, ground-based spectrographs provide high resolving power that separates the emission from different lines and molecules and reveals the gas kinematics in high detail, but they can only observe in limited spectral windows of higher telluric transmission within the $L$, $M$, and $N$ bands (Figure \ref{fig: spec_overview} and Table \ref{tab:water_surveys}).

\begin{deluxetable*}{c c c c c c c c c}
\tabletypesize{\small}
\tablewidth{0pt}
\tablecaption{\label{tab:water_surveys} Summary of high-resolution water spectra observed in disks with four ground-based instruments.}
\tablehead{\colhead{$\lambda$} & \colhead{Instrument} & \colhead{R} & \colhead{\# lines} & \colhead{$E_{u}$} & \colhead{$n_{\rm{crit}}$} & \colhead{Sample} & \colhead{Detections} & \colhead{References} \\ 
\colhead{($\mu$m)} & \colhead{} & \colhead{} & \colhead{} & \colhead{(K)} & \colhead{(cm$^{-3}$)} & \colhead{(\# disks)} & \colhead{(\# disks)} & \colhead{}}
\tablecolumns{9}
\startdata
2.9--2.98 & CRIRES & 95,000 & $\approx25$ & 8,000--9,500 & $10^{15}$--$10^{16}$ & 46 & 16  & 1 \\
\\
4.52--5.24 & iSHELL & 60,000 & $\approx40$ & 4,500--9,500 & $\approx 10^{13}$ & 60 & 10   & this work, 2 \\
\\
12.23--12.87 & VISIR & 30,000 & 2--11 & 3,200--7,000 & $\approx 10^{11}$ & 64 & 10   & this work \\
\\
12.37--12.46 & TEXES & 100,000 & 3--7 & 3,600--6,000 & $\approx 10^{11}$ & 9 & 7 &  3\\
\enddata
\tablecomments{Critical densities $n_{\rm{crit}}$ = $A_{ul}$/$C_{ul}$ are estimated using molecular data adopted in RADEX (see Section \ref{sec: rot_diagr_models}). As collision temperature, we use the excitation temperatures from Table \ref{tab: slab_results}. Since upper level energies $E_{u}$ only up to 7200~K are included in \citet{faure08}, as collisional rates $C_{ul}$ for the 2.9--2.98~$\mu$m lines we take those of lines near 2.7~$\mu$m with same range in $A_{ul}$ but $E_{u}$ of 6000--7200~K.}
\tablerefs{1: \citet{banz17,banz15,mandell12,fedele11,salyk22}; 2: \citet{banz22}; 3: \citet{najita18,salyk19}. Additional high-resolution water spectra from CSHELL and NIRSPEC have been published in \citet{carr04,salyk08,doppmann11}.}
\end{deluxetable*}

In Figure \ref{fig: spec_overview} we report a water emission model for quick guidance on the structure of the spectrum across wavelengths. The model is described below in Section \ref{sec: rot_diagr_models}, and in this figure we use a temperature of 700~K, column density of $10^{18}$~cm$^{-2}$ and line FWHM of 15~km/s. We visually scale the model to approximately match the observed emission in DR~Tau just for line identification (in Section \ref{sec: slab_fits} we will present actual model fits to the data). Purely rotational lines populate a large part of the spectrum at wavelengths longer than 10~$\mu$m, while ro-vibrational transitions in the bending and stretching modes present more compact bands at 4--9~$\mu$m and 2.5--3.5~$\mu$m respectively. We include in the figure a sky transmission spectrum as obtained from the ESO SkyCalc application, setup for the Paranal site (altitude of 2640 m) and using airmass = 1 and precipitable water vapor (PWV) = 2.5 mm. Comparison of Earth's transmission to the water emission spectrum in the top two panels in Figure \ref{fig: spec_overview} demonstrates that ground-based instruments enable observations of the water spectrum only where it is relatively weak (the tails of ro-vibrational bands and a few rotational lines near 12.4~$\mu$m). 

Using observations from space or from the ground, previous work has therefore focused on datasets or samples that were limited one way or the other, resulting in an incomplete view of the distribution and excitation of water in inner disks. Some studies analyzed large samples (up to $\approx100$ objects) of spectrally-unresolved rotational water emission observed with Spitzer-IRS at 10--37~$\mu$m or Herschel-PACS/HIFI at 50--540~$\mu$m; due to the low resolving power and blending of emission lines, Spitzer spectra only provided degenerate model solutions for the properties (temperature and column density) of multiple molecules and no direct kinematic information on the radial location of emitting regions \citep{cn11,salyk11_spitz,pascucci13}, while Herschel observations provided very low detection rates of a few lines only \citep[e.g.][]{rivmarichalar12,du17}. 

Other studies analyzed small samples of high-resolution spectra from the ground, mostly biased towards highly accreting disks that had been observed with Spitzer to have strong water emission; these spectra provided first kinematic information on the water emission but were very limited in sensitivity and by strong telluric absorption \citep{pont10b,salyk19} and, in the case of the 2.9~$\mu$m water band, by absorption in stellar photospheres \citep{banz17}. The properties of water emission as observed in different samples and different wavelengths have ranged from high temperatures in very optically thick regions in the $L$ band \citep[T $\approx$ 1500~K and N $\approx 10^{20}$ cm$^{-2}$, found in only 3 disks so far;][]{carr04,doppmann11,salyk22}, to moderate temperatures and opacity in a narrow window in the $N$ band \citep[T $\approx$ 500--700~K and N $\approx 10^{18}$ cm$^{-2}$,][]{najita18,salyk19}, down to cooler temperatures at mid- and far-infrared wavelengths \citep[300--600~K,][]{cn11,salyk11_spitz,rivmarichalar12,liu19}. Apart from those analyzing Spitzer spectra, previous results were always based on small samples of 1--10 bright sources.

Supporting the scenario of water emitting from a radial range of disk regions, the combined analysis of spectrally-unresolved mid- and far-infrared water spectra has suggested that water emission should come from disk surfaces from the dust sublimation radius out to 10~au or even larger radii \citep[e.g.][]{antonellini15,blevins16,notsu16,woitke18}, where different emission lines probe a range of disk regions and layers depending on their upper level energy and Einstein-$A$ coefficient. However, a unified picture of water in inner disks across different emission bands, including the ro-vibrational lines in the near-infrared and calibrated on spectrally-resolved water line kinematics, is still lacking. JWST now covers both the rotational lines at $>10$~$\mu$m ($E_{u} \approx$ 1000--6000~K) and the ro-vibrational lines from the bending mode ($E_{u} \approx$ 4000--10000~K) simultaneously and at similar resolving power (R = 1500--3700 across MIRI wavelengths), providing unprecedented leverage on the excitation of water across inner disk radii.
One goal of this paper is to support the analysis of JWST spectra by providing spectrally-resolved line kinematics at multiple wavelengths from a suite of high-resolution ground-based spectrographs (Table \ref{tab:water_surveys}). A comprehensive view of water in disks will require the combination of multiple instruments from ground and space at least until a space-based infrared telescope with high resolving power (R~$>30000$) may be built \citep[e.g.][]{pont18,kamp21}.

\begin{deluxetable}{c l c c c c c}
\tabletypesize{\small}
\tablewidth{0pt}
\tablecaption{\label{tab: lines} List of prominent H$_2$O lines covered in this work.}
\tablehead{\colhead{Wavelength} & Transition (upper-lower levels) & $A_{ul}$ & $E_{u}$ &  \\ \colhead{ ($\mu$m)} & (level format: $v_1 v_2 v_3~~J _{\:K_a \: K_c}$) &  (s$^{-1}$) &  (K)}
\tablecolumns{4}
\startdata
2.90813 & 001-000 \: $14_{\:0\:14} - 15_{\:0\:15}$ & 48.4 &  8341 \\
2.90829 & 001-000 \: $11_{\:4\:8} - 12_{\:4\:9}$ & 42.5 &  8004 \\
2.90911 & 001-000 \: $10_{\:6\:4} - 11_{\:6\:5}$ & 31.8 &  8031 \\
2.90998 & 001-000 \: $12_{\:2\:10} - 13_{\:2\:11}$ & 48.8 &  8177 \\
2.91012 & 001-000 \: $13_{\:2\:12} - 14_{\:2\:13}$ & 48.8 &  8293 \\
2.92726 & 001-000 \: $15_{\:1\:15} - 16_{\:1\:16}$ & 49.5 &  8744 \\
2.92726 & 001-000 \: $15_{\:0\:15} - 16_{\:0\:16}$ & 48.9 &  8744 \\
2.92780 & 001-000 \: $12_{\:3\:9} - 13_{\:3\:10}$ & 51.6 &  8388  \\
2.92912 & 001-000 \: $13_{\:3\:11} - 14_{\:3\:12}$ & 48.3 &  8583 \\
2.92920 & 001-000 \: $14_{\:1\:13} - 15_{\:1\:14}$ & 48.7 &  8698  \\
2.93096 & 001-000 \: $11_{\:6\:6} - 12_{\:6\:7}$ & 33.6 &  8411 \\
2.94665 & 001-000 \: $16_{\:0\:16} - 17_{\:0\:17}$ & 48.9 & 9172 \\
2.94861 & 001-000 \: $14_{\:2\:12} - 15_{\:2\:13}$ & 48.9 &  9012 \\
\hline
4.72810 & 010-000 \: $9_{\:7\:2} - 8_{\:6\:3}$ ($\star$) & 2.4 &  5074 \\
4.72942 & 010-000 \: $11_{\:6\:5} - 10_{\:5\:6}$ ($\star$) & 2.1 &  5515 \\
4.79063 & 001-000 \: $10_{\:6\:5} - 9_{\:5\:4}$ ($\star$) & 2.6 &  5129 \\
4.84063 & 010-000 \: $9_{\:6\:3} - 8_{\:5\:4}$ & 2.6 &  4778 \\
4.88205 & 010-000 \: $20_{\:3\:18} - 19_{\:2\:17}$ & 10.6 &  9179 \\
4.91324 & 010-000 \: $22_{\:2\:21} - 21_{\:1\:20}$ & 14.3 &  9864 \\
4.95058 & 020-010 \: $9_{\:5\:4} - 8_{\:4\:5}$ & 4.7 & 6884 \\
4.96548 & 010-000 \: $16_{\:4\:13} - 15_{\:3\:12}$ & 6.7 & 7329 \\
4.99774 & 020-010 \: $6_{\:6\:1} - 5_{\:5\:0}$ & 19.0 &  6341 \\
5.00798 & 010-000 \: $15_{\:4\:12} - 14_{\:3\:11}$ & 6.2 & 6814 \\
5.03885 & 010-000 \: $19_{\:1\:18} - 18_{\:2\:17}$ ($\star$) & 14.5 & 8103 \\
5.06240 & 010-000 \: $16_{\:3\:14} - 15_{\:2\:13}$ ($\star$) & 9.6 & 6975 \\
5.06429 & 010-000 \: $16_{\:2\:14} - 15_{\:3\:13}$ & 9.5 & 6974 \\
5.07548 & 020-010 \: $7_{\:5\:2} - 6_{\:4\:3}$ & 5.9 & 6286 \\
5.10815 & 010-000 \: $12_{\:4\:9} - 11_{\:3\:8}$ ($\star$) & 4.2 & 5425 \\
5.11080 & 010-000 \: $5_{\:3\:13} - 14_{\:2\:12}$ ($\star$) & 9.1 & 6485 \\
5.13018 & 010-000 \: $11_{\:4\:8} - 10_{\:3\:7}$ & 3.7 & 5018 \\
5.16471 & 010-000 \: $20_{\:1\:20} - 19_{\:0\:19}$ ($\star$) & 18.2 &  8074 \\
5.16709 & 010-000 \: $14_{\:2\:12} - 13_{\:3\:11}$ & 8.6 &  6019 \\
5.22140 & 010-000 \: $15_{\:1\:14} - 14_{\:2\:13}$ & 12.4 &  6105 \\
5.22307 & 010-000 \: $13_{\:2\:11} - 12_{\:3\:10}$ & 8.0 &  5578 \\
\hline
12.2481 & 000-000 \: $17_{\:5\:13} - 16_{\:2\:14}$ & 7.8 &  5795 \\
12.2654 & 000-000 \: $18_{\:7\:12} - 17_{\:4\:13}$ & 12.3 &  6954 \\
12.2772 & 000-000 \: $12_{\:8\:5} - 11_{\:5\:6}$ & 0.5 &  4048 \\
12.2876 & 000-000 \: $9_{\:9\:1} - 8_{\:6\:2}$ & 0.02 &  3202 \\
12.3757 & 000-000 \: $16_{\:4\:13} - 15_{\:1\:14}$ & 4.2 &  4948 \\
12.3962 & 000-000 \: $17_{\:4\:13} - 16_{\:3\:14}$ & 7.7 &  5781 \\
12.4070 & 000-000 \: $16_{\:3\:13} - 15_{\:2\:14}$ & 4.2 &  4945 \\
12.4448 & 000-000 \: $11_{\:8\:3} - 10_{\:5\:6}$ & 0.3 &  3629  \\
12.4535 & 000-000 \: $13_{\:7\:6} - 12_{\:4\:9}$ & 1.2 &  4213  \\
12.8319 & 000-000 \: $10_{\:8\:2} - 9_{\:5\:5}$ & 0.2 &  3243  \\
12.8702 & 000-000 \: $10_{\:8\:3} - 9_{\:5\:4}$ & 0.2 &  3243  \\
\enddata
\tablecomments{Line properties are from HITRAN \citep{hitran20}. ($\star$) = lines used for stacking.}
\end{deluxetable}

\section{Spectra included in this work}  \label{sec: data}
Table \ref{tab:water_surveys} reports the samples of water spectra available from four high-resolution ground-based instruments and some properties of the water emission they cover. Each dataset is described in the next sections instrument by instrument; the data newly presented in this work come from two separate surveys performed with VISIR and iSHELL, while the CRIRES water spectra \citep{banz17} and TEXES water spectra \citep{najita18,salyk19} were published before. All the reduced, science-ready spectra from CRIRES, VISIR, and iSHELL are available on \url{www.spexodisks.com} \citep[][and Wheeler et al., in prep.]{perez21_spexodisks}. A list of the most prominent water lines covered in these spectra is reported in Table \ref{tab: lines}. A gallery of portions of the water spectra at multiple wavelengths object by object is included in Appendix \ref{app: extra_plots}. Throughout this paper, spectra and spectral lines are shown in the heliocentric reference frame as corrected for the barycentric velocity at the time of observation of each target.

\begin{figure*}
\centering
\includegraphics[width=1\textwidth]{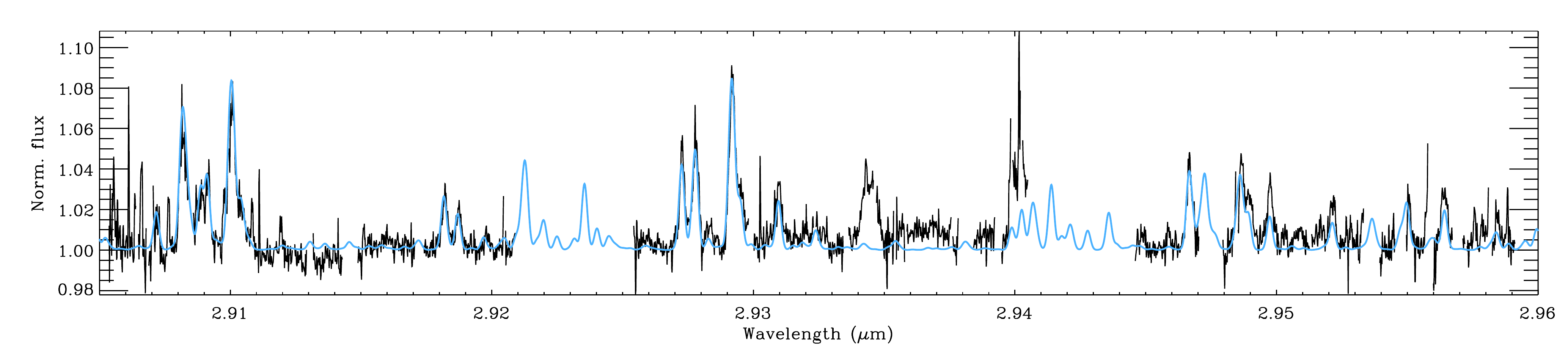} 
\caption{Example of a CRIRES $L$-band spectrum, for the disk of DR~Tau \citep[from][]{banz17}. A water emission spectrum is shown in light blue, using the LTE slab model fit described in Section \ref{sec: slab_fits}. The prominent feature near 2.935~$\mu$m is an OH doublet.}
\label{fig: crires_spec}
\end{figure*}

\subsection{VLT-CRIRES spectra at 2.9~$\mu$m} \label{sec: CRIRES_data}
Water ro-vibrational emission spectra at 2.9--2.98~$\mu$m are included as observed with the Cryogenic Infrared Echelle Spectrometer \citep[CRIRES,][]{crires} on the Very Large Telescope (VLT) as part of a survey from 2007-2008 \citep[ESO Large Program 179.C-0151;][]{pont11,brown13}. The water spectra were obtained for $\approx 50\%$ of the sample in that survey (35 out of 69 disks, mostly T~Tauri stars) and were presented and analyzed in \citet{banz17}, except for the very different spectrum with a much larger number of high-excitation lines observed in VV~CrA~S that is published in \citet{salyk22}. Additional spectra for 11 Herbig Ae/Be stars were presented in \citet{fedele11}. 

The T~Tauri spectra included in this work were taken with a resolving power of $\approx 95000$ or $\approx 3$~km/s, and most of them required correction for stellar photospheric absorption on top of telluric correction. The observed water emission lines were found to be generally as broad as the broad component of CO $M$-band emission (FWHM up to $\approx100$~km/s in some disks), and in a few cases to include also weak narrower central emission similar to the kinematic structure of CO \citep[for details, see][]{banz17}. An example $L$-band spectrum from CRIRES is shown in Figure \ref{fig: crires_spec}, showing $\approx$~15--20 prominent water lines that in some cases are blends of two nearby transitions \citep[the blending of lines increases with line broadening, and is generally more severe than in the example shown here; see Figure 14 in][]{banz17}.

\subsection{IRTF-iSHELL spectra at 5~$\mu$m} \label{sec: iSHELL_data}
Water ro-vibrational emission spectra at 5~$\mu$m are included as observed with iSHELL \citep{ishell,ishell22} at the NASA Infrared Telescope Facility (IRTF), as part of an ongoing $M$-band survey of protoplanetary disks \citep{banz22}. iSHELL covers a wide portion of the observable $M$ band at 4.5--5.25~$\mu$m in one shot, with only narrow gaps between the echelle orders at longer wavelengths, providing simultaneous observations of CO, \ce{H2O}, and HI \citep{banz22}. Most spectra were taken with the 0.75" slit and a resolving power of $\approx 60000$ or $\approx 5$~km/s, while the narrower slit with $\approx 92000$ or $\approx 3.3$~km/s was used only for the brightest disks (mostly around Herbig Ae/Be stars, where water is not detected). The iSHELL spectra included here were reduced with Spextool v5.0.3 \citep{spextool} and have been presented in \citet{banz22} or obtained in January-July 2022, for a total current sample of 60 disks. Multi-epoch observations have been obtained for some disks and show variability \citep[e.g. Figure 8 in][]{banz22}; in this work we only include one epoch, and we will study the variability in a future publication.

\begin{figure*}
\centering
\includegraphics[width=1\textwidth]{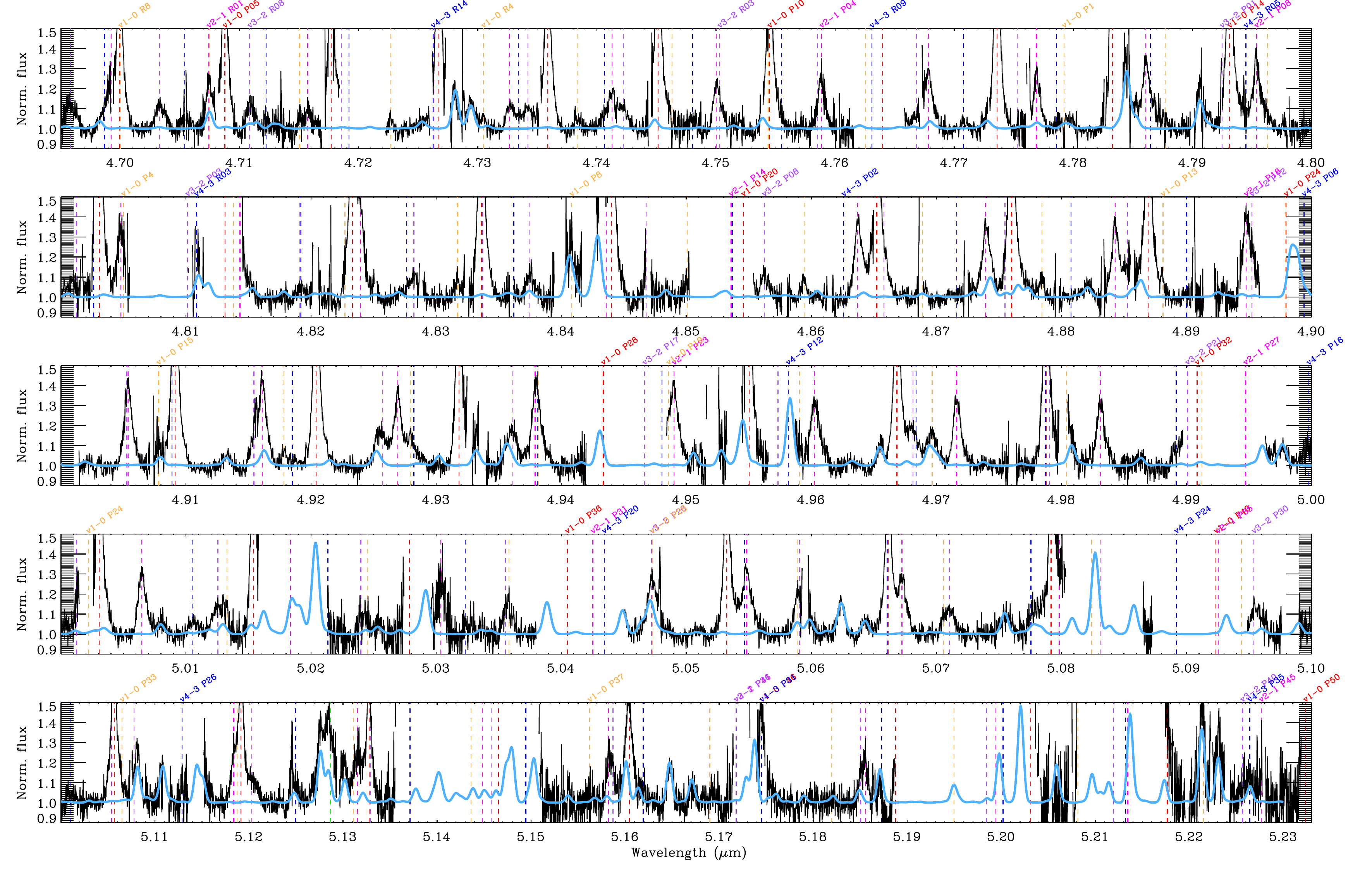} 
\caption{Example of part of an iSHELL $M$-band spectrum, for the disk of FZ~Tau. A water emission spectrum is shown in light blue, using the LTE slab model fit described in Section \ref{sec: slab_fits}. CO lines are marked in other colors ($^{13}$CO in orange, all other colors for $^{12}$CO) and labelled for reference, but the plot zooms-in on the water lines for better visualization of these and cuts off the peaks of $v = 1-0$ CO lines. HI emission from the Hu~$\delta$ line blended with water near 5.13~$\mu$m is marked in light green.}
\label{fig: ishell_spec}
\end{figure*}

The $M$-band water lines have been discovered in the disk of AS~205~N and reported for the first time in \citet{banz22}, and in this new work we include all the current detections from the rest of the survey. An example is presented in Figure \ref{fig: ishell_spec}. In the currently available $M$-band spectra, a total of $\approx 40$ ro-vibrational water lines are observed (mostly from $v=1$ and some from $v=2$, see Table \ref{tab: lines}), some of which are blended with CO or \ce{HI} emission. In this work, we make a weighted average of eight water lines that are not blended with CO to produce a stacked velocity profile with higher S/N; these lines are marked in Table \ref{tab: lines}.
At the S/N obtained in these spectra, the $M$-band water lines are consistent with having the same velocity profile; future higher-sensitivity spectra could be used to test for any change in ro-vibrational line profiles as a function of $E_{u}$.

\subsection{VLT-VISIR spectra at 12.4~$\mu$m} \label{sec: VISIR_data}
\begin{figure*}
\centering
\includegraphics[width=1\textwidth]{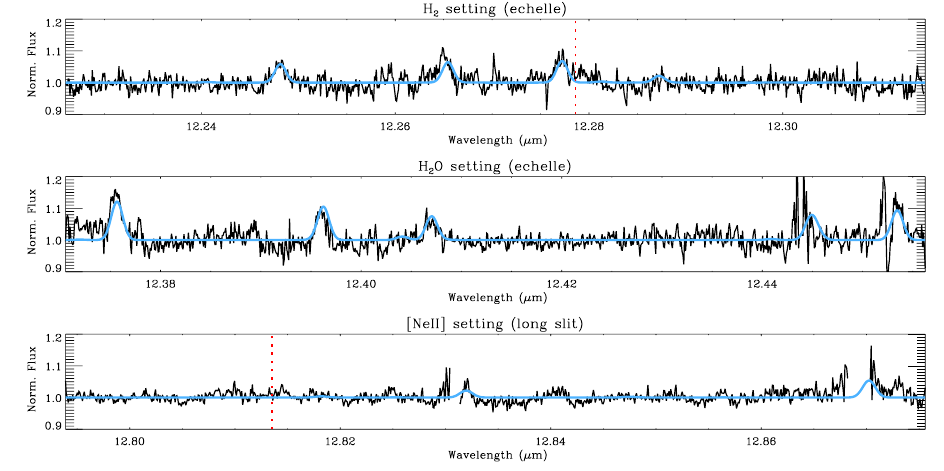} 
\caption{Example of all three spectral settings from the VISIR survey, for the disk of CrA-IRS2 (CHLT~1). Up to 11 water lines are covered, and up to 7 detected in this survey. Water lines near 12.85~$\mu$m are from levels with upper level energies 3200~K, and could not be corrected from the deep and broad telluric lines (see Appendix \ref{app: visir_survey}). A water emission spectrum is shown in light blue, using the LTE slab model fit described in Section \ref{sec: slab_fits}. The positions of [NeII] and H$_2$ lines are marked with dotted lines (neither is detected in this spectrum).}
\label{fig: visir_spec}
\end{figure*}

Water rotational emission spectra at 12.2--12.9~$\mu$m were obtained within a Large Program with the European Southern Observatory (ESO) Very Large Telescope (VLT) Imager and Spectrometer for the Mid-Infrared \citep[VISIR,][]{visir}, and are presented and analyzed in this work for the first time.
This survey used VISIR at the VLT right after its upgrade \citep{visirUP} and over semesters between August 2016 and August 2018 (program IDs 095.C-0203 and 198.C-0104, PI: K. Pontoppidan). A spectrum is shown in Figure \ref{fig: visir_spec} as an example.

The survey included three settings: two echelle settings centered at 12.27\,$\mu$m and 12.41\,$\mu$m to cover emission from H$_2$O lines and the 0-0 S(2) H$_2$ line at 12.2786\,$\mu$m (and possibly OH), and one long-slit setting centered at 12.84\,$\mu$m to include the [NeII] line at 12.8136\,$\mu$m and two water lines with upper level energy 3200~K. The central wavelengths of all settings were set to maximize the number of H$_2$O emission lines to be observed, a total of eleven (Table \ref{tab: lines}). We used the 0.75" slit providing a resolving power R $\approx30,000$ (or 10 km/s). The data reduction is based on a custom pipeline initially made for VISIR 1 data \citep{banz14} and adapted to the upgraded VISIR. Appendix \ref{app: visir_survey} reports more details on the observations and reduction.

The total numbers of spectra obtained in each setting, including multiple epochs and/or slit orientations for a given target, are: 60 spectra in the H$_2$O setting, 16 spectra in the H$_2$ setting, and 45 spectra in the [NeII] setting. Out of 64 targets, 22 were observed in at least two settings, and 4 targets were observed in all three settings (CrA-IRS2, SCrA, TCrA, VVCrA; see Appendix \ref{app: visir_survey}). To increase S/N, we have combined by weighted average spectra from multiple epochs and/or slit orientations obtaining one spectrum per object per setting. In the case of DR~Tau, we averaged three epochs of data obtained earlier in program 088.C-0666 at higher S/N \citep{banz14}.
The [NeII] detections obtained in this survey have been presented and analyzed in \citet{pascucci20}; \ce{H2O} and [NeII] detections overlap in two disks only, RU~Lup and VW~Cha, but the lines have very different kinematic structure where [NeII] only traces gas at high blue-shifted velocities of -50 to -200~km/s in these two systems.
No \ce{H2} detections are reported from this survey, consistent with previous low detection rates \citep{carmona08,bitner08}.

\subsection{Gemini-TEXES spectra at 12.4~$\mu$m} \label{sec: TEXES_data}
Additional water rotational emission spectra at 12.2--12.4~$\mu$m are included as obtained with TEXES on Gemini \citep{texes} with a resolving power of 100,000 or 3~km/s and previously presented in \citet{najita18} and \citet{salyk19}. In this work we consider only the highest S/N among these spectra, obtained in 4 disks (see Section \ref{sec: h2o_kinematics}).

\section{Sample}
The sample included in this work is the combination of samples from the different programs described above as obtained with each instrument. The total is 85 protoplanetary disks around pre-main sequence stars with temperatures of 3000--20000~K and masses of 0.4--5~M$_{\odot}$ (i.e. including both T~Tauri and Herbig Ae/Be stars) in nearby star-forming regions (mostly within $\approx 200$~pc). Accretion luminosities are in the range of 0.01--200~L$_{\odot}$ and millimeter disk radii of 10--200~au, including a number of disks with inner dust cavities \citep[here traced using the infrared index $n_{13-30}$, see][]{brown07,furlan09,banz20} and some disks in wide multiple systems (See Section \ref{sec: sample_binary}). The sample is listed in full in the Appendix and sample properties are visualized in Section \ref{sec: h2o_trends}. Out of $\approx 70$ disks planned for observations with JWST-MIRI in Cycle 1, 30 have high-resolution spectra that are included in this work.

\subsection{Wide multiple systems} \label{sec: sample_binary}
There are some wide multiple systems in the sample \citep[e.g.][]{prato03,manara19,panic21}: AS~205 (separation 1.3"), DoAr~24E (GSS~31, 2.2"), S~CrA (1.3"), VV~CrA (1.9"), DK~Tau (2.4"), UY~Aur (0.9"), T~Tau (0.7"), KK~Oph (1.6"). In good seeing conditions, we aligned the slit along the system axes so that the main binary components could be extracted. 
AS~205 A is the brighter component in the N, AS~205 B the fainter component in the S. 
DoAr~24E A is the brighter component in the N, but it becomes fainter than the B component in the S in the $L$ band and at longer wavelengths \citep{prato03,brown13}. Therefore, in our data DoAr~24E B (S) is the brighter component.
SCrA A is the brighter component in the NW, SCrA B is the fainter component in the SE; VVCrA A is the brighter component in the S, VVCrA B is the fainter component in the N \citep{sullivan19}.
DK~Tau~A, UY~Aur~A, T~Tau~A, and KK~Oph~A are the brighter components in the N in each system \citep[e.g.][]{manara19,panic21}.

\section{ANALYSIS} \label{sec: analysis}

\begin{figure*}
\centering
\includegraphics[width=1\textwidth]{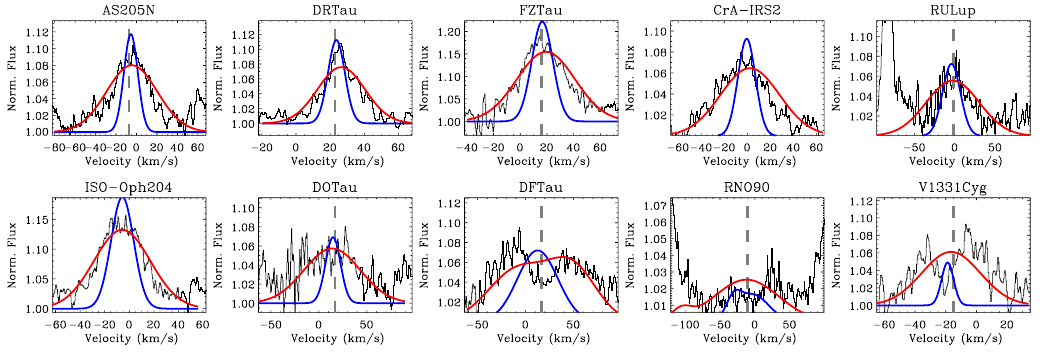} 
\caption{Gallery of H$_2$O detections in iSHELL spectra near 5~$\mu$m, using the stacked line profile (see Section \ref{sec: iSHELL_data}). CO line components are scaled to match the relative strength of water lines for visual comparison, with the BC in red and the NC in blue. Spectral lines are shown in heliocentric velocity and stellar RVs are marked with vertical dashed lines, where available.}
\label{fig: H2O_iSHELL_profiles_fit}
\end{figure*}

\begin{figure*}
\centering
\includegraphics[width=1\textwidth]{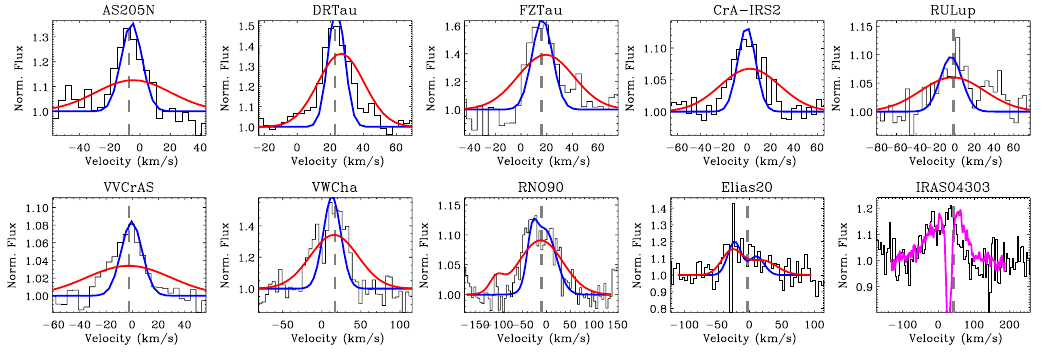} 
\caption{Gallery of H$_2$O detections in VISIR spectra, using the 12.396~$\mu$m line that is the least affected by telluric absorption (see Figure \ref{fig: settings}). CO components and stellar RVs are marked as in Figure \ref{fig: H2O_iSHELL_profiles_fit}. Due to the deep and broad absorption, CO components are not extracted in IRAS~04303 and the full observed line profile is shown instead.}
\label{fig: H2O_VISIR_profiles_fit}
\end{figure*}

\begin{figure*}
\centering
\includegraphics[width=1\textwidth]{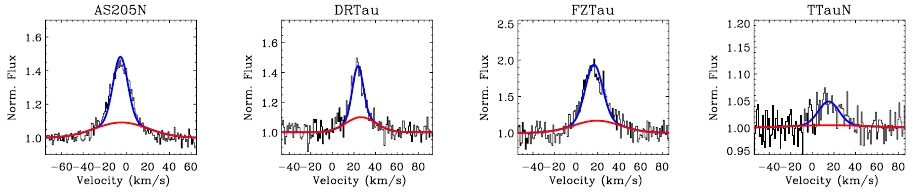} 
\caption{Two-Gaussian fits to the 12.396~$\mu$m water line in TEXES spectra using the BC and NC CO components in each disk.}
\label{fig: TEXES_analysis}
\end{figure*}

\subsection{\ce{H2O} line kinematics} \label{sec: h2o_kinematics}
The first point we address in this work is about the line kinematics and emitting regions of water as spectrally-resolved at multiple wavelengths, addressing questions 1) and 4) listed in Section \ref{sec: intro}. Previous work found that 2.9~$\mu$m lines are dominated by a broad emission component that matches the width of the broad component in CO emission lines in the same systems \citep{banz17}. At 12.4~$\mu$m, instead, line velocities were found to be narrower and match the narrow component in CO emission in a few disks where data was available \citep{banz17,salyk19}. With the new surveys and larger sample included in this work, it is now possible to expand the comparison between CO emission components and water emission lines on a larger number of detections and, for the first time, including water emission in the $M$ band around 5~$\mu$m.

Figures \ref{fig: H2O_iSHELL_profiles_fit} and \ref{fig: H2O_VISIR_profiles_fit} show a comparison of the \ce{H2O} line profiles using the CO line profiles observed in each object as reference. Line profiles from the fundamental $v=1-0$ lines have been observed with VLT-CRIRES and IRTF-iSHELL \citep[][]{pont11,brown13,banz22}, and have been decomposed into broad BC and narrow NC velocity components as described in \citet{bp15,banz22}. Gaussian fits to these two components are then scaled to match the relative strength of water lines to visualize the broadening of water relative to each CO component.
The 10 iSHELL spectra where \ce{H2O} is detected show that the kinematic profile of \ce{H2O} emission at 5~$\mu$m is broader than the NC but narrower than the BC in at least 6/10 cases. In the other 4 cases (DO~Tau, DF~Tau,  RNO~90, and V1331~Cyg), water lines are as broad as the BC. In DF~Tau, and tentatively in RNO~90 too, water shows a double-peak shape that matches the width of the BC. 
In V1331~Cyg, water is as broad as the BC and possibly has a narrower blue-shifted absorption (see also Section \ref{sec: wat_absorpt}); its near-infrared spectrum is overall more complex due to an uncommonly high excitation of both CO and water \citep[see the $L$-band water spectrum in][]{doppmann11}, and will be analyzed in detail in a future work. Additional tentative detections are found in three more disks (UY~Aur~A, Elias~24, and HD~35929), which are all included in the plots in the Appendix \ref{app: extra_plots}.

The 10 \ce{H2O} lines detected in the VISIR spectra show that the kinematic profile of \ce{H2O} emission at 12.4~$\mu$m is rather well matched by the narrow CO component in 6/10 cases, within the uncertainty of the much lower pixel sampling and resolution of the VISIR spectra. The comparison between CO and \ce{H2O} lines is ambiguous in two cases (RU~Lup and Elias~20), due to the low S/N of the water spectra. The case of VW~Cha shows a broader water line with potentially a blue-shifted absorption, which will be discussed in Section \ref{sec: wat_absorpt}. The case of IRAS~04303 shows a very broad water line, similarly broad as the CO line; in this case, Figure \ref{fig: H2O_VISIR_profiles_fit} shows the complete CO line profile as observed, rather than decomposed into BC and NC as in the other objects, to illustrate the presence of a broad, deep, and blue-shifted absorption component in the CO line \citep[which is detected up to P38, i.e. this is another one of the highly excited blue-shifted absorption spectra reported in][]{banz22}.

In Figure \ref{fig: TEXES_analysis} we show fits to the 12.396~$\mu$m water emission line observed in the four highest quality TEXES spectra from \citet{najita18,salyk19}. Water was detected in three more disks in \citet{salyk19}, DoAr~44, HL~Tau, and RW~Aur, but with too low S/N to apply this analysis. Given the higher resolution and S/N of these spectra, instead of using BC and NC individually as in Figures \ref{fig: H2O_iSHELL_profiles_fit} and \ref{fig: H2O_VISIR_profiles_fit}, we fit the full line shape with a two-Gaussian model based on the measured FWHM of BC and NC as done for the 2.9~$\mu$m lines in \citet{banz17}, to test whether there is evidence for both components contributing to the observed water line profile. The analysis of this small sample confirms that water rotational lines at 12.4~$\mu$m overall match well the shape of the NC, but it also shows that water lines include weak broad wings from a broader velocity component similar to the BC.

All together, the picture emerging from the kinematics of spectrally-resolved water emission observed between 2.9 and 12.4~$\mu$m is that water lines are broader at shorter wavelengths in the ro-vibrational bands and narrower at longer wavelengths in the purely rotational transitions. This suggests that ro-vibrational lines are excited predominantly (or exclusively) in an inner, hotter region as compared to the rotational lines, a region that may match that of the BC in CO. The rotational lines, instead, are dominated by an emission component that matches the NC in CO, at least down to upper level energies of $\sim4000$~K as covered in the VISIR and TEXES spectra; lower energy levels at longer wavelengths covered by JWST-MIRI remain to be spectrally-resolved, and might be even narrower by being excited in outer disk regions (See Section \ref{sec: discuss}). The fact that water lines at different wavelengths should have different FWHM reflecting excitation from a wide range of emitting regions (broader to narrower from higher to lower level energies) has been the fundamental expectation of model predictions for some time \citep[e.g.][]{blevins16,woitke16}, and this multi-wavelength high-resolution dataset directly confirms it for the first time.
We will combine the picture emerging from line kinematics to that emerging from line excitation in Section \ref{sec: slab_fits}.

\begin{figure}
\centering
\includegraphics[width=0.45\textwidth]{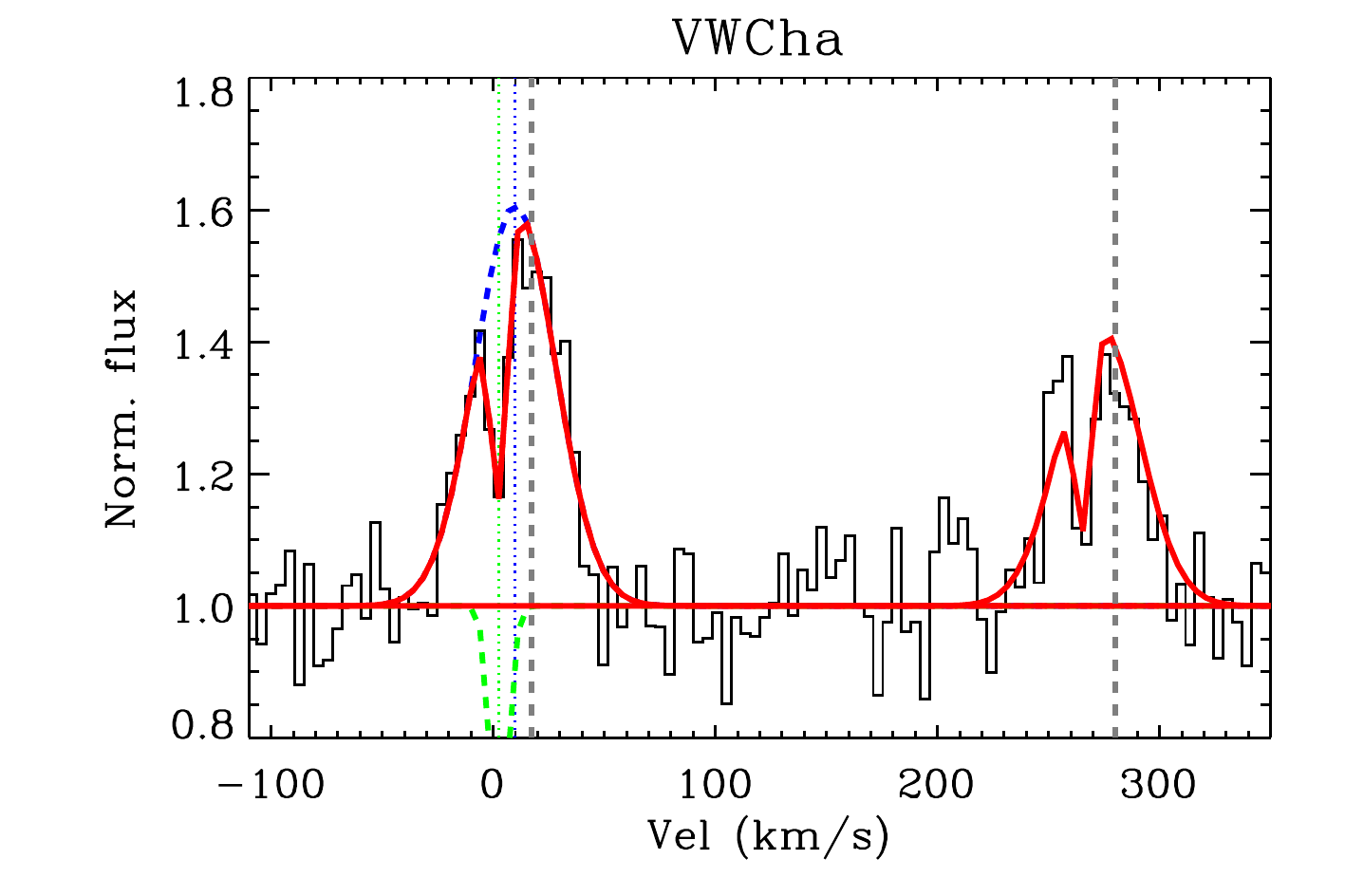} 
\caption{Emission and absorption components in VW~Cha, as observed with VISIR in two water lines at 12.3962 and 12.407~$\mu$m (with velocity axis centered on the stronger line). Relative to the stellar RV (dashed vertical line near each water transition), the Gaussian component in emission peaks at -7.3~km/s, and the absorption at -14.5~km/s. The absorption line width is consistent with being unresolved (10~km/s).}
\label{fig: VWCha_wind}
\end{figure}

\subsubsection{Blue-shifted absorption in \ce{H2O} spectra} \label{sec: wat_absorpt}
Two peculiar cases to note in the context of water line profiles and their kinematic structure are those of VW~Cha and V1331~Cyg (Figures \ref{fig: H2O_iSHELL_profiles_fit} and \ref{fig: H2O_VISIR_profiles_fit}). The water spectrum observed with VISIR in VW~Cha shows a potential narrow absorption line blue-shifted from the peak of the emission line (Figure \ref{fig: VWCha_wind}). This emission+absorption structure is observed similarly in both water lines that are covered and detected in this object, and the absorption line is much narrower than the telluric lines (Appendix \ref{app: visir_survey}); we cannot attribute this feature to any artifact possibly present in the VISIR spectra at the time of observation, and we present it here as a potential first detection of a blue-shifted absorption feature in mid-infrared water spectra from disks. It should be noted that highly blue-shifted [NeII] emission at -150 and -40 km/s has been observed in VW~Cha in this same VISIR survey \citep{pascucci20}, demonstrating the presence of an outflow that could also be linked to the absorption feature in the water lines (although if that is the case, [NeII] would trace an atomic part of the outflow at higher velocities). 

A unified scenario for emission and absorption in the context of an inner disk wind has been proposed to explain blue-shifted narrow absorption lines that are observed on top of CO emission lines in other disks \citep{pont11,banz22}, conditions that are found since earlier stages of disk evolution with additional complexity due to envelopes and multiple absorption components \citep{herczeg11}. 
A similar blue-shifted absorption is possibly detected in the $M$-band water spectrum of V1331~Cyg (Figure \ref{fig: H2O_iSHELL_profiles_fit}), a known source of a large-scale jet and molecular outflow \citep{mundt98,wu04}. In this case, the blue-shifted water absorption is very similar to the blue-shifted CO NC in its centroid and FWHM. 
Additional epochs of high-resolution and high-sensitivity observations are required to further investigate these potential absorption components in infrared water spectra.

\begin{figure*}[ht]
\includegraphics[width=1\textwidth]{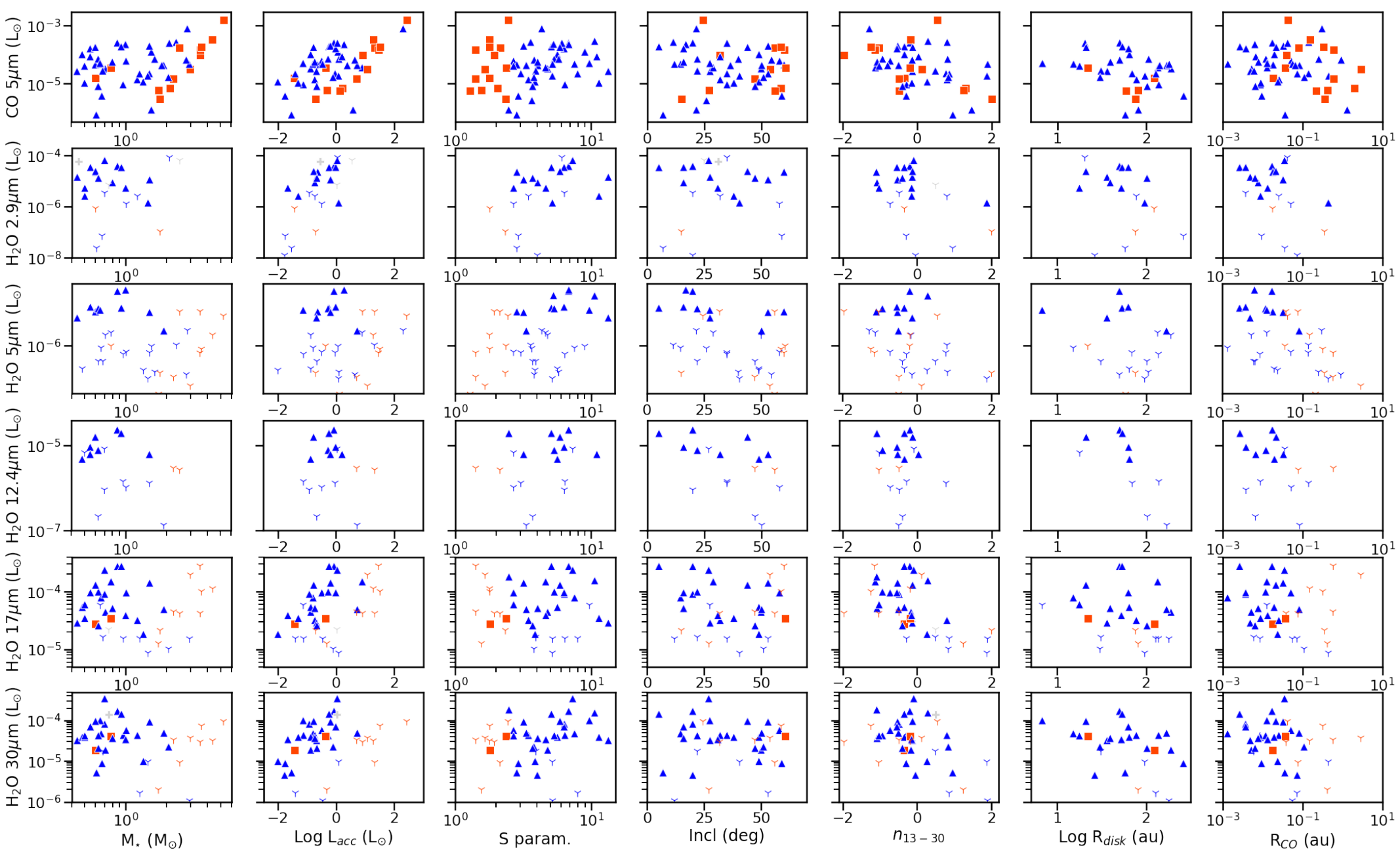} 
\caption{Overview of line luminosity for CO and water lines between 2.9 and 30~$\mu$m, as a function of stellar and disk parameters (Section \ref{sec: h2o_trends}). Datapoints are classified according to the kinematic shape of the CO line using a line shape parameter $S$ \citep{banz22}: ``triangular" lines with $S > 2.5$ associated to disk+wind emission (marked with blue triangles), and double-peak lines with $S < 2.5$ associated to purely Keplerian motion (marked with orange squares). Upper limits are marked with thin symbols. Line flux measurements included in this figure are reported in Appendix \ref{app: fluxes}.}
\label{fig:corr_grid}
\end{figure*}

\subsection{\ce{H2O} line detections and trends} \label{sec: h2o_trends}
A second point that we investigate in this work is the excitation of water transitions from different energy levels as a function of stellar and disk properties, addressing questions 1) and 3) from Section \ref{sec: intro}. Previous work identified the following trends from spectrally-unresolved water emission observed at 12--17~$\mu$m: 1) water detections are much higher in K and M stars as compared to earlier types \citep{pont10} and lower in M-dwarf disks \citep{pascucci13}, 2) water detections are lower in disks with inner dust cavities \citep{salyk11_spitz,banz17}, 3) water line fluxes correlate with stellar luminosity in T~Tauri stars \citep{salyk11_spitz} and, more strongly, with their accretion luminosity \citep{banz20}, 4) water line fluxes anti-correlate with the disk dust radius as spatially-resolved with mm interferometry \citep{banz20}. As noted before, the analysis of water spectra is intrinsically a multi-dimensional problem where multiple star/disk properties affect the observed emission \citep[see e.g. discussion in][]{banz20}.

We provide in Figure \ref{fig:corr_grid} an overview of these and other trends by considering spectrally-resolved water lines at 2.9, 5, and 12.4~$\mu$m, and spectrally-unresolved water lines at 17 and 30~$\mu$m (from Spitzer-IRS), including the $M$-band spectrally-resolved CO lines for reference. All the line flux measurements included in this figure are reported in Appendix \ref{app: fluxes}. The line luminosity for each row of plots in the figure is respectively as measured in: the low-$J$ $v=1-0$ CO lines \citep{banz22}, the three water lines around 2.9285~$\mu$m \citep{banz17}, the 5~$\mu$m stacked line (Figure \ref{fig: H2O_iSHELL_profiles_fit}), the 12.396~$\mu$m line (Figure \ref{fig: H2O_VISIR_profiles_fit}), and the few prominent water features around 17.25~$\mu$m and 30.7~$\mu$m included in the plots in Appendix \ref{app: extra_plots}. 
In this work we do not re-analyze correlations that have already been presented and discussed in previous work, but we find it useful to include the full grid of plots in Figure \ref{fig:corr_grid} to combine in one place tracers, samples, and trends that have been previously considered only separately, for future reference. In this section, we briefly describe the general trends and any notable differences between different tracers, and remark that a comprehensive view of this kind will be increasingly valuable in future work to correctly interpret individual spectra and samples from JWST observations within the broader context of the global structure and evolution of molecular gas in inner disks (see guidelines in Section \ref{sec: guidelines}). 

Since CO lines provide the highest detection rates in molecular lines tracing inner disks \citep[e.g.][]{salyk11,brown13,vdplas15} and a fundamental reference to interpret other molecular tracers, each datapoint in Figure \ref{fig:corr_grid} is classified according to the kinematic shape of its CO line using a line shape parameter $S$ = FW10\%/FW75\% (the ratio of the line width at 10\% and 75\% of the peak flux). Lines identified as ``triangular" have $S > 2.5$ (with broad wings and a much narrower line center, associated to disk+wind emission) and double-peak lines have $S < 2.5$ \citep[with typical line profiles associated to purely Keplerian, symmetric disk emission; see more on the definition and discussion of the line shape parameter in][]{banz22}.

\paragraph{Trends in CO luminosity}
For reference to trends observed in water emission, we first describe trends in the CO line luminosity $L_{\rm{CO}}$. $L_{\rm{CO}}$ as measured in this sample shows trends that are sometimes different or even opposite in the two types of lines (top of Figure \ref{fig:corr_grid}): in particular, in triangular lines $L_{\rm{CO}}$ presents a large scatter without an obvious trend as a function of $M_{\star}$, while in double-peak lines there is a strong positive correlation above 1~M$_{\odot}$ (and similarly is observed in  $L_{\rm{acc}}$). Other notable trends in $L_{\rm{CO}}$ are with disk inclination, infrared index $n_{13-30}$, and millimeter dust disk radius $R_{\rm{disk}}$; the latter, excluding the weakest objects that are disks with inner cavities, presents an interesting anti-correlation similar to that found in water in \citet{banz20}. The trends observed in $L_{\rm{CO}}$ are analyzed and discussed in Perez Chavez et al., in prep.

\paragraph{Trends in 2.9~$\mu$m water luminosity}
The 2.9~$\mu$m water sample was biased toward disks around T~Tauri stars with moderate to high accretion rates \citep{banz17}, resulting in perceived high detection rates ($\sim45\%$, Table \ref{tab:water_surveys}). The relatively small sample in this case does not allow to report strong trends, apart from a general agreement with those observed in CO.

\paragraph{Trends in 5~$\mu$m and 12.4~$\mu$m water luminosity}
The samples at 5~$\mu$m and 12.4~$\mu$m are larger but provided detection rates of only $\sim20\%$; despite that, some differences from the trends observed in CO emerge clearly. Water detections are exclusively found in objects that have a triangular CO line shape. These objects have in common a moderate accretion luminosity (0.1-1~L$_{\odot}$), T$_{\rm{eff}} < 6000$~K, detection of jets and disk winds, and CO emitting from within the dust sublimation radius \citep[see more discussion of triangular lines in][]{banz22}. The upper limits measured in the 5~$\mu$m water lines in some disks around intermediate-mass stars appear in stark contrast with their high CO luminosity, suggesting a different C/O ratio in these disks. 

\paragraph{Trends in 17~$\mu$m and 30~$\mu$m water luminosity}
Water line detections and luminosity at 17~$\mu$m and 30~$\mu$m, tracing lower-excitation transitions at 2400--3300~K and 1800~K respectively, are still predominantly detected in disks with a triangular line shape. An important difference with the higher-excitation water lines at shorter wavelengths is the detection in a Herbig Ae disk \citep[HD~163296, one of the few intermediate-mass stars that has a triangular line shape in CO, which has been associated with an inner disk wind, see][]{heinbert16_var,banz22} and in a few disks that have an inner dust cavity (DoAr~44, TW~Hya, SR~9, SU~Aur). Detection of water in Spitzer-IRS spectra of these disks was previously reported in \citet{fedele12,salyk11_spitz,banz17}. These cases demonstrate that rotationally-cold water emission can still be observed in some disks with dust cavities and disks around Herbig Ae stars (plots of multi-wavelength water spectra are included for reference in Appendix \ref{app: extra_plots}).

\paragraph{General trends in water luminosity}
Across different energy levels (and wavelengths), the water luminosity is in general higher in T~Tauri disks than in Herbig disks. The non-detection of water in Herbig disks reflects a true decrease in luminosity and is not simply a S/N issue. In T~Tauri disks, the water luminosity increases with accretion luminosity and with line shape parameter $S$, and decreases with disk inclination, infrared index $n_{13-30}$, millimeter disk radius $R_{disk}$, and radius of CO emission $R_{\rm{CO}}$ \citep[estimated from the line width at 10\%, as a tracer of the inner emitting radius for warm molecular gas; see][for details]{banz22}.
The higher detection rates of water in spectral lines with larger $S$ values (e.g. AS~205~N and DR~Tau) as observed at high-resolution, therefore, is not simply due to the higher line-to-continuum contrast. Broader lines with lower $S$ (e.g. in RNO~90) are more blended with nearby lines (e.g. CO, in $M$-band spectra) and do require a higher S/N per pixel to be detected, implying that their detection rate will be lower. However, two trends suggest an intrinsically decreasing water (and CO) luminosity due to 1) the $S$ parameter itself, where triangular lines with $S > 5$ have a median luminosity that is $\approx 5$ times higher than triangular lines with $S < 5$, and 2) the viewing angle, where disks with $incl < 25$~deg have a median luminosity a few times higher than triangular lines with $incl \approx$~30--50~deg. Both trends suggest that the geometry of emitting regions as viewed under different inclinations determines the observed properties of molecular spectra \citep[see also Figure 6 in][]{banz22}.

\begin{figure*}[ht]
\includegraphics[width=1\textwidth]{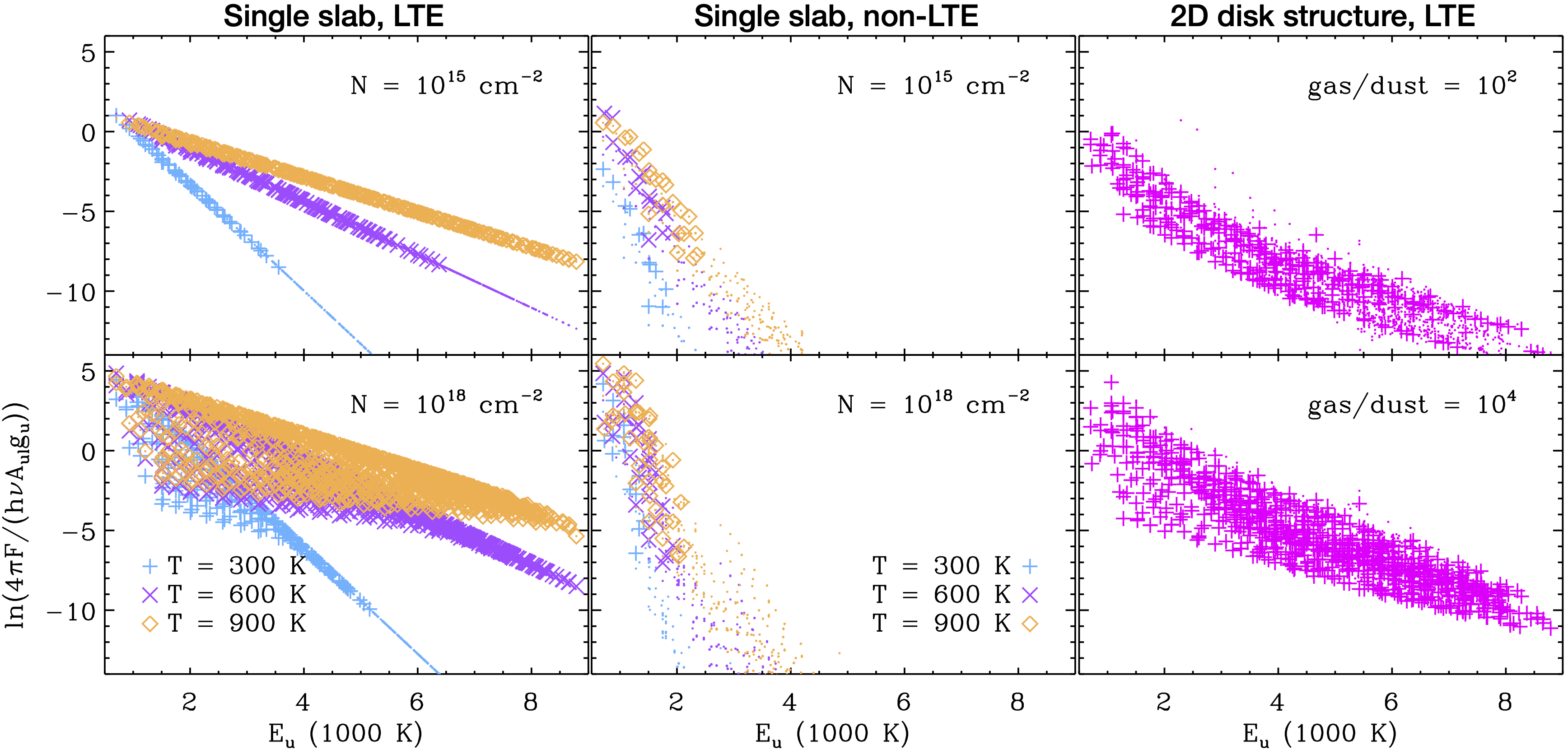} 
\caption{Rotation diagrams of infrared water rotational lines (Section \ref{sec: rot_diagr_models}). Three different models are used to illustrate the effects of different excitation temperatures, line opacity, and LTE/non-LTE excitation: a slab model in LTE \citep{banz12}, a slab model in non-LTE using RADEX \citep{vdt07}, and a two-dimensional disk structure in LTE using RADLite \citep{pont09}. In each plot and model, larger symbols mark transitions that are stronger than 1\% of the strongest line, to simulate a detection threshold; weaker lines are marked with small dots.}
\label{fig: rot_diagr_models}
\end{figure*}

\subsection{Water rotation diagrams} \label{sec: rot_diagr_models}
The third point we address in this work is the relative excitation of water spectra across different bands and transitions, addressing questions 1) and 3) from Section \ref{sec: intro}. To do so, we describe the use of rotation diagrams of infrared water emission, which for the first time can be analyzed from spectrally-resolved emission covering upper level energies between 4000 and 9500~K.
The rotation diagram technique has been thoroughly described for space applications in \citet{GL99}, but has never been systematically applied to water emission from protoplanetary disks due to the lack of spectrally-resolved line fluxes over a broad range in Einstein-$A$ coefficients $A_{ul}$ and in upper level energies $E_{u}$. While this is not necessary with linear molecules in the simplest excitation conditions, covering a large range in $E_{u}$ and $A_{ul}$ is essential in the case of an asymmetric top polyatomic molecule like water, especially in circumstellar disks where the emission is not optically thin and non-LTE conditions may sub-thermally excite lines with high $A_{ul}$ \citep{meijerink09,herczeg12,kamp13}. Below, we summarize previous explorations by \citet{banz13} to describe how different conditions (line opacity, LTE/non-LTE excitation) produce distinct features in water rotation diagrams.

\subsubsection{LTE case}
Rotation diagrams are defined such that, in conditions of optically thin emission at a single excitation temperature and in local thermal equilibrium (LTE), the fluxes from lines at frequency $\nu$ observed from an unresolved point source form a straight line as \citep[see][]{larsson02}:
\begin{equation}\label{eqn:rotdiagr}
ln\left( \frac{4 \pi F}{h \nu g_{u} A_{ul}}\right) = ln\left( \frac{N \Omega}{Q(T)}\right) - \frac{E_{u}}{T} \, ,
\end{equation}
where $F$ is the integrated line flux, $\Omega$ is the solid angle $A/d^2$ ($A$ the emitting area of the source and $d$ the distance to the source), $g_{u}$ is the statistical weight of the upper level, $Q(T)$ is the partition function. Throughout this paper, we use $E_{u}$ in units of K and cgs units for the rotation diagram.
The excitation temperature $T$ and the column density $N$ can be easily derived from the slope and intercept of a linear fit using Equation \ref{eqn:rotdiagr}. 
This technique is useful, however, even when the emission becomes optically thick and/or the excitation deviates from LTE. In fact, in these cases the observed molecular line fluxes would produce specific curvatures and/or a large vertical spread in line fluxes in the diagram departing from the linear behavior given by the LTE optically thin case, as described in the following. 

\begin{figure*}[ht]
\includegraphics[width=1\textwidth]{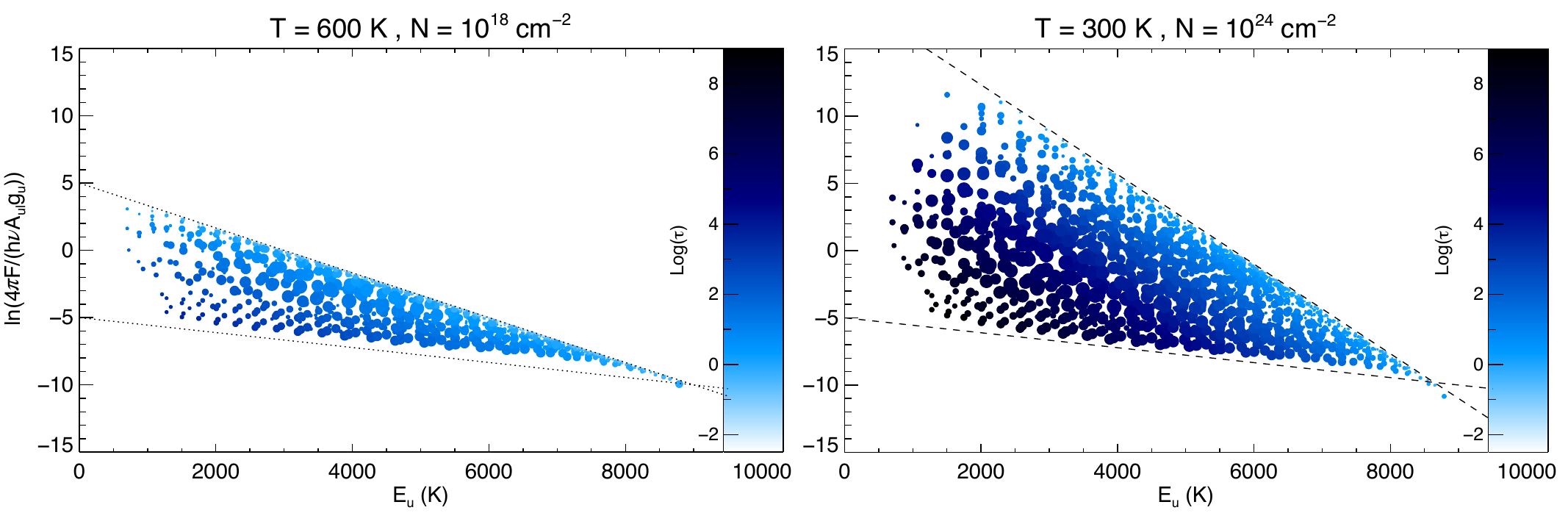} 
\caption{Similar to Figure \ref{fig: rot_diagr_models}, but including two more properties: the line intensity proportional to dot sizes, and the line opacity $\tau$ (proportional to $A_{ul}$) in blue-color scale. Moderate to high column densities produce optically thick emission and increase the vertical spread of lines in the diagram. The model to the right assumes a very high column density to better illustrate the effect of the spread in line opacities.}
\label{fig: lte_mod_fancy}
\end{figure*}

The basic properties of rotation diagrams of infrared water emission are displayed in Figure \ref{fig: rot_diagr_models} using a single-slab model in LTE \citep[described in][]{banz12}, with the temperature $T$ and the column density $N$ as free parameters. For this demonstration, we adopt a reference distance of 140~pc and a thermal line broadening of 1~km/s, and we fix the slab surface area $A$ to a disk with radius 1~au.
In the optically thin case ($N=10^{15}$ cm$^{-2}$, top left in the figure), water lines form a straight line in the diagram, with slope set by the temperature. It is worth noting that higher $T$ increasingly populate transitions from higher energy levels, as indicated by the larger symbols in the figure. When the column density is increased, instead, line opacities increase and lines spread over the diagram following a specific pattern. When a line gets optically thick ($\tau\gg1$) it ``freezes" on the diagram, i.e. its intensity is only weakly dependent on the column density. This happens first to those lines with large $A_{ul}$ (the low-energy, bottom left corner of the diagram, see also Figure \ref{fig: lte_mod_fancy}), as the line opacity is proportional to $A_{ul}$. The optically thin lines, instead, still follow Equation \ref{eqn:rotdiagr} and can rise in the diagram together with $N$. 

Therefore, unlike the case of a linear molecule like CO where a curve is produced, an increase in optical depth for the non-linear water molecule spreads lines vertically in the diagram. The most optically thin lines (those with the lowest $A_{ul}$ at any given bin in $E_{u}$) define the upper edge of the spread, while the most optically thick ones progressively set the middle and lower edge (Figure \ref{fig: lte_mod_fancy}). In these conditions, the vertical spread in the diagram is primarily sensitive to the column density. However, as Figure \ref{fig: lte_mod_fancy} shows, line intensities are larger at the center of the diagram, indicating that low-sensitivity spectra may only detect part of the full rotation diagram (as it typically happens with ground-based instruments, see Section \ref{sec: slab_fits}).

\subsubsection{Non-LTE case}
To explore non-LTE excitation, we simulate the emission from a slab of gas using the RADEX code \citep{vdt07}, which accounts for the volume density of the collisional partner H$_{2}$. RADEX uses molecular data from the LAMDA database \citep{lamda} complemented with data from \citet{tennyson01,faure08,dubernet09}. To better illustrate the effects of non-LTE excitation we adopt $n$(H$_{2}$) = $10^{\rm 6}$ cm$^{-3}$, where water emission is far from being thermalized by collisions \citep[critical densities for infrared water lines are at least $n$(H$_{2}$) = $10^{\rm 8}$ cm$^{-3}$;][]{meijerink09}. We remark that this is not meant to be an exhaustive exploration but rather a quick reference to the LTE case; a more extensive discussion of non-LTE excitation can be found in \citet{meijerink09}. 

Overall, the rotation diagram shape produced in non-LTE is different from the LTE case (see Figure \ref{fig: rot_diagr_models}): it displays larger spread or curvature and it produces steeper slopes due to sub-population of upper levels, which if observed in real data could be interpreted as rotationally cold emission in a ``quasi-thermal" behavior \citep[see also][]{herczeg12}. It is worth noting in this case that deviations from the LTE case become increasingly apparent at higher energy levels, because critical densities generally increase with $E_{u}$.

\begin{figure*}[ht]
\includegraphics[width=1\textwidth]{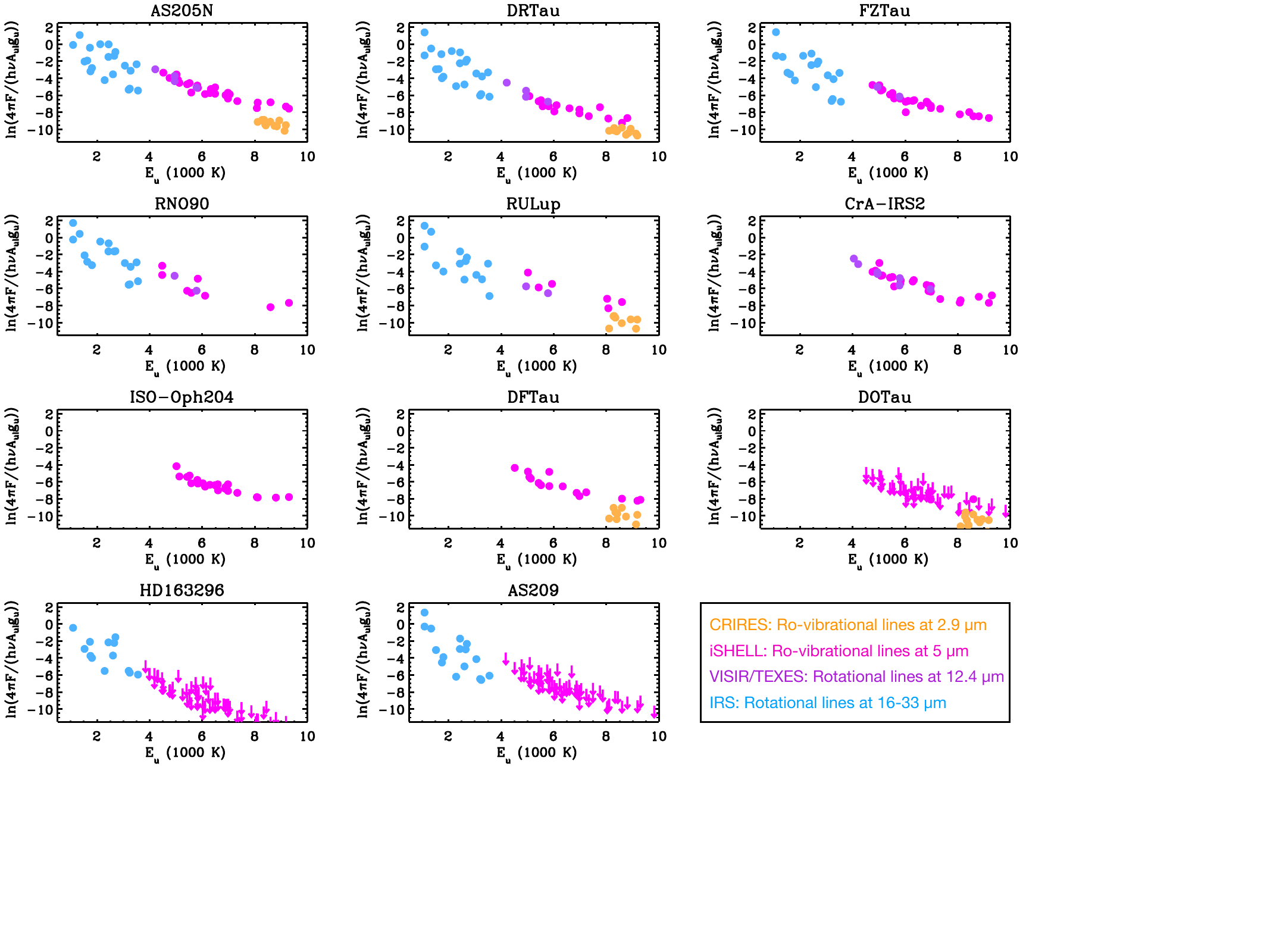} 
\caption{Rotation diagram of water emission as observed at multiple wavelengths in a selection of disks with water detections. Line flux measurements are reported in Appendix \ref{app: fluxes}. The last two plots at the bottom are examples of rotationally-cold water detected in Spitzer-IRS spectra (see also the multi-wavelength plots in Appendix \ref{app: extra_plots}).}
\label{fig: rot_diagr_grid}
\end{figure*}

\subsubsection{Disk structure in LTE}
To explore a more realistic disk structure, we adopt the two-dimensional RADLite code \citep{pont09}. RADLite is a raytracer for infrared molecular emission from circumstellar disks, based on the dust temperature and density structure calculated self-consistently using the RADMC code \citep{radmc}. RADLite accounts for dust and gas opacities, and here we set the gas temperature equal to the dust temperature and LTE excitation for simplicity. In modeling the water emission, RADLite takes into account the effects of a snow line set by the temperature and density structure of the disk \citep{blevins16}. We set the water abundance to solar oxygen values ($\approx 10^{-4}$ relative to hydrogen) inward of the snow line (where $T>170$ K), and a low, constant value of $10^{-9}$ per hydrogen outside of it to simulate freeze-out onto dust grains. For illustration, as a reference model we assume a 0.01 $M_{\odot}$ flared disk around a 1 $M_{\odot}$ star and gas-to-dust ratios (GTD) of $10^2$, to mimic ISM conditions, and as high as $10^4$ to mimic dust settling as in \citet{meijerink09}.

The most distinct feature that can be observed in rotation diagrams of water emission from RADLite, as compared to the slab models, is a curvature due to the contribution of a range in temperature and density from different disk radii (Figure \ref{fig: rot_diagr_models}, plots to the right). Models also show that the gas-to-dust ratio determines the spread of lines in the rotational diagram, by regulating the column of water observable above the dust continuum. Even in this case, the vertical spread in the rotation diagram is linked to the spread in observed line opacities, similarly to slab models, and indicative of the observed water column density. The shape and spread of lines in the rotational diagram of water vapor emission can therefore be useful to generally evaluate the contributions from different effects, provided that the observations cover a large enough range in $E_{u}$ and $A_{ul}$.

\begin{deluxetable*}{l c c c c c c c c c c c}
\tabletypesize{\footnotesize}
\tablewidth{0pt}
\tablecaption{\label{tab: slab_results} Emission properties from LTE slab model fits at different wavelengths.}
\tablehead{\colhead{Object} & \multicolumn{3}{|c|}{H$_{2}$O 2.9\,$\mu$m} & \multicolumn{3}{c|}{H$_{2}$O 5\,$\mu$m} & \multicolumn{3}{c|}{H$_{2}$O 12.4\,$\mu$m} & \multicolumn{1}{c|}{CO 5\,$\mu$m - BC} & \multicolumn{1}{c}{CO 5\,$\mu$m - NC}
\\ 
 & FWHM & $T$ & log~$N$ & FWHM & $T$ & log~$N$ & FWHM & $T$ & log~$N$ & FWHM  & FWHM 
\\
 & (km/s) & (K) & (cm$^{-2}$) & (km/s) & (K) & (cm$^{-2}$) & (km/s) & (K) & (cm$^{-2}$) & (km/s) & (km/s)  }
\tablecolumns{12}
\startdata
AS~205~N & 28 & 1200 & 8.2e17 & 40 & 1140 & 3.4e18 & 23 & 780 & 2.0e18 & 58  & 16  \\
CrA-IRS~2 & -- & -- & -- & 45 & 970 & 2.5e18 & 33 & 790 & 2.0e18 & 60  & 18  \\
DF~Tau & (80) & -- & -- & 100 & 1880 & 2.1e18 & -- & -- & -- & 113   & 49  \\
DO~Tau & (70) & -- & -- & (70) & nc & nc & -- & -- & -- & 70   & 19   \\
DR~Tau & 27 & 970 & 1.7e18 & 22 & 1160 & 1.2e18 & 16 & nc & nc & 33  & 12  \\
FZ~Tau & -- & -- & -- & 40 & 1060 & 2.5e18 & 24 & 730 & 6.0e16 & 51  & 19  \\
IRAS~04303 & -- & -- & -- & -- & -- & -- & 100 & 585 & 2.3e18 & (200) &  (90)  \\
ISO-Oph204 & -- & -- & -- & 40 & 1160 & 1.1e18 & -- & -- & -- & 54  &  22   \\
RNO~90 & -- & -- & -- & (100) & nc & nc & 83 & (810) & (7.4e18) & 110   & 62   \\
RU~Lup & (65) & -- & -- & (50) & nc & nc & (56) & nc & nc & 77  &  26   \\
VV~CrA~S & -- & -- & -- & -- & -- & -- & 25 & 840 & 2.2e18 & 76 &  19   \\
VW~Cha & (80) & -- & -- & -- & -- & -- & 44 & 695 & 6.3e17 & 72  &  26   \\
\enddata
\tablecomments{Slab models corresponding to the best-fit values reported in this table are shown for a few examples in Figures \ref{fig: crires_spec}, \ref{fig: ishell_spec}, and \ref{fig: visir_spec}. ``nc": model fit not converged, typically for the scarcity of line flux detections. Values in parentheses have an uncertainty larger than 20\%. Typical uncertainties for the best-fit results are of the order of 100~K in $T$ and a factor of a few in $N$. Measurements of CO line widths are given in the last four columns for reference to water emission properties (see Figure \ref{fig: concl_figure}) and are obtained from iSHELL spectra, except for VV~CrA~S and VW~Cha that are from CRIRES spectra. The range of CO emitting radii for BC and NC, assuming a purely Keplerian interpretation of the measured FWHM, can be found in \citet{banz22}.}
\end{deluxetable*}

\subsection{Observed rotation diagrams and LTE slab fits} \label{sec: slab_fits}
After briefly describing general properties of the technique in the previous section, we now illustrate in Figure \ref{fig: rot_diagr_grid} for the first time rotation diagrams of spectrally-resolved infrared water emission at energies of 4000--9500~K, for the small sample of disks where the emission is detected in this energy range (see Section \ref{sec: h2o_kinematics}). Previous attempts to retrieve line fluxes over a similarly large range were done by de-blending Spitzer-IRS spectra in \citet{banz13}, but were strongly limited by the low resolution of the data and blending between lines from different levels and molecules. Here, we simply measure and include in Figure \ref{fig: rot_diagr_grid} the total flux from $\sim20$ lines as observed in Spitzer-IRS spectra to extend the observed rotation diagram down to 1000~K, but we warn the reader that this is done for illustration only and that all the measured fluxes are blends of multiple transitions. This problem will be solved with JWST-MIRI spectra, as discussed below (Section \ref{sec: guidelines}).

The largest portion of the observed rotation diagrams is provided by $M$-band spectra, which cover the largest number of water lines from a single high-resolution spectrum (Section \ref{sec: data}) and energies between 4000~K and 9500~K. In this work, these lines are observed in the iSHELL spectra, which are flux calibrated using WISE W2 photometry at 4.6~$\mu$m \citep{allwise}. The high-energy end of these lines overlaps with lines extracted from $L$-band spectra previously observed with CRIRES (flux calibrated using WISE W1 photometry at 3.4~$\mu$m), while the lower-energy end overlaps with the rotational lines observed with VISIR and TEXES (flux calibrated using Spitzer-IRS, where available, or WISE W3 photometry at 12~$\mu$m). Overall, spectral lines from different instruments overlap well in the rotation diagram of each object apart from the $L$-band lines, which look under-excited as compared to the $M$-band lines in all disks. While real variability and the non-simultaneous flux calibration may offset the emission as observed with different instruments at different times, the apparent systematic offset between ro-vibrational bands suggests a difference in their excitation (see Section \ref{sec: discuss}).

Considering line fluxes from different instruments, the observed rotation diagrams show an overall curvature that is reminiscent of what is found from the 2D disk model in Figure \ref{fig: rot_diagr_models} as due to emission from a range of temperatures at different disk radii. These curvatures could not be seen before in previous work that only covered $< 10$~lines in a narrow portion of the diagram at 4000--6000~K \citep{pont10b,salyk19}, and are now most evident in the $M$-band spectra alone, as well as in combination with the Spitzer-IRS spectra. In addition to a curvature, lines from the Spitzer-IRS spectra spread vertically as expected in case of large water column densities (Figure \ref{fig: rot_diagr_models}). The vertical spread is more visible than in the case of spectra from ground-based instruments, because these have more stringent sensitivity limits and only detect stronger lines close to the upper edge of the rotational spread (Figure \ref{fig: lte_mod_fancy}). By better de-blending line fluxes at mid-infrared wavelengths and detecting weaker lines with lower $A_{ul}$, the analysis of JWST spectra will reveal the shape and spread in rotation diagrams in higher detail (see Section \ref{sec: discuss}).

\subsubsection{LTE slab fits}
To investigate the excitation of water in different bands and energies, we fit the measured line fluxes with the same LTE slab model adopted above in Section \ref{sec: rot_diagr_models}. The fit is done on the line fluxes directly by simulating spectra with the resolving power and observed line FWHM from each instrument, and minimizing the chisquare between measured and model line fluxes over the same individual spectral window around each line (the flux included within twice the FWHM of each line). The model includes line opacity and accounts for line blending as observed at the resolution of each instrument, rather than assuming optically thin emission to perform a linear regression fit in the rotation diagram. 

\begin{figure*}[ht]
\includegraphics[width=1\textwidth]{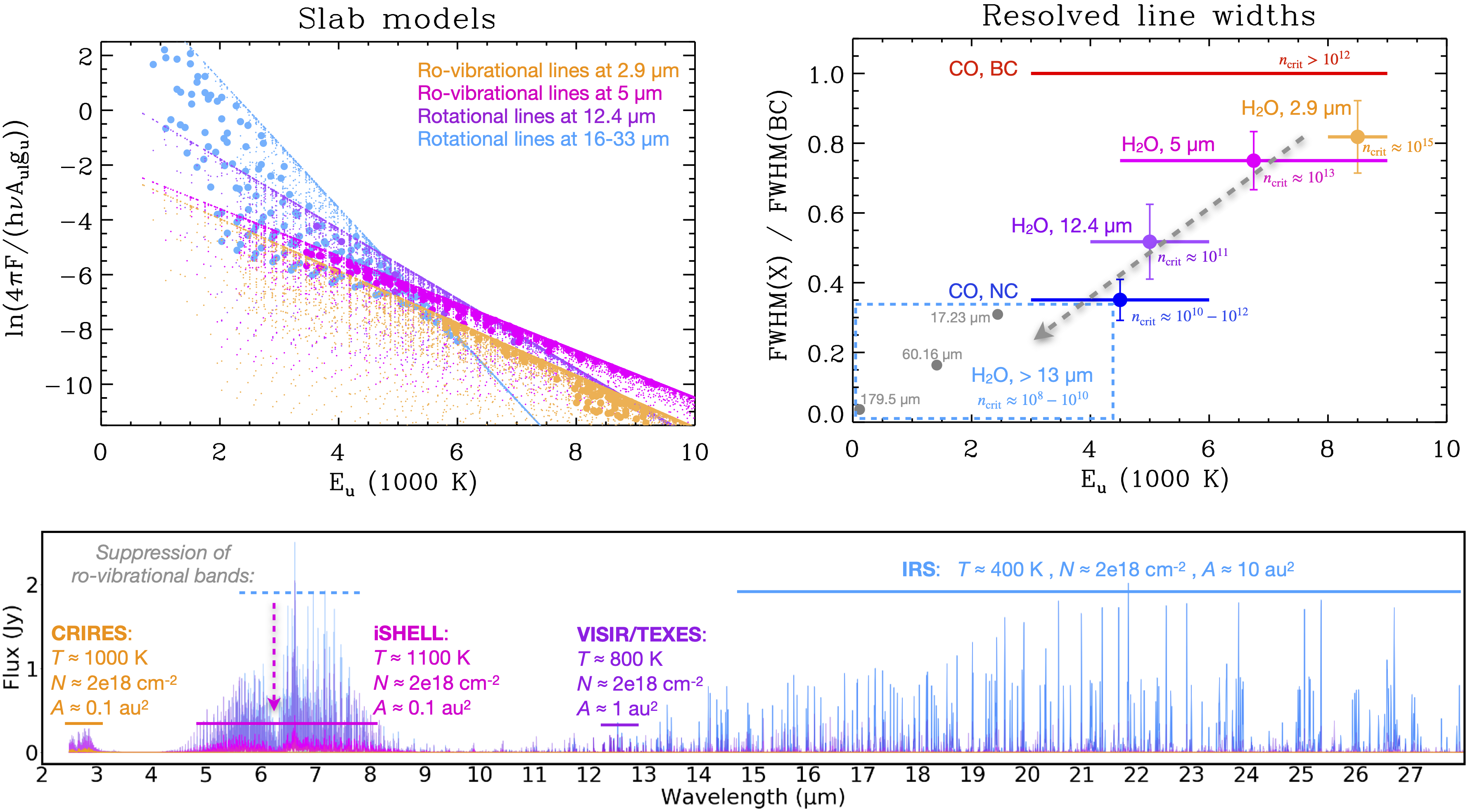} 
\caption{Overview of excitation and kinematics of water emission as spectrally-resolved at multiple wavelengths in this work. \textit{Top left and bottom}: representative LTE slab models from results in Table \ref{tab: slab_results}; the overall curvature in the observed rotation diagrams (Figure \ref{fig: rot_diagr_grid}) can be approximated by a series of slab model fits to the different wavelengths covered by different instruments (Section \ref{sec: T_grad}). Non-LTE excitation may cause the suppression of ro-vibrational bands as compared to slab model fits to the rotational lines (see Section \ref{sec: non-LTE}). \textit{Top right}: observed gradient in line widths as a function of upper level energy for water and CO (Section \ref{sec: FWHM_grad}); grey datapoints at 100--2400~K are model predictions for three water lines in RNO~90 using RADLite, from \citet{blevins16}. Critical densities $n_{\rm{crit}}$ are reported for reference (see Table \ref{tab:water_surveys}).}
\label{fig: concl_figure}
\end{figure*}

Table \ref{tab: slab_results} reports the fit results as obtained from spectra at different wavelengths, which are illustrated above in a few examples in Figures \ref{fig: crires_spec}, \ref{fig: ishell_spec}, and \ref{fig: visir_spec}. We report results in terms of $T$ and $N$, which are most sensitive to the measured flux ratios between lines of different energy and Einstein-$A$ coefficients (especially in the moderately optically thick conditions found in the results); in terms of emitting area A, the parameter that most relies on the overall flux calibration, we find similar values of $\approx 0.1$~au$^2$ for the 2.9~$\mu$m and 5~$\mu$m ro-vibrational lines, and $> 1$~au$^2$ for the 12.4~$\mu$m and 16--33~$\mu$m rotational lines, indicating widely different emitting areas for different transition bands.

The $L$-band spectrum can be fitted in two disks only, due to the narrow spectral range available, the multiple spectral gaps introduced by telluric and photospheric lines, and severe blending of lines that are typically broad at these wavelengths \citep[see][]{banz17}. The best-fit excitation temperature is around 1000~K \citep[close to the 900~K previously estimated in a few disks in][]{mandell12} and the column density about $10^{18}$~cm$^{-2}$, indicating moderately optically thick emission. Here we measure the full line flux as dominated by a broad component, even though in a few cases there could be a weak contribution from a narrow component in the line peak \citep[see Figure 15 in][]{banz17}.

The $M$-band spectrum can be fitted in most disks where water is detected, but does not converge in case of very broad and very faint emission (DO~Tau, RNO~90, RU~Lup). The best-fit excitation temperature is around 1100~K and the column density about a few $10^{18}$~cm$^{-2}$, except for the case of DF~Tau where $T$ is possibly as high as 1900~K. The similar results found from the $L$-band and $M$-band spectra suggest that the ro-vibrational bands share a similar excitation, with the stretching mode slightly sub-thermally excited (which could explain the lower line fluxes in rotation diagrams in Figure \ref{fig: rot_diagr_grid}).

At 12.4~$\mu$m, the fit converges only in spectra where at least 3 lines are detected, while it does not converge in RNO~90 and RU~Lup with only 2 lines. The best-fit excitation temperature is overall around 700--800~K and consistent with what has been found in previous work fitting the 12.4~$\mu$m lines \citep{najita18,salyk19}, and the column density about a few $10^{18}$~cm$^{-2}$, again indicating moderate optical depth.

At longer wavelengths observed with Spitzer-IRS, temperatures are lower in the range of $\approx450$~K, as found in previous work \citep{salyk11_spitz}. As a comparison test, we have independently measured line fluxes at 16--33~$\mu$m from the IRS spectrum of DR~Tau and find a best fit with $T$ = 430~K and $N$ = 2.6e18 cm-2, a result that is consistent with that found previously in \citet{salyk11_spitz}.

\section{Discussion} \label{sec: discuss}
\subsection{The picture emerging from spectrally-resolved data} \label{sec: discuss_1}
The overall picture emerging from this analysis is summarized in Figure \ref{fig: concl_figure}, by combining results from Section \ref{sec: analysis} in terms of line excitation and kinematics as spectrally-resolved at multiple wavelengths. 
LTE slab model fits show that the infrared water spectrum may probe moderately optically thick emission from hotter inner regions exciting the higher energy levels, to colder outer regions populating the lower energy levels, qualitatively supporting general expectations from previous models (Section \ref{sec: intro}). The wide range of observable/observed inner disk water column densities ($10^{17}$--$10^{20}$~cm$^{-2}$) produced by previous works modeling water spectra from inner disks \citep{najita11,du14,walsh15,agundez18,woitke18} or estimated from spectrally-unresolved disk observations \citep{cn11,salyk11_spitz,liu19} demonstrates that this parameter, and more fundamentally the water vapor abundance as a function of disk radius and disk height (down to the disk midplane), are still to date critical unknowns \citep[see discussion in][]{bosman22}.

The similar column density found in slab fits to the spectra at different wavelengths (a few times $10^{18}$~cm$^{-2}$) across the sample in Table \ref{tab: slab_results} is therefore particularly interesting: recent models that include chemical heating and UV-shielding propose that this is the column density that ends up being probed in the disk surface due to the combination of the most prominent lines becoming optically thick, and the gas temperature steeply falling off at higher densities \citep{bosman22}. The results presented above suggest that water spectra across infrared wavelengths may trace gas where the right density for excitation is met at any given disk radius, resulting in a similar column density across radii but different excitation temperatures.

\subsubsection{A gradient in excitation temperature} \label{sec: T_grad}
The overall curvature observed in rotation diagrams (Figure \ref{fig: rot_diagr_grid}) can be approximated by a series of slab models fitted to the different wavelengths covered by different instruments (i.e. different ranges of energy levels), as reported in Table \ref{tab: slab_results}. In the rotation diagram in Figure \ref{fig: concl_figure}, similarly to Figure \ref{fig: rot_diagr_models}, we show with small dots each model including all transitions between 2.5 and 33~$\mu$m, and then highlight with larger dots the stronger transitions that can be observed in the settings covered by each instrument (the color code follows that of Figure \ref{fig: rot_diagr_grid}). This way, it can be seen that while all models present a large spread of lines due to the moderately large column density (Section \ref{sec: rot_diagr_models}), each instrument ends up observing only a fraction of stronger lines close to the upper edge of the rotational diagram (as observed in Figure \ref{fig: rot_diagr_grid}). The rotational lines at $>16$~$\mu$m are an exception, by tracing a larger part of the vertical spread in the rotation diagram thanks to the wider spectral coverage and higher sensitivity obtained from space.
 
The same slab models are illustrated in the lower part of Figure \ref{fig: concl_figure}, marking which spectral range each model is actually sensitive to due to the coverage of each instrument. For guidance, we report representative values for $T$, $N$, and $A$ for each model from the fits in Section \ref{sec: slab_fits}. The data demonstrate that no single slab model can reproduce the whole spectrum, an issue that was noticed early on from fits to the rotational lines in Spitzer spectra \citep{cn11,salyk11_spitz,liu19} and that this analysis now extends to observations of ro-vibrational bands at $< 10$~$\mu$m.

\subsubsection{Non-LTE vibrational excitation} \label{sec: non-LTE}
Further, it now becomes clear that the vibrational excitation is not in LTE: the fits to rotational lines at $> 12$~$\mu$m (from VISIR, TEXES, and Spitzer) strongly over-predict the ro-vibrational bands at 2.7~$\mu$m and 6.5~$\mu$m that are observed to be much weaker (using iSHELL and CRIRES). A similar effect is possibly seen also between the two vibrational bands, where the fit to the bending mode over-predicts the emission in the stretching modes. This effect has already been seen in the excitation of $M$-band CO emission: while an LTE model can reproduce quite well the CO spectrum in both BC and NC in the $v = 1-0$ lines, it over-predicts the emission observed from the higher vibrational bands, suggesting sub-thermal vibrational excitation \citep[see Figure 11 in][]{banz22}. Water seems to have a similar excitation: slab models can reproduce quite well the emission as observed over different limited spectral ranges and in different bands (Section \ref{sec: slab_fits}), suggesting that the rotational excitation is close to LTE \citep[as expected by models, e.g.][]{woitke18,bosman22}, but globally they cannot reproduce the vibrational excitation between the the ground state and higher states, implying non-LTE vibrational excitation. 

The same conclusion has been reached recently in thermo-chemical modeling of water in inner disks in \citet{bosman22}, who proposed that observing weaker emission in the 6.5~$\mu$m band as compared to the excitation of the rotational lines would be a signature of non-LTE excitation in low-density gas. This argument now suggests that it is the emitting layer of the rotational lines to be at gas densities lower than the critical density of ro-vibrational bands ($\approx 10^{13}$~cm$^{-3}$), because it is the slab model fit to the rotational lines that strongly over-predicts ro-vibrational emission at $<9$~$\mu$m. This is once again consistent with what is observed in CO: it is the NC (which matches the shape of water rotational lines) to show evident sub-thermal excitation, while the BC can be matched by an LTE model \citep[Figure 11 in][]{banz22}. 

The higher excitation temperature, larger FWHM, and smaller emitting area found for the ro-vibrational bands suggest that these must be excited in an inner, denser disk region at higher temperature. The excitation of different energy levels and bands, therefore, globally corresponds to the gradient in critical densities: from $10^{13} - 10^{15}$~cm$^{-3}$ for the water ro-vibrational bands and the high-$J$ ($E_{u} = 6000$--9000~K) $v = 1-0$ CO lines detected in BC, down to $10^{10} - 10^{11}$~cm$^{-3}$ for the water rotational lines near 12~$\mu$m and the CO NC at $E_{u} = 3000 - 6000$~K \citep[for CO critical densities, see][]{thi13,woitke16}. Water lines at longer wavelengths have lower critical densities of $10^{8} - 10^{10}$~cm$^{-3}$ \citep{meijerink09}.
These conditions are in accord with detailed laboratory pump-probe experiments that have determined the excitation/de-excitation of vibrational modes in water (and CO) to be orders of magnitude less efficient than the collisional processes that determine the equilibration of the rotational and translational degrees of freedom \citep{finzi77}.

\subsubsection{A gradient in FWHM and emitting regions} \label{sec: FWHM_grad}
The top right of Figure \ref{fig: concl_figure} summarizes the picture emerging from the spectrally-resolved line kinematics, using values reported in Table \ref{tab: slab_results}. We make this figure in a way to generalize the results to any other disk (see Section \ref{sec: guidelines}), assuming that the results from the $\approx10$ disks where water is detected may reflect general properties of Class II disks that do not have an inner dust cavity. For each line tracer X, in Figure \ref{fig: concl_figure} we report the median value of the sample distribution of the observed FWHM as normalized to the FWHM of the CO BC in each disk (FWHM(X)/FWHM(BC)). In this way, FWHM differences due to different viewing angles across the sample are removed, putting the figure in the general reference frame of the innermost CO gas observed in any disk (the BC). Each tracer in the figure has an horizontal bar that covers the range in energy levels covered at different wavelengths with the different instruments, and a vertical bar that is the median absolute deviation of the distribution of values across the sample from Table \ref{tab: slab_results}. For reference to the \ce{H2O} lines, we report also the CO BC (at a $y$-axis value of 1 by definition, by being normalized to itself in each disk), which is typically detected up to $E_{u} \approx 9000$~K in this sample, and the CO NC, which instead gets weaker more rapidly than BC at higher energy levels and is more representative of levels up to $E_{u} < 6000$~K \citep{banz22}. 

\begin{figure*}[ht]
\centering
\includegraphics[width=0.8\textwidth]{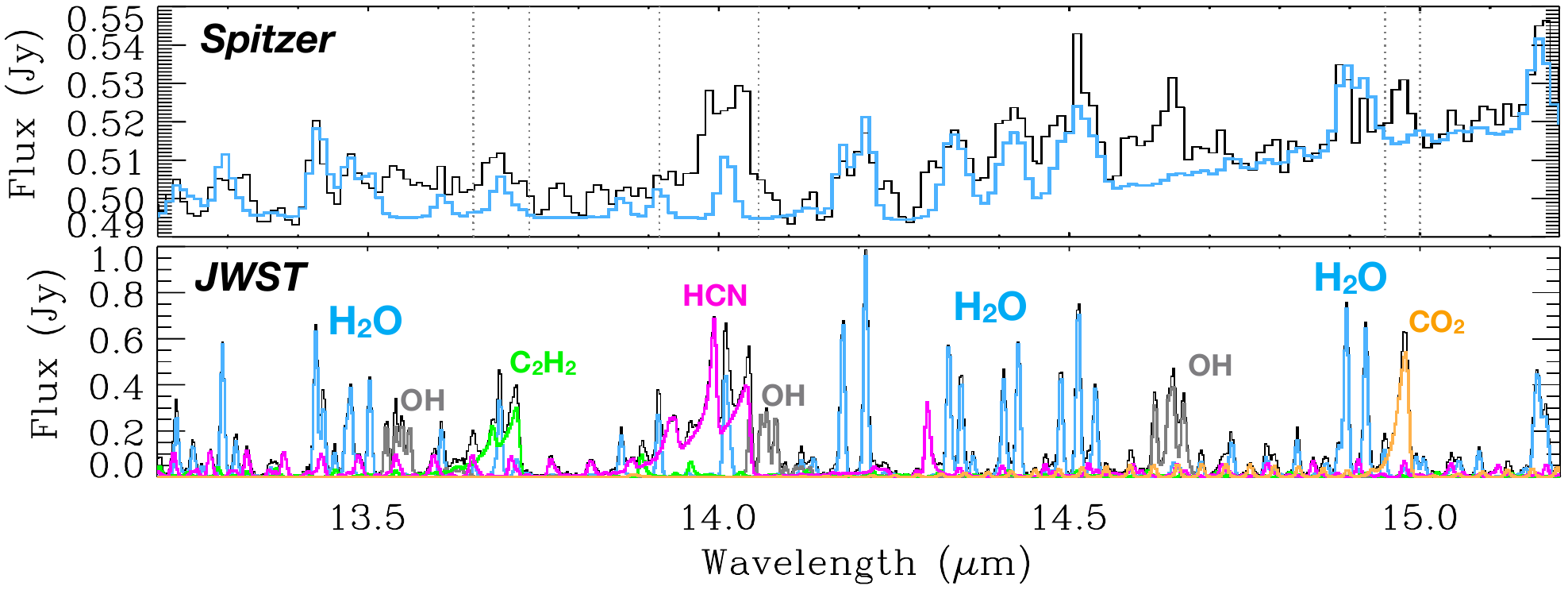} 
\caption{Example of a portion of a Spitzer-IRS spectrum (black, upper plot), compared to an LTE slab model of water emission (in light blue) from \citet{banz20}. The lower plot shows how the same water model would be observed at the resolution of JWST-MIRI, assuming S/N=100; representative models for other molecules previously identified in Spitzer spectra are included for visualization of the improved de-blending between different lines and molecular spectra.}
\label{fig: spitz_compar}
\end{figure*}

As reported above in Section \ref{sec: h2o_kinematics}, the spectrally-resolved water ro-vibrational lines at 2.9~$\mu$m and 5~$\mu$m have FWHM similar to that of the CO BC, while the water rotational lines at 12.4~$\mu$m are instead much narrower and similar to the CO NC, defining the global trend visible in the top right of Figure \ref{fig: concl_figure}. While the rotational lines at $> 13$~$\mu$m remain to be spectrally-resolved, we add in the figure the expected FWHM values for three lines (again divided by the FWHM of the BC) from levels down to 100~K as reported in \citet{blevins16} using RADLite to model mid- and far-infrared water emission in RNO~90. These previous model predictions, which assumed a radially decreasing temperature profile in a Keplerian disk, show a line FWHM increasing with $E_{u}$ due to higher energy lines tracing progressively hotter inner regions, consistent with the global trend emerging from spectrally-resolved water emission at higher $E_{u}$ in this work.
In the rest of this section, we discuss some important implications of this global picture for the analysis of spectrally-unresolved JWST spectra.

\begin{figure*}[ht]
\centering
\includegraphics[width=0.8\textwidth]{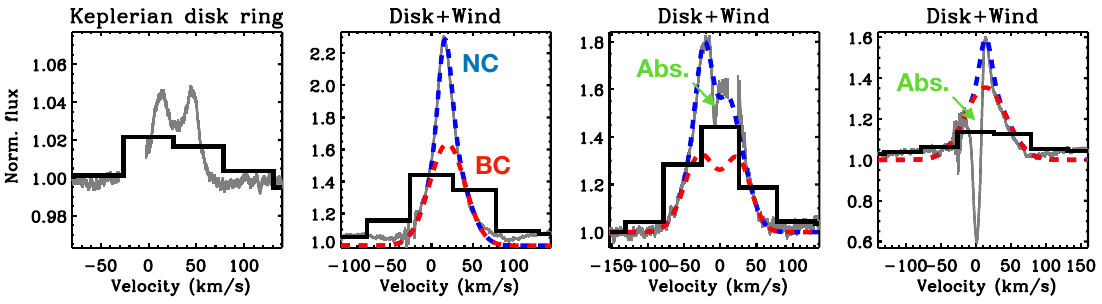} 
\caption{Gallery of representative CO line shapes from $M$-band spectra of inner disks observed with iSHELL \citep[data in grey, from][and this work]{banz22}, illustrating the loss of kinematic information and blending of different components when observed at the resolution of MIRI (in black, note how absorption and emission may cancel each other out in MIRI spectra).}
\label{fig: CO_profiles}
\end{figure*}

\subsection{JWST-MIRI spectra: opportunities and challenges} \label{sec: guidelines}
A total of $\sim 70$ Class II protoplanetary disks are going to be observed with JWST spectrographs in Cycle 1, most of them with the Mid-Infrared Instrument \citep[MIRI,][]{miri} providing R~$\approx$~1500--3700 (80--200~km/s) between 4.9 and 28~$\mu$m \citep{miri_res}. The wide spectral coverage of MIRI includes the ro-vibrational bending mode and rotational lines of water vapor emission in disks for the first time at a similar resolution in the two bands \citep[Spitzer-IRS did cover the water bending mode but only with the low resolution module with R $\approx$ 60--120,][]{sargent14}. The higher resolving power of MIRI will de-blend many of the emission features previously observed with Spitzer (see example in Figure \ref{fig: spitz_compar}). Fundamental expectations from MIRI spectra are therefore a much improved characterization of the water (and OH) spectrum and its excitation across near- and mid-infrared wavelengths, a better characterization of the emission from some organic molecules (although their ro-vibrational bands near 14~$\mu$m will still be severely unresolved, see Figure \ref{fig: spitz_compar}), and the discovery of features from additional molecules previously unidentified in Spitzer spectra. 

MIRI will not provide, instead, measurements of the gas kinematics and the distinction between double-peak (Keplerian) and triangular line shapes, between BC and NC (where present), and between emission and absorption components observed in inner disks (Figure \ref{fig: CO_profiles}). At most, MIRI is expected to partially resolve the wings of emission lines that are broader than $\approx 100$~km/s (Figure \ref{fig: CO_profiles}).
By combining advantages and disadvantages, we describe in the following a number of guidelines for future work to support the analysis of space data and propose how the combination with ground data may be used to obtain a comprehensive understanding of molecular spectra from inner disks.

\subsubsection{The emitting regions and excitation of H$_2$O}
From the data and analysis included in this and previous work, it is clear that infrared water spectra observed from inner disks trace a range of emitting regions producing different line widths and different excitation (Figure \ref{fig: concl_figure}). This picture is naturally expected by disk models (Section \ref{sec: water_overview}) and provided good fits to rotational lines using slab models that include a distribution of temperatures \citep{liu19}. Yet, the absence of spectrally-resolved observations, the several uncertainties in the physical and thermal structure of inner disks (both radially and vertically), and the possibility of non-LTE excitation made previous thermo-chemical modeling of infrared water spectra challenging and degenerate \citep[e.g.][]{meijerink09,kamp13,antonellini15,woitke18,bosman22}. 
Without direct support from spectrally-resolved line kinematics, the modeling of MIRI spectra will still have to essentially rely on fitting line fluxes from different $E_{u}$ and $A_{ul}$ to reconstruct the radial and vertical distribution of water vapor across disk radii. The analysis included in this work identifies some useful guidelines:
\begin{itemize}
    \item Rotation diagrams of observed water spectra will reveal large spread and curvature across energy levels (Figures \ref{fig: rot_diagr_grid} and \ref{fig: concl_figure}), indicative of optically thick emission (current results suggest a column density of a few $\times 10^{18}$~cm$^{-2}$) from a range of excitation temperatures ($\approx$~300--1100~K); MIRI spectra will be sensitive to detect a larger part of the spread of observed rotation diagrams and better measure the observable column density by detecting weak lines with low $A_{ul}$ (Figure \ref{fig: lte_mod_fancy}).
    
    \item Single-temperature slab models in LTE can be expected to provide good fits to JWST spectra over narrow ranges of $E_{u}$ and within the same band (e.g. the ro-vibrational bending mode around 6.5~$\mu$m), but the rotational spectrum at $> 10$~$\mu$m, which overall spans a large range in $E_{u}$ from a larger emitting area, will require accounting for gradients in excitation conditions \citep[at least a temperature radial gradient, e.g.][]{liu19}.
    
    \item Using the spectrally-resolved line shape and FWHM from ro-vibrational CO spectra as observed from the ground, which are available or currently being obtained for all JWST targets in Cycle 1 (see \url{www.spexodisks.com}), Figure \ref{fig: concl_figure} can provide general guidance on the expected FWHM for water lines as a function of $E_{u}$, which models can use to determine the emitting regions in disks. For a limited number of disks observed from the ground, this can be done by directly using spectrally-resolved water emission at multiple wavelengths (Table \ref{tab: slab_results}).
    
    \item The ro-vibrational bands at 2.7 and 6.5~$\mu$m show different temperature and non-LTE vibrational excitation as compared to the rotational lines at $> 10$~$\mu$m, and trace different disk regions as shown by their different FWHM (Section \ref{sec: analysis}); modeling of JWST spectra will need to account for these differences in excitation conditions.
    
    \item The infrared water spectrum can be expected to first approximation to be rotationally in LTE, but not vibrationally, similarly to what is observed in CO spectra (Section \ref{sec: discuss_1}); detailed analysis of MIRI spectra might reveal more subtle non-LTE effects in the rotational excitation too \citep{meijerink09}.
    
    \item Obtaining a good model fit to the observed water spectra will be important for the correct analysis of emission from other molecules too, e.g. the organics at 13--15~$\mu$m that are blended with water (Figure \ref{fig: spitz_compar}).
\end{itemize}

\subsubsection{CO- and water-rich inner disk winds?}
The kinematic structure of $M$-band CO lines observed at high resolution from most T~Tauri disks can be decomposed into two velocity components, BC and NC, that show very different FWHM and vibrational excitation \citep{bast11,bp15,banz22}. Of these components, NC has been associated with a disk wind due to a number of observed properties, most notably the common presence of blue-shifted asymmetries \citep{brown13,banz22} and an asymmetric spectro-astrometric signal detected in a few disks \citep{pont11}. The similar properties of NC across T~Tauri disks, once projection effects at different viewing angles are considered, and the presence of blue-shifted absorption components suggest that the NC may trace a slow molecular inner disk wind in T~Tauri disks in general \citep[see discussion in][]{banz22}. 

The similarity between the profile of NC in CO lines and that of water lines at 12.4~$\mu$m (Figures \ref{fig: H2O_VISIR_profiles_fit} and \ref{fig: TEXES_analysis}) therefore raises the question of whether water may trace the same molecular inner disk wind as CO. MHD disk wind models find that water and CO should be present in a low-velocity molecular region of inner disk winds \citep[e.g.][]{panoglou12,yvart16,wang19}, suggesting that the infrared molecular spectra could indeed trace a wind. The narrow absorption tentatively detected in water lines in two disks (Section \ref{sec: wat_absorpt}) may also suggest water-rich outflowing gas, similarly to the absorption observed in CO \citep{banz22}. The current dataset does not show a correspondence between CO absorption and water absorption in lines at $E_{u} \approx 4000$--6000~K (Figure \ref{fig: H2O_VISIR_profiles_fit} and Appendix \ref{app: extra_plots}), but water lines at $>13~\mu$m are yet to be spectrally-resolved to test whether any absorption may be present at lower $E_{u}$. The quality and number of high-resolution water spectra do not allow a definitive answer yet on water, but a few useful guidelines can be formulated for future work to study molecular inner disk winds at infrared wavelengths:

\begin{itemize}
    \item Absorption components, where present, will be blended to emission at the resolution of MIRI (Figure \ref{fig: CO_profiles}); MIRI spectra will observe, in these situations, weaker emission lines that will be particularly affected in case of deep absorption. These situations can be expected to be more frequent at high disk inclinations when the line of sight intercepts gas at or above the disk surface \citep[see Sections 4.6 and 5.3 in][]{banz22}; high-resolution CO spectra from the ground will support the identification of disks with line absorption, but it is still unclear how much that may affect the water spectra and over which energy levels (Section \ref{sec: wat_absorpt}).
    
    \item In absence of direct measurements of gas kinematics from spectrally-resolved water emission lines in individual disks, models may have to explore the distinction between disk emission and any emission from a wind from the excitation of water and CO spectra, from estimates of their density and temperature. Advancements in models to simulate molecular line profiles from molecular inner disk winds in protoplanetary disks \citep[e.g.][]{panoglou12,wang19,tabone20,rab22} may also provide more guidance in the future.
\end{itemize}

\subsubsection{Measuring \ce{H2O}/CO ratios in inner disks}
Another fundamental expectation from the analysis of JWST spectra is the possibility to characterize inner disk chemistry and its connections to planet formation. Models expect that the excitation of infrared molecular lines and bands will reflect chemical signatures (e.g. different C/O ratios) as well as dynamical processes (e.g. inner disk oxygen enrichment from icy pebble drift) in individual as well as in ratioed molecular tracers \citep[e.g.][]{najita11,bosman17,woitke18}. Some trends possibly related to different C/O ratios due to the drift of icy solids have been measured from water and HCN spectra observed with Spitzer \citep[][]{najita13,banz20}, but their interpretation is still tentative. A fundamental problem is still how to reliably estimate using disk models the column densities from observed spectra that only trace an upper disk layer, and in turn the elemental C/O ratio deeper down in the disk midplane \citep[e.g.][]{walsh15,woitke18,bosman22}.

MIRI covers both CO and water in a single spectrum \citep[e.g. Figure \ref{fig: spec_overview}; see first detection in][]{yang22}, avoiding uncertainties in flux calibration and variability that currently affect data obtained from the ground at different times and with different instruments (Section \ref{sec: analysis}). However, MIRI only covers the high-$J$ levels of $^{12}$CO lines at $>4.9$~$\mu$m (P25 and up, or $E_{u} > 4700$~K) and of $^{13}$CO (P15 and up, or $E_{u} > 3500$~K). These ranges of higher $J$ levels lead to large degeneracies in estimates of $T$ and $N$ because they cover a region in the rotation diagram that can be rather flat even when the emission is optically thick, while strong curvatures are only visible at lower $J$ levels \citep[see e.g. Figure 11 in][]{banz22}.
Measuring \ce{H2O}/CO column density ratios in inner disks will therefore be in principle possible from MIRI spectra, but it requires the identification of CO and \ce{H2O} lines that trace the same disk region and layer (Figure \ref{fig: concl_figure}); gaining a more global view of C/O as a function of disk radius will however require to model a radial density profile in CO and \ce{H2O} as observed at multiple wavelengths. Even here, the available spectrally-resolved data analyzed in this work provide some guidelines:

\begin{itemize}
    \item While ro-vibrational CO lines, when observed at high resolution, can generally be separated into two discrete components that show different kinematics and excitation (BC and NC), infrared water lines seem to show a more continuous range of line shapes and widths that can only partly be matched by BC and NC or their combination (Section \ref{sec: h2o_kinematics}), making it hard to identify specific lines from different molecules that trace the same disk region.
    
    \item Nonetheless, the rotational water lines with $E_{u} \approx 4000$~K, close to that of the 12.4~$\mu$m lines that almost entirely match the NC line shape (Figure \ref{fig: TEXES_analysis}), should provide a good match to the NC in CO; by considering the measured FWHM as a function of $E_{u}$ (Figure \ref{fig: concl_figure}); these lines could therefore be used to empirically measure \ce{H2O}/CO(NC) ratios and their dependence on different stellar and disk properties and their evolution (Figure \ref{fig:corr_grid}). The question remains of which region in a disk (or wind) layer these specific lines would trace.
    
    \item When observing CO in MIRI spectra, it should be kept in mind that only the high-$J$ lines at $E_{u} > 4700$~K are covered \citep{yang22}, and that these are typically dominated by the BC and are not a good match to the rotational water lines at $>10$~$\mu$m. To measure empirical \ce{H2O}/CO ratios from MIRI spectra alone, water lines from the ro-vibrational band at 6.5~$\mu$m and similar $E_{u} \approx 5000$~K may need to be used.
\end{itemize}

\section{Summary \& conclusions} \label{sec: conclusions}
In this work, we presented the largest dataset of spectrally-resolved water emission from Class II protoplanetary disks collected to date, as observed with multiple high-resolution (R~$\sim$~30000--100000) spectrographs at infrared wavelengths that are now covered by spectrographs on JWST (from 2.9~$\mu$m with NIRSpec to 5~$\mu$m and 12.4~$\mu$m with MIRI), for a total of $\approx160$ spectra from a sample of 85 disks (30 of which are targets of JWST programs in Cycle 1). We presented an analysis of the kinematics and excitation of water emission from ro-vibrational and rotational lines in different bands observable from the ground, with the goal of obtaining a comprehensive view of the distribution and excitation of water in inner disks.

A correct interpretation of water spectra from disks requires observations across IR wavelengths, supporting the key role that JWST is going to play in this field. This analysis shows that MIRI spectra will observe vertical spread and curvatures in rotation diagrams of water emission, by tracing a range of emitting regions in a moderately optically thick upper disk layer with excitation temperatures of 300--1100~K. We provided in Section \ref{sec: discuss} a list of guidelines to support the analysis of the large number of disk spectra to be observed with JWST, including how ground-based high-resolution data will be fundamental to provide the missing kinematic information.

The main findings of this work are:
\begin{itemize}
    \item ro-vibrational water lines at 2.9~$\mu$m and 5~$\mu$m ($E_{u}$ of 4500--10000~K) have larger FWHM and higher excitation temperature (1000--1100~K), while rotational lines near 12.4~$\mu$m ($E_{u}$ of 4000--6000~K) have narrower FWHM and lower excitation temperature (800~K); LTE slab model fits suggest that a similar column density is observed in all water bands (a few $\times 10^{18}$~cm$^{-2}$, Section \ref{sec: slab_fits});
    \item the new 5~$\mu$m spectra from iSHELL for the first time directly demonstrate that slab model fits to the rotational lines at $> 10$~$\mu$m strongly over-predict the ro-vibrational emission bands at $< 9$~$\mu$m; this is interpreted as evidence for non-LTE vibrational excitation, as discussed in \citet{bosman22};
    \item the picture emerging from combining line kinematics and excitation is in good agreement with the range in critical density for the different bands: 1) an inner, hotter, and denser disk region exciting the ro-vibrational water bands and the broad CO component, and 2) a progressively cooler disk surface (and low-velocity region in an inner disk wind) exciting the rotational lines and the narrow CO component (with density $< 10^{12}$~cm$^{-3}$ so as to sub-thermally excite the ro-vibrational lines); the relative contribution of disk vs wind emission across the different radial regions remains to be clarified.
\end{itemize}

These results are valid for a small sample of $\approx10$ objects where water is detected at high S/N from the ground (Table \ref{tab: slab_results}), which is mostly composed of Class II disks with high accretion, outflows and/or winds (as observed in e.g. forbidden optical emission lines), and without a large inner dust cavity (as observed with millimeter interferometry). We also remind the reader that the column density estimates obtained from slab models represent only the observed layer \citep[which could be just 1--10\% of the total column density, see e.g.][]{bosman22}, and assume LTE excitation.
Water spectra observed with JWST, supported with line kinematics from higher-resolution ground-based observations, will enable a comprehensive characterization of excitation conditions and of the density and distribution of water (and other molecules) in inner disks on a larger population. Analyses of large samples of de-blended (or fully spectrally-resolved) molecular spectra will be fundamental to study the properties of molecules in inner disks and translate global trends currently observed in line luminosities (Figure \ref{fig:corr_grid}) into trends in physical and chemical conditions that are evolving in inner disks, to inform models of planet formation and disk dispersal. We therefore invite the community to make use of the large database of infrared spectra available on \url{www.spexodisks.com}, with currently $> 1000$ spectra from 6 spectrographs, to support and complement the analysis of spectrally-unresolved JWST spectra in the coming years.

\acknowledgments
This work includes data gathered at the Infrared Telescope Facility, which is operated by the University of Hawaii under contract 80HQTR19D0030 with the National Aeronautics and Space Administration.
This work includes observations made with ESO telescopes at the Paranal Observatory under programs 179.C-0151, 093.C-0432, 095.C-0203, 088.C-0666, and 198.C-0104.
This work includes observations made with the Spitzer Space Telescope, which is operated by the Jet Propulsion Laboratory, California Institute of Technology under a contract with NASA.
This work includes observations obtained at the international Gemini Observatory, a program of NSF’s NOIRLab, which is managed by the Association of Universities for Research in Astronomy (AURA) under a cooperative agreement with the National Science Foundation on behalf of the Gemini Observatory partnership: the National Science Foundation (United States), National Research Council (Canada), Agencia Nacional de Investigación y Desarrollo (Chile), Ministerio de Ciencia, Tecnología e Innovación (Argentina), Ministério da Ciência, Tecnologia, Inovações e Comunicações (Brazil), and Korea Astronomy and Space Science Institute (Republic of Korea).
The authors wish to recognize and acknowledge the very significant cultural role and reverence that the summit of Maunakea has always had within the indigenous Hawaiian community. We are grateful to the Hawaiian community to have the opportunity to conduct observations from this mountain.

The authors warmly thank all the observatory staff that supported observations at the VLT, IRTF, and Gemini telescopes, additional colleagues involved in the VISIR survey (H.U. K\"aufl, M.F. Sterzik, E. Flaccomio, G. Meeus, R. Alexander, C. Dullemond, S. Steendam), and the anonymous reviewer for suggestions that helped improving this paper.
A.B. acknowledges support from NASA/STScI GO grant JWST-GO-01640.
G.J.H. is supported by general grant 12173003 from the National Natural Science Foundation of China.
A.C. acknowledges funding from the French ANR under contract number ANR\-18\-CE31\-0019 (SPlaSH), and support from the French National Research Agency in the framework of the Investissements d'Avenir program (ANR-15-IDEX-02), through the funding of the ``Origin of Life" project of the Grenoble-Alpes University.
E.v.D. has been funded by the European Research Council (ERC) under grant
101019751 MOLDISK and the Deutsche Forschungsgemeinschaft (DFG,
German Research Foundation) - 325594231, FOR 2634/2.

\facilities{IRTF, VLT, \textit{Spitzer}}

\software{{\tt mpfit} \citep{mpfit} , {\tt matplotlib} \citep{matplotlib} , {\tt seaborn} \citep{seaborn}}

\newpage
\appendix

\section{Multi-wavelength spectral plots} \label{app: extra_plots}
\begin{figure*}
\centering
\includegraphics[width=1\textwidth]{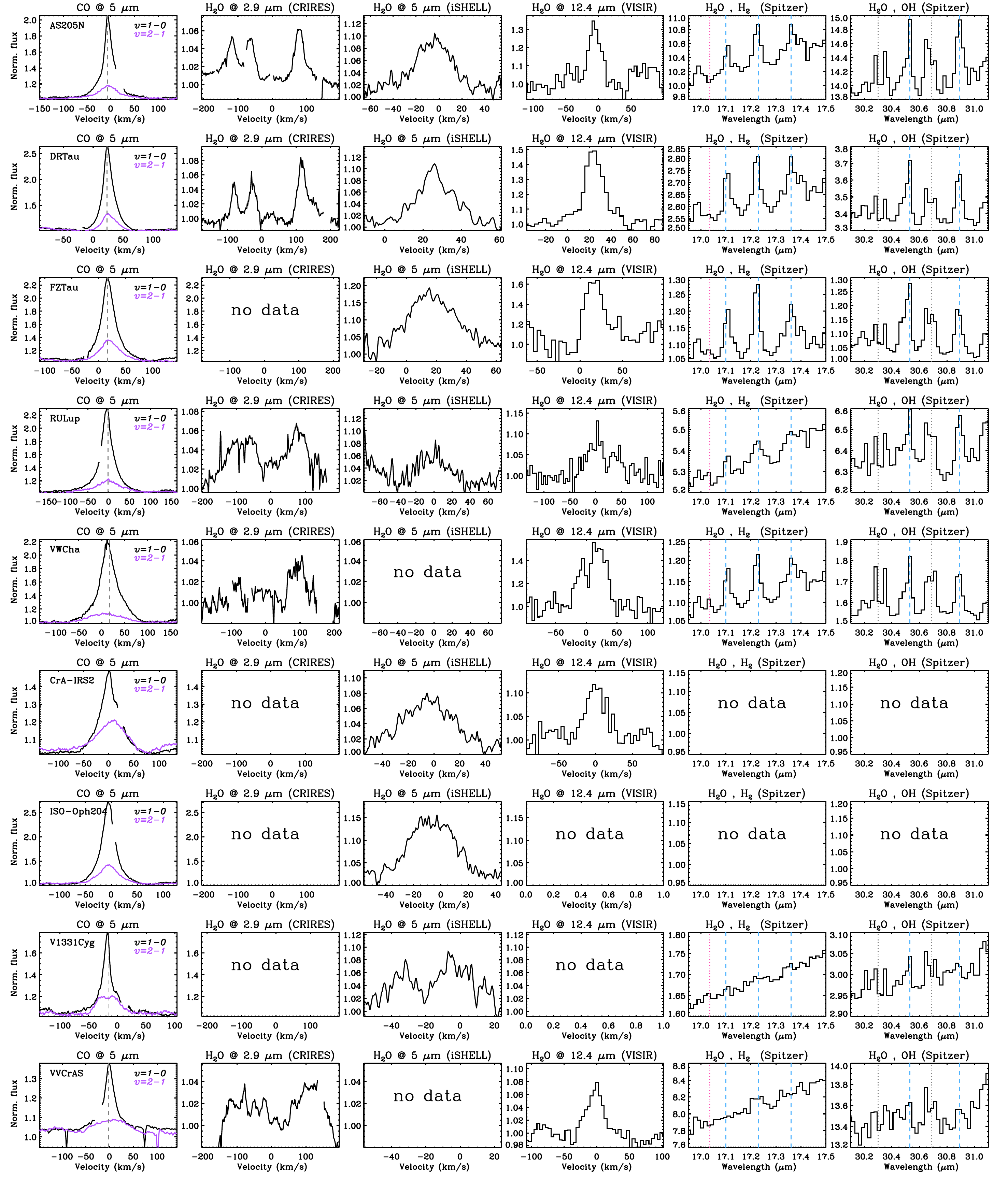} 
\caption{Multi-wavelength infrared spectra included in this work. The stellar RV is showed with a dashed line on top of CO lines, where available. High-resolution spectral lines are shown in velocity space and continuum-normalized (the $L$-band lines are shown in reference to 2.9283~$\mu$m), those from Spitzer spectra are shown in wavelength and in calibrated flux units of Jy. \ce{H2O} emission is marked with dashed lines while \ce{H2} and OH emission is marked with dotted lines in the Spitzer spectra.}
\label{fig: WATER_multiwl_1}
\end{figure*}

\begin{figure*}
\centering
\includegraphics[width=1\textwidth]{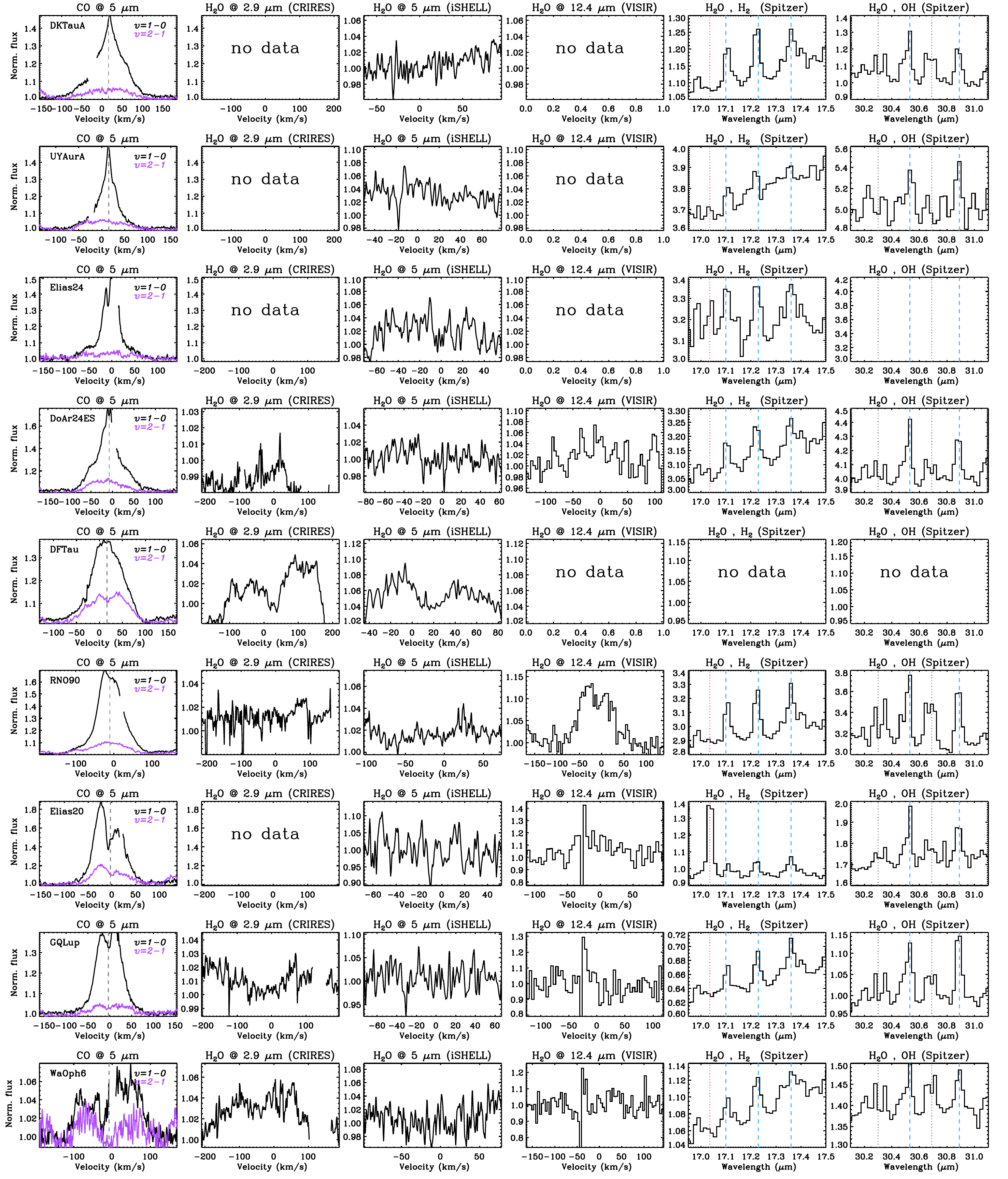} 
\caption{Same as Figure \ref{fig: WATER_multiwl_1}, for additional disks.}
\label{fig: WATER_multiwl_2}
\end{figure*}

\begin{figure*}
\centering
\includegraphics[width=1\textwidth]{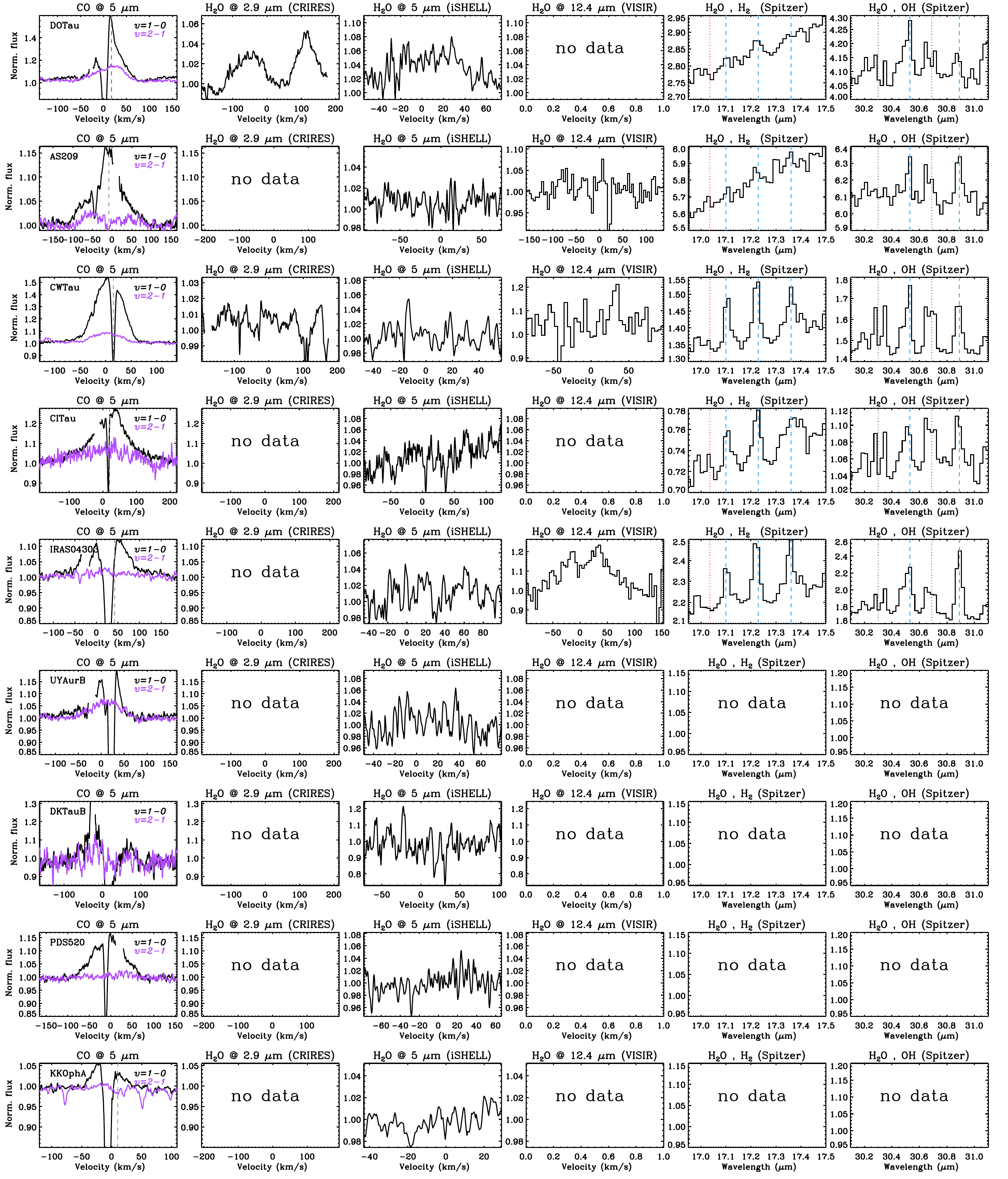} 
\caption{Same as Figure \ref{fig: WATER_multiwl_1}, for disks with absorption in CO lines.}
\label{fig: WATER_multiwl_3}
\end{figure*}

\begin{figure*}
\centering
\includegraphics[width=1\textwidth]{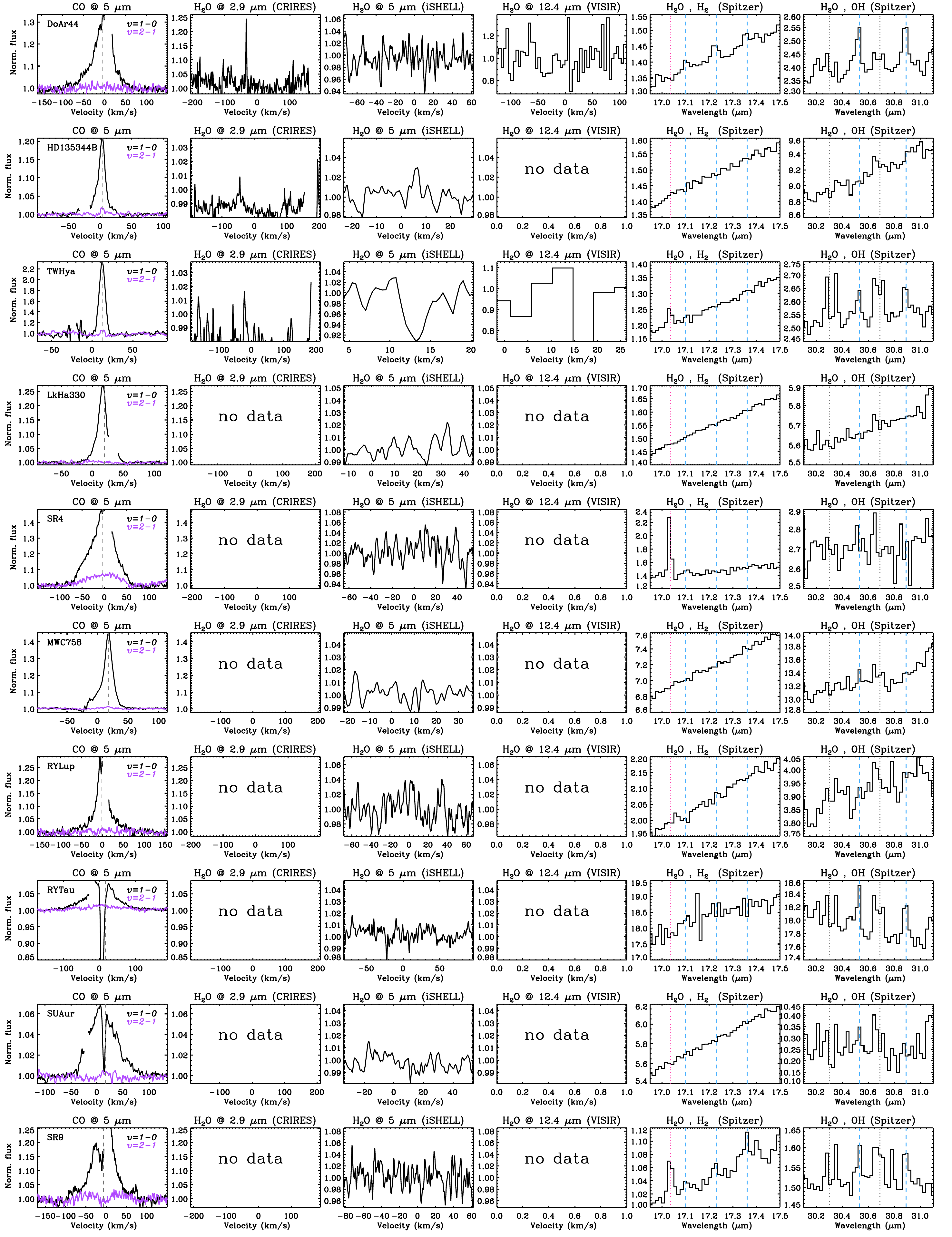} 
\caption{Same as Figure \ref{fig: WATER_multiwl_1}, for disks with inner dust cavities.}
\label{fig: WATER_multiwl_cav}
\end{figure*}

\begin{figure*}
\centering
\includegraphics[width=1\textwidth]{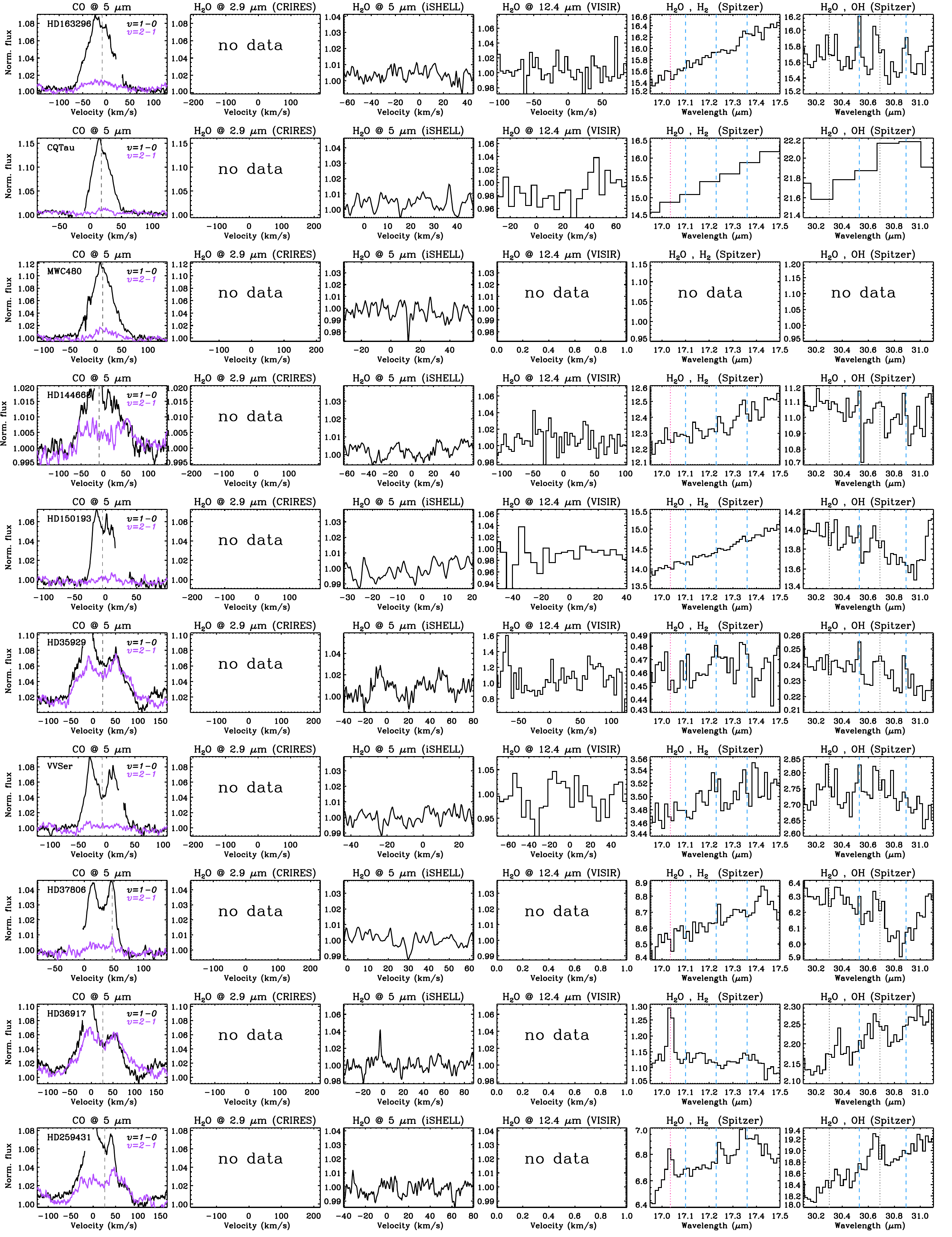} 
\caption{Same as Figure \ref{fig: WATER_multiwl_1}, for selected disks around Herbig Ae/Be stars.}
\label{fig: WATER_multiwl_herbigs}
\end{figure*}

Figures \ref{fig: WATER_multiwl_1} to \ref{fig: WATER_multiwl_herbigs} illustrate portions of spectra in disks that have CRIRES, iSHELL, and/or VISIR spectra of infrared water emission, as included in this work. All spectra can be visualized interactively on \url{www.spexodisks.com}.

\section{A survey of mid-infrared water spectra with VLT-VISIR} \label{app: visir_survey}
\subsection{Data reduction}
With VISIR in echelle mode, the telescope chops off the slit due to the short slit of 4.1"; the nod B positions are therefore used only for background subtraction. 
Mechanical oscillations of the grating between sequential chops and nods, and sometimes a difficult guiding on weak sources, introduce small shifts (between fractions of a pixel to a few pixels) in both the spatial and spectral directions on the detector. These are corrected by cross-correlation of the position of one echelle order (for the spatial direction) and telluric lines (for the spectral direction), during combination of individual chops and nods. This procedure ensures the correct combination of the PSF as observed in different nods, and avoids producing an artificial PSF broadening and residuals from the shift of telluric lines. A similar procedure for the spectral direction was previously implemented for VISIR 1 data \citep{carmona08,banz14}.

Observing $N$-band water spectra from the ground is very challenging due the variability of Earth's atmospheric conditions, especially in the precipitable water vapor (PWV). Apart from exceptional conditions, the sky typically changes significantly between sequential nods and causes residuals from variable telluric lines in the spectra. In this VISIR survey, we carefully monitored the observing conditions nod by nod, and reject individual nods where the sky varied by more than five sigma from the mean sky variation in each observing block. Any remaining sky residuals were removed by averaging the signal over 10--15 pixels at the two sides of the PSF pixel by pixel in the spectral direction, fitting this average with a Savitzky-Golay polynomial smoothing filter, and then subtracting the fit from the PSF before extracting the 1D spectrum. We extensively tested this correction, which in previous work was performed using a simple median subtraction \citep{pont10b,banz14}, and found that it gives the best results in removing telluric residuals while minimizing noise addition to the spectral PSF. The 1D spectrum is then extracted using optimal extraction.

After the detector upgrade in 2016, strong high-frequency fringing was found to affect VISIR high-resolution spectra. In connection to that, the sensitivity was also found to be a factor $\approx 10$ lower (Figure \ref{fig: visir_sensitivity}). In April 2017, the entrance window of VISIR was replaced to solve this issue. This replacement removed completely the high-frequency fringing in the echelle mode, and suppressed it in the long-slit mode; however, the low-frequency fringing that was affecting VISIR 1 was still present in all modes (see e.g. Figure \ref{fig: settings}). Since VISIR observations do not include flats, but fringing was found to not be stable enough to be removed simply by division with the standard, we fitted and removed fringes in each spectrum after extraction of the 1D spectrum and prior to telluric correction. Some fringe residuals remained when fringes varied between a science target spectrum and its telluric standard spectrum and in the spectra of binary systems, where the two components were observed in a different position on the detector as compared to the telluric standard.

Telluric correction of the extracted 1D spectra was performed using observations of several bright asteroids and a few bright stars (Alpha Cen, Alpha Car, Alpha CMa = Sirius, Theta Sco = HR6553). Telluric observations were interspersed during science observations in order to ensure a good match (difference $< 0.1$) in airmass and PWV with the science targets. This was achieved in most but not all cases. Differences in both airmass and PWV between science targets and telluric standards are compensated during reduction using telluric water and CO$_2$ lines. We have tested using the airmass and PWV values recorded in fits headers, but these average values do not fully capture real differences and variability happening during observations. We therefore scaled telluric water lines until they match those observed in the science targets to find the best PWV correction, and do the same with a CO$_2$ line to find the best airmass correction. These correction factors were applied to the telluric spectra as an exponential correction of the form $f_{corr} = f_{obs}^{(A_{science}/A_{telluric})}$ as in previous work \citep{pont10b,banz14}.
The science and telluric spectra were then normalized to their continua, aligned by cross-correlation of telluric lines, and then divided after applying the airmass and PWV correction to the telluric spectrum. 

Wavelength calibration of individual spectra was done by cross-correlation in each setting with an atmospheric model set on Paranal conditions (same model as above in Section \ref{sec: water_overview}). We fine-tuned the wavelength solution on low-PWV spectra that allowed us to use all the observed telluric lines out to the detector edges (Figure \ref{fig: settings}), and then we applied that solution to all other spectra after shifting it by cross-correlation of sky lines in each spectrum.

\subsection{VISIR 2 performance} \label{sec: VISIR_performance}
We measured the spectral resolution of VISIR from unresolved O$_3$ lines detected in telluric absorption spectra. We used several asteroid observations obtained during the survey and consistently measure a FWHM of 10\,km/s (corresponding to R $\sim30,000$) in four O$_3$ lines detected in the 12.84\,$\mu$m setting. In the other two settings, O$_3$ lines are much weaker and possibly detected only in a few spectra; these measurements are less consistent and provide FWHM in the range 5--13.6\,km/s, suggesting a similar resolution of R $\approx30,000$ also in echelle mode at 12.2--12.5\,$\mu$m.

\begin{figure}
\centering
\includegraphics[width=0.45\textwidth]{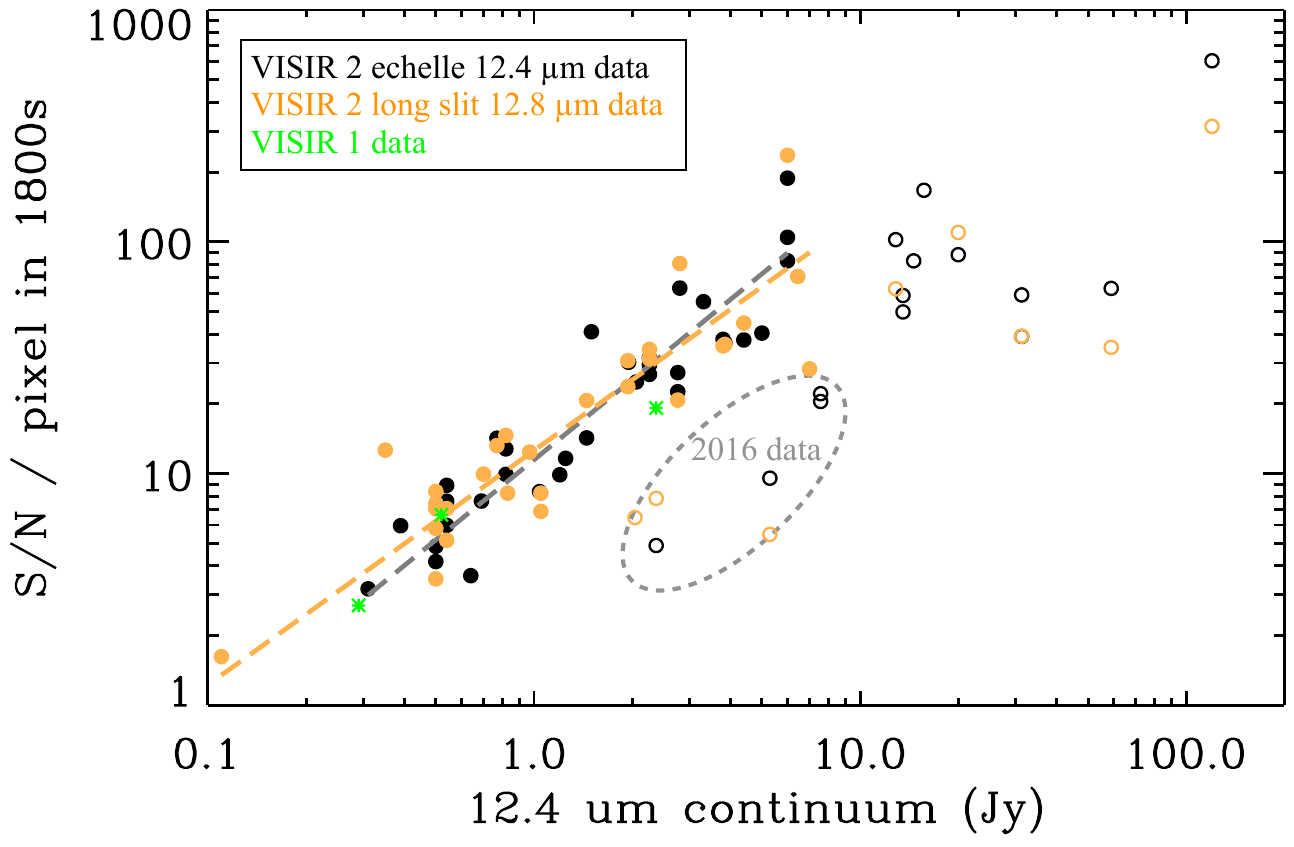} 
\caption{VISIR~2 sensitivity estimated from this survey, as compared to previous data from VISIR~1 \citep{pascucci09,banz14}. Linear fits to the echelle and long-slit modes are shown with dashed lines. A lower sensitivity is recorded in 2016 after the detector upgrade but before the entrance window was replaced in 2017, as well as in brighter sources. }
\label{fig: visir_sensitivity}
\end{figure}

\begin{figure*}
\centering
\includegraphics[width=1\textwidth]{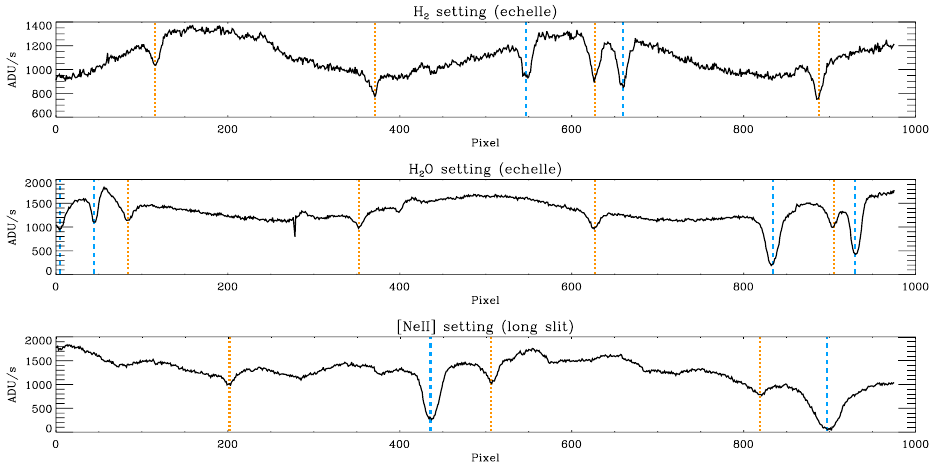} 
\caption{Spectral settings covered in the VISIR survey, shown as extracted from the combined 2D spectra (but before any correction/calibration is applied). The target is CrA-IRS2, which is one of the four targets observed in all three settings. Telluric water absorption is marked with dashed lines, telluric CO$_2$ absorption is marked with dotted lines. The echelle settings are affected by lower-frequency fringes, the long-slit setting by higher-frequency fringes.}
\label{fig: settings}
\end{figure*}

In terms of sensitivity, we find that VISIR~2 observations in the echelle and long-slit settings utilized here delivered a similar S/N per pixel (Figure \ref{fig: visir_sensitivity}). Using as reference spectra of DR~Tau, T~Cha, and Sz~73 taken in 2008--2012 \citep{pascucci09,banz14}, the new sensitivity is $\approx60\%$ higher than VISIR~1 observations, after the different pixel sampling has been accounted for (VISIR~2 has a pixel/wavelength ratio larger by 60\% than VISIR~1).

\begin{figure}
\centering
\includegraphics[width=0.45\textwidth]{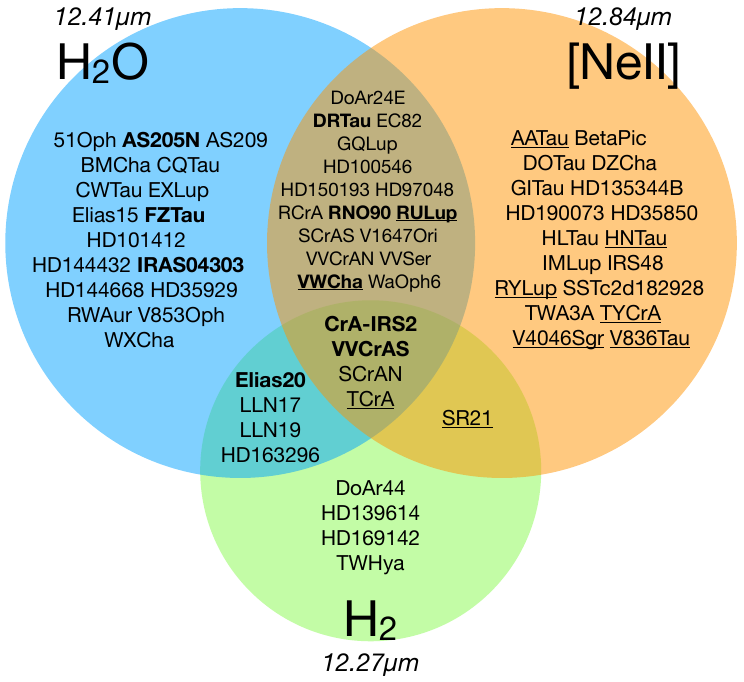} 
\caption{Venn diagram showing spectral settings and sample obtained in the VISIR survey included in this work. Detections are shown in boldface for water (presented in this work), and underlined for [NeII] \citep[presented in][]{pascucci20}. No detections are reported for \ce{H2}.}
\label{fig: Venn_diagr}
\end{figure}

\section{Sample and line flux measurements} \label{app: fluxes}
Table \ref{tab: fluxes} reports the full sample of 85 disks included in this work based on the available high-resolution spectra of infrared water emission. Line fluxes are measured from the CRIRES, iSHELL, and VISIR spectra as available from the different surveys (Section \ref{sec: data}) and as explained in Section \ref{sec: h2o_trends}. Line fluxes are measured from continuum-normalized spectra because absolute flux calibration of high-resolution infrared spectra from the ground is typically uncertain due to flux losses through narrow slits, which may vary from night to night even for the same instrument due to variable sky conditions. The continuum-normalized line fluxes $F_{cn}$ are turned into the absolute line luminosity shown in Figure \ref{fig:corr_grid} using the measured distance and continuum fluxes $F_c$, as $L = 4 \pi$~dist$^2 F_{cn} \times F_c$. As continuum flux $F_c$, we take the W1 and W2 fluxes from WISE for the 2.9~$\mu$m and 5~$\mu$m water lines, and the continuum flux as measured at 12.3~$\mu$m from Spitzer-IRS spectra for the 12.4~$\mu$m water line.
Table \ref{tab: fluxes_5um} reports line flux measurements for transitions covered in the $M$-band iSHELL spectra in four disks that have the largest number of detections. Line flux measurements for the entire iSHELL survey are available upon request to the author.

We report here the SIMBAD name of some objects, for easier identification: CrA-IRS~2 is CHLT~1 (2MASS J19014156-3658312), EC~82 is HBC~672 (2MASS J18295688+0114463), Elias 20 and 24 are Elia 2-20 and 2-24, IRAS~04303 is IRAS 04303+2240 (2MASS J04331907+2246342), IRS~48 is IRAS 16245-2423 (2MASS J16273718-2430350), ISO-Oph~204 is 2MASS J16315211-2456156, SR~21 is Elia 2-30, SR~9 is Elia 2-34, SR~4 is Elia 2-13.

Sample properties used in this work have been collected and reported in \citet{banz22} for most of the sample. We include here references used in that paper, which are also used for the rest of the sample included in this work: distances are from \citet{bj18}; stellar and accretion properties are from \citet{guzmandiaz21,wichittanakom20,herczeg14,salyk13,dsharp,fang18,simon16}; the infrared index $n_{13-30}$ is measured from Spitzer-IRS spectra as in \citet{banz20}; WISE fluxes are taken from the AllWISE Data Release \citep{allwise}; references for the stellar RV, the disk inclination, and the millimeter dust disk radius $R_{\rm{disk}}$ are: \citet{alecian13,dunkin97};  \citet{fang18,banz19};  \citet{liu11,vdmarel16,wolfer21}; \citet{hartmann86};  \citet{sullivan19_rv};  \citet{acke08,brittain07}; \citet{long18,long21}; \citet{huang18,kurtovic18};  \citet{labdon19};  \citet{pinilla18,tripathi17}; \citet{margi19}; \citet{pont11};  \citet{brown12}.

\startlongtable
\begin{deluxetable*}{l c c c c c c c c c c c c c}
\tabletypesize{\footnotesize}
\tablewidth{0pt}
\tablecaption{\label{tab: fluxes} Sample and line flux measurements.}
\tablehead{\colhead{} & \colhead{} & \colhead{} & \multicolumn{3}{|c|}{2.9\,$\mu$m} & \multicolumn{5}{c|}{5\,$\mu$m} & \multicolumn{3}{c}{12.4\,$\mu$m} \\
\colhead{Object} & \colhead{Dist} & \colhead{$S$} & \colhead{$F_c$} & \colhead{$F_{cn}$} & \colhead{err} & \colhead{$F_c$} & \colhead{$F_{cn}$} & \colhead{err} & \colhead{$F_{cn}$(CO)} & \colhead{err(CO)} & \colhead{$F_c$} & \colhead{$F_{cn}$} & \colhead{err}
 }
\tablecolumns{14}
\startdata
51Oph & 125. & -- & 11.2 & -- & -- & 35.7 & -- & -- & -- & -- & -- & -3.02e-16 & 5.66e-17 \\
AATau & 137. & 1.8 & 0.3 & 4.41e-15 & 4.41e-15 & 0.3 & -- & -- & 8.10e-14 & 5.99e-16 & 0.3 & -- & -- \\
ABAur & 155. & 3.6 & 14.8 & -- & -- & 36.7 & 3.73e-17 & 2.31e-16 & 9.69e-15 & 1.53e-17 & -- & -- & -- \\
AS205N & 127. & 7.0 & 4.4 & 1.67e-14 & 6.18e-15 & 6.8 & 8.73e-15 & 1.12e-16 & 7.31e-14 & 2.69e-16 & 7.2 & 6.35e-15 & 6.25e-16 \\
AS209 & 121. & 6.4 & 0.8 & -- & -- & 0.8 & 1.58e-15 & 3.46e-15 & 2.56e-14 & 3.31e-16 & 3.4 & 9.06e-16 & 1.16e-15 \\
CITau & 158. & 4.8 & 0.6 & -- & -- & 0.7 & 3.77e-15 & 9.91e-15 & 6.38e-14 & 6.44e-16 & 0.5 & -- & -- \\
CQTau & 149. & 9.2 & 2.3 & -- & -- & 2.9 & 3.21e-16 & 6.20e-16 & 6.81e-15 & 8.39e-17 & 4.9 & 3.85e-16 & 9.74e-16 \\
CrA-IRS2 & 150. & 6.9 & 2.7 & -- & -- & 6.5 & 7.17e-15 & 9.23e-17 & 4.91e-14 & 2.04e-16 & -- & 2.65e-15 & 4.10e-16 \\
CWTau & 132. & 2.7 & 1.5 & 1.56e-15 & 1.56e-15 & 1.8 & 6.95e-16 & 3.73e-15 & 6.29e-14 & 2.09e-16 & 1.2 & 1.55e-15 & 6.51e-15 \\
DFTau & 125. & 2.9 & 1.2 & 3.81e-14 & 4.95e-15 & 1.4 & 1.24e-14 & 1.90e-16 & 6.70e-14 & 1.87e-16 & -- & -- & -- \\
DKTauA & 128. & 7.6 & 1.1 & -- & -- & 1.3 & 1.32e-15 & 4.41e-15 & 7.27e-14 & 4.07e-16 & -- & -- & -- \\
DKTauB & 128. & 15.1 & -- & -- & -- & -- & -3.44e-16 & 2.86e-14 & -2.52e-15 & 3.08e-14 & -- & -- & -- \\
DoAr24EN & 137. & -- & -- & -- & -- & -- & -- & -- & -- & -- & -- & -9.42e-16 & 1.10e-15 \\
DoAr24ES & 137. & 6.4 & -- & -- & -- & -- & 8.12e-16 & 4.34e-15 & 9.60e-14 & 4.12e-16 & 2.6 & 5.74e-16 & 1.05e-15 \\
DoAr44 & 145. & 3.5 & 0.6 & 6.26e-15 & 1.66e-14 & 0.6 & 5.66e-18 & 5.72e-15 & 3.69e-14 & 7.91e-16 & 0.6 & -- & -- \\
DOTau & 139. & 13.5 & 0.9 & 2.60e-14 & 2.09e-15 & 1.2 & 8.17e-15 & 2.74e-16 & 3.80e-14 & 4.76e-16 & 2.0 & -- & -- \\
DRTau & 195. & 5.1 & 1.4 & 1.98e-14 & 2.98e-15 & 1.9 & 4.65e-15 & 1.01e-16 & 8.26e-14 & 1.38e-16 & 1.9 & 8.43e-15 & 4.36e-16 \\
EC82 & 399. & -- & -- & -- & -- & -- & -- & -- & -- & -- & 1.5 & -2.20e-16 & 4.90e-15 \\
Elias20 & 138. & 5.6 & 1.0 & -- & -- & 1.6 & 2.50e-16 & 1.08e-14 & 1.08e-13 & 6.04e-16 & 1.0 & 8.27e-15 & 7.24e-15 \\
Elias24 & 136. & 4.6 & 0.7 & -- & -- & 1.3 & 3.03e-15 & 5.00e-15 & 4.88e-14 & 6.01e-16 & 2.8 & -- & -- \\
EXLup & 157. & 11.4 & 0.3 & 1.16e-14 & 5.47e-16 & 0.3 & -- & -- & 4.48e-14 & 5.07e-16 & -- & 3.85e-15 & 3.08e-14 \\
FNTau & 131. & 4.0 & 0.3 & 7.71e-17 & 1.54e-16 & 0.3 & -- & -- & 1.05e-14 & 3.28e-16 & 0.7 & -- & -- \\
FZTau & 130. & 5.5 & 1.1 & -- & -- & 1.3 & 1.45e-14 & 1.87e-16 & 1.03e-13 & 3.49e-16 & 1.0 & 1.43e-14 & 1.02e-15 \\
GQLup & 151. & 2.4 & 1.1 & -- & -- & 1.0 & 1.32e-15 & 8.90e-15 & 4.68e-14 & 1.18e-16 & 0.5 & -9.79e-16 & 5.34e-15 \\
HD100546 & 108. & -- & 6.4 & -- & -- & 7.1 & -- & -- & -- & -- & -- & 1.70e-16 & 1.06e-15 \\
HD101412 & 410. & -- & 1.1 & -- & -- & 1.7 & -- & -- & -- & -- & 2.4 & 6.08e-16 & 2.79e-15 \\
HD135344B & 134. & 5.2 & 3.2 & 7.67e-16 & 3.03e-16 & 3.5 & 6.28e-17 & 1.12e-15 & 9.21e-15 & 1.95e-16 & 0.9 & -- & -- \\
HD141569 & 111. & 2.0 & 1.2 & -- & -- & 0.6 & 7.34e-17 & 1.13e-15 & 2.86e-14 & 2.69e-16 & -- & -- & -- \\
HD142666 & 146. & 1.3 & 2.5 & -- & -- & 3.0 & 2.35e-17 & 4.98e-16 & 2.86e-15 & 1.22e-16 & -- & -- & -- \\
HD143006 & 166. & 1.6 & 1.4 & -- & -- & 1.4 & -1.34e-18 & 7.55e-16 & 4.77e-15 & 2.12e-16 & 0.7 & -- & -- \\
HD144432S & 677. & 2.8 & -- & -- & -- & -- & -- & -- & 1.14e-14 & 1.57e-16 & -- & -2.73e-16 & 1.08e-15 \\
HD144668 & 160. & 2.1 & 32.5 & -- & -- & 61.7 & 1.69e-16 & 6.36e-16 & 3.57e-15 & 7.82e-17 & 12.0 & 2.76e-16 & 1.16e-15 \\
HD150193 & 150. & 1.4 & 5.1 & -- & -- & 6.5 & 3.02e-17 & 5.18e-16 & 3.34e-15 & 1.24e-16 & 10.8 & 3.86e-16 & 4.40e-16 \\
HD163296 & 101. & 3.3 & 9.8 & -- & -- & 11.6 & 6.83e-16 & 4.32e-17 & 1.17e-14 & 8.68e-17 & 11.5 & 3.60e-17 & 7.95e-16 \\
HD169142 & 114. & 2.5 & 1.0 & -- & -- & 0.9 & 3.99e-18 & 6.69e-16 & 3.30e-15 & 8.01e-17 & -- & -- & -- \\
HD179218 & 258. & 1.7 & 3.6 & -- & -- & 4.8 & 7.92e-18 & 2.25e-16 & 3.13e-15 & 1.78e-17 & 15.6 & -- & -- \\
HD190073 & 872. & 4.3 & 4.0 & -- & -- & 5.6 & 1.93e-17 & 1.23e-16 & 5.80e-15 & 1.28e-17 & 5.2 & -- & -- \\
HD259431 & 640. & 2.4 & 5.1 & -- & -- & 9.6 & 5.25e-17 & 1.09e-15 & 1.29e-14 & 6.29e-17 & 6.2 & -- & -- \\
HD35929 & 377. & 1.9 & 1.1 & -- & -- & 1.1 & 1.72e-15 & 1.79e-15 & 1.96e-14 & 2.35e-16 & 0.6 & -3.32e-15 & 7.06e-15 \\
HD36917 & 445. & 1.8 & 2.2 & -- & -- & 2.7 & 1.14e-16 & 1.76e-15 & 1.91e-14 & 9.91e-17 & 1.2 & -- & -- \\
HD37806 & 397. & 1.4 & 4.5 & -- & -- & 6.7 & 1.92e-17 & 5.75e-16 & 4.56e-15 & 8.50e-17 & 8.0 & -- & -- \\
HD97048 & 184. & -- & 5.0 & -- & -- & 2.8 & -- & -- & -- & -- & -- & 3.48e-16 & 1.17e-14 \\
HTLupA & 154. & 5.3 & 1.5 & -- & -- & 1.6 & -- & -- & 2.22e-14 & 1.45e-16 & -- & -- & -- \\
IMLup & 158. & 2.9 & 0.5 & 1.80e-16 & 1.80e-16 & 0.4 & -- & -- & 1.16e-14 & 3.79e-16 & 0.5 & -- & -- \\
IRAS04303 & 148. & 2.5 & 0.9 & -- & -- & 1.4 & 1.55e-15 & 4.00e-15 & 9.69e-15 & 3.58e-16 & 2.1 & 1.38e-14 & 8.42e-16 \\
IRS48 & 125. & 1.5 & 1.5 & -- & -- & 2.2 & 2.33e-16 & 1.38e-15 & 2.62e-15 & 4.15e-17 & -- & -- & -- \\
ISO-Oph204 & 167. & 5.3 & 1.0 & -- & -- & 2.0 & 1.33e-14 & 1.76e-16 & 1.32e-13 & 4.01e-16 & -- & -- & -- \\
KKOphA & 221. & 7.5 & 5.1 & -- & -- & 6.8 & -7.50e-18 & 1.55e-15 & -4.43e-15 & 1.40e-15 & -- & -- & -- \\
KKOphB & 221. & -- & -- & -- & -- & -- & -3.27e-15 & 9.50e-15 & -- & -- & -- & -- & -- \\
LkHa330 & 308. & 5.2 & 1.1 & -- & -- & 1.4 & 1.85e-18 & 8.74e-16 & 1.16e-14 & 1.82e-16 & 0.5 & -- & -- \\
MWC297 & 408. & -- & 895.1 & -- & -- & 950.9 & 8.41e-17 & 1.07e-16 & -- & -- & -- & -- & -- \\
MWC480 & 155. & 4.1 & 3.5 & -- & -- & 4.6 & 9.96e-18 & 9.02e-16 & 1.23e-14 & 6.94e-17 & -- & -- & -- \\
MWC758 & 155. & 5.8 & 4.1 & -- & -- & 4.9 & 5.18e-17 & 7.54e-16 & 1.92e-14 & 8.99e-17 & 3.2 & -- & -- \\
PDS156 & 398. & 3.6 & 1.6 & -- & -- & 1.6 & 1.46e-17 & 2.38e-15 & 2.01e-14 & 2.74e-16 & -- & -- & -- \\
PDS520 & 375. & 9.0 & 1.1 & -- & -- & 1.4 & 1.59e-16 & 5.04e-15 & 2.08e-14 & 5.27e-16 & -- & -- & -- \\
RCrA & 95. & -- & 395.4 & -- & -- & 544.7 & -- & -- & -- & -- & -- & 4.05e-18 & 2.10e-16 \\
RNO90 & 117. & 3.2 & 2.7 & 9.59e-15 & 1.25e-15 & 3.9 & 4.88e-15 & 1.09e-16 & 1.20e-13 & 1.71e-16 & 2.2 & 6.74e-15 & 5.96e-16 \\
RULup & 159. & 6.4 & 1.2 & 3.50e-14 & 4.91e-15 & 1.7 & 8.35e-15 & 2.27e-16 & 1.20e-13 & 3.17e-16 & 4.4 & 2.62e-15 & 3.44e-16 \\
RWAur & 163. & -- & 1.0 & -- & -- & 1.4 & -- & -- & -- & -- & 1.5 & -6.09e-15 & 4.82e-15 \\
RYLup & 158. & 8.0 & 1.0 & -- & -- & 1.3 & 5.66e-18 & 5.20e-15 & 2.08e-14 & 5.50e-16 & 0.9 & -- & -- \\
RYTau & 444. & 3.8 & 6.2 & -- & -- & 7.8 & 5.13e-16 & 1.77e-15 & 8.96e-15 & 1.43e-16 & 10.1 & -- & -- \\
SCrAN & 152. & 7.3 & -- & -- & -- & -- & -- & -- & 4.47e-14 & 6.07e-17 & 28.0 & 4.09e-16 & 7.65e-16 \\
SCrAS & 168. & -- & -- & -- & -- & -- & -- & -- & -- & -- & -- & 9.37e-16 & 3.90e-15 \\
SR21 & 138. & 2.3 & 1.1 & 1.57e-16 & 1.62e-16 & 1.1 & 2.79e-16 & 8.88e-16 & 4.70e-15 & 1.10e-16 & 2.0 & -- & -- \\
SR4 & 134. & 3.9 & 1.0 & -- & -- & 0.8 & 8.38e-16 & 5.72e-15 & 6.22e-14 & 7.00e-16 & 0.8 & -- & -- \\
SR9 & 130. & 3.9 & 0.7 & -- & -- & 0.7 & 6.07e-16 & 6.87e-15 & 3.30e-14 & 7.60e-16 & 0.5 & -- & -- \\
SUAur & 158. & 3.8 & 2.6 & -- & -- & 2.8 & -1.20e-18 & 1.10e-15 & 7.65e-15 & 1.14e-16 & 2.2 & -- & -- \\
TCrA & -- & -- & 0.5 & -- & -- & 1.0 & -8.77e-17 & 9.01e-16 & -- & -- & -- & 3.08e-16 & 8.11e-16 \\
TTauN & 144. & 6.1 & 2.2 & 5.77e-14 & 4.64e-14 & 10.0 & -- & -- & 3.82e-14 & 1.04e-16 & -- & -- & -- \\
TTauS & 144. & -- & 2.2 & 4.45e-14 & 3.56e-14 & 10.0 & -- & -- & -- & -- & -- & -- & -- \\
TWCha & 184. & 4.6 & 0.2 & 2.38e-14 & 1.84e-16 & 0.2 & -- & -- & 6.85e-14 & 1.26e-15 & 0.1 & -- & -- \\
TWHya & 60. & 2.8 & 0.5 & 4.29e-16 & 4.36e-16 & 0.3 & 2.42e-20 & 1.46e-15 & 2.50e-14 & 1.89e-16 & 0.5 & -- & -- \\
UYAurA & 155. & 10.9 & 1.1 & -- & -- & 1.6 & 7.44e-15 & 2.29e-16 & 4.22e-14 & 3.15e-16 & -- & -- & -- \\
UYAurB & 155. & 3.7 & 1.1 & -- & -- & 1.6 & 6.73e-16 & 6.73e-15 & 2.14e-15 & 6.48e-15 & -- & -- & -- \\
V1331Cyg & 587. & 10.7 & 0.4 & -- & -- & 0.6 & 3.49e-15 & 1.75e-16 & 4.58e-14 & 4.48e-16 & 1.0 & -- & -- \\
V1647Ori & 451. & -- & 0.9 & -- & -- & 1.9 & -- & -- & -- & -- & -- & 8.21e-16 & 1.77e-15 \\
V853Oph & 83. & -- & 0.4 & -- & -- & 0.4 & -- & -- & -- & -- & 0.5 & 4.65e-15 & 1.50e-14 \\
V892Tau & 117. & 3.8 & 2.7 & -- & -- & 3.9 & 9.67e-17 & 4.02e-16 & 3.93e-14 & 3.31e-17 & -- & -- & -- \\
VVCrAN & 149. & -- & 2.7 & -- & -- & 3.5 & -- & -- & -- & -- & -- & 9.65e-16 & 8.90e-16 \\
VVCrAS & 149. & 11.0 & 2.7 & -- & -- & 3.5 & -- & -- & 3.52e-14 & 4.22e-17 & 5.7 & 1.56e-15 & 1.14e-16 \\
VVSer & 403. & 1.8 & 2.6 & -- & -- & 3.8 & 4.17e-17 & 5.46e-16 & 9.48e-15 & 1.09e-16 & 3.7 & -1.17e-15 & 1.09e-15 \\
VWCha & 190. & 5.9 & 1.2 & 1.67e-14 & 8.77e-16 & 1.3 & -- & -- & 1.23e-13 & 2.30e-16 & 0.8 & 1.80e-14 & 1.25e-15 \\
VZCha & 191. & 4.0 & 0.4 & 1.87e-14 & 2.16e-16 & 0.4 & -- & -- & 1.11e-13 & 6.31e-16 & 0.4 & -- & -- \\
WaOph6 & 123. & 3.7 & 1.0 & 2.64e-14 & 1.85e-15 & 1.0 & 7.77e-16 & 7.30e-15 & 1.71e-14 & 5.01e-16 & 0.8 & 5.34e-16 & 2.89e-15 \\
WXCha & 190. & 2.7 & 0.4 & 1.13e-14 & 3.77e-16 & 0.4 & -- & -- & 8.33e-14 & 1.21e-15 & 0.3 & 1.92e-14 & 2.41e-14 \\
\enddata
\tablecomments{
Line fluxes and errors $F_{cn}$ are in units of erg s$^{-1}$ cm$^{-2}$ Jy$^{-1}$, and continuum fluxes $F_{c}$ are in Jy. Unless indicated in parentheses, the line flux and error refers to the water line at spectral wavelengths indicated at the top of the table (see Section \ref{sec: h2o_trends} for the identification of lines at each wavelength). The CO line shape parameter $S$ is reported in the third column for reference \citep[for its definition, see][]{banz22}.
}
\end{deluxetable*} 

\begin{deluxetable*}{l c c c c c c c c c}
\tabletypesize{\small}
\tablewidth{0pt}
\tablecaption{\label{tab: fluxes_5um} Water line flux measurements from $M$-band iSHELL spectra.}
\tablehead{\colhead{Line ID} & \colhead{Wavelength} & \multicolumn{2}{|c|}{AS~205~N} & \multicolumn{2}{c|}{DR~Tau} & \multicolumn{2}{c|}{FZ~Tau} & \multicolumn{2}{c|}{CrA-IRS~2} \\
\colhead{$v$~(u-l)~$J_{\:K_a \: K_c}$~(u-l)} & \colhead{($\mu$m)} & \colhead{$F_{cn}$} & \colhead{err} & \colhead{$F_{cn}$} & \colhead{err} & \colhead{$F_{cn}$} & \colhead{err} & \colhead{$F_{cn}$} & \colhead{err} }
\tablecolumns{10}
\startdata
1-0 \: $13_{\:7\:6} - 12_{\:6\:7}$ & 4.52474 & 9.06e-15 & 1.77e-14 & 2.15e-15 & 1.13e-14 & 1.90e-15 & 1.79e-14 & 2.43e-16 & 1.85e-14 \\
1-0 \: $9_{\:9\:0} - 8_{\:8\:1}$ & 4.53470 & 4.82e-15 & 1.67e-14 & 1.66e-15 & 1.18e-14 & 1.46e-14 & 1.70e-14 & 3.86e-15 & 1.86e-14 \\
1-0 \: $10_{\:8\:3} - 9_{\:7\:2}$ & 4.57264 & 3.42e-15 & 8.60e-15 & 3.60e-15 & 9.28e-15 & 4.98e-15 & 1.50e-14 & 3.80e-15 & 1.10e-14 \\
1-0 \: $12_{\:7\:6} - 11_{\:6\:5}$ & 4.57608 & 3.88e-15 & 4.67e-15 & 2.62e-15 & 6.17e-15 & 3.43e-15 & 8.04e-15 & 3.81e-15 & 4.96e-15 \\
1-0 \: $13_{\:6\:7} - 12_{\:5\:8}$ & 4.61395 & 1.84e-16 & 2.25e-15 & 1.93e-16 & 1.39e-15 & 3.08e-16 & 2.62e-15 & 2.35e-16 & 1.53e-15 \\
1-0 \: $11_{\:7\:4} - 10_{\:6\:5}$ & 4.62229 & 5.23e-15 & 3.66e-16 & 1.64e-15 & 2.59e-16 & 5.70e-15 & 4.23e-16 & 3.15e-15 & 3.71e-16 \\
1-0 \: $9_{\:7\:2} - 8_{\:6\:3}$ & 4.72810 & 1.01e-14 & 1.07e-15 & 5.31e-15 & 5.28e-16 & 1.47e-14 & 1.36e-15 & 7.74e-15 & 1.10e-15 \\
1-0 \: $11_{\:6\:5} - 10_{\:5\:6}$ & 4.72942 & 7.34e-15 & 7.45e-16 & 3.48e-15 & 3.22e-16 & 1.16e-14 & 5.67e-16 & 7.14e-15 & 8.04e-16 \\
1-0 \: $10_{\:6\:5} - 9_{\:5\:4}$ & 4.79063 & 7.67e-15 & 7.78e-16 & 4.37e-15 & 4.57e-15 & 1.73e-14 & 1.39e-15 & 8.73e-15 & 7.73e-16 \\
1-0 \: $18_{\:4\:15} - 17_{\:3\:14}$ & 4.87733 & 5.81e-15 & 6.52e-15 & 1.09e-15 & 1.80e-15 & 6.88e-15 & 5.21e-16 & 1.90e-15 & 6.31e-15 \\
1-0 \: $20_{\:3\:18} - 19_{\:2\:17}$ & 4.88205 & 4.09e-15 & 3.31e-16 & 1.55e-15 & 1.81e-15 & 5.43e-15 & 5.02e-16 & 3.00e-15 & 2.44e-16 \\
2-1 \: $7_{\:6\:1} - 6_{\:5\:2}$ & 4.93554 & 3.59e-15 & 4.89e-15 & 1.80e-16 & 2.21e-15 & 3.89e-15 & 4.18e-15 & 3.50e-15 & 4.93e-15 \\
2-1 \: $9_{\:5\:4} - 8_{\:4\:5}$ & 4.95058 & 3.14e-15 & 4.41e-16 & 1.79e-16 & 3.15e-15 & 3.94e-15 & 6.09e-15 & 2.42e-15 & 3.40e-16 \\
1-0 \: $16_{\:4\:13} - 15_{\:3\:12}$ & 4.96548 & 3.98e-15 & 3.05e-16 & 2.41e-15 & 3.19e-16 & 8.20e-15 & 5.14e-16 & 2.41e-15 & 2.30e-16 \\
1-0 \: $15_{\:4\:12} - 14_{\:3\:11}$ & 5.00798 & 1.02e-15 & 2.51e-15 & 6.17e-16 & 1.86e-15 & 5.29e-15 & 5.52e-16 & 3.50e-15 & 2.52e-16 \\
1-0 \: $17_{\:3\:15} - 16_{\:2\:14}$ & 5.01522 & 1.69e-16 & 3.35e-15 & 1.76e-16 & 5.71e-15 & 2.85e-16 & 1.02e-14 & 2.20e-16 & 2.92e-15 \\
1-0 \: $14_{\:4\:11} - 13_{\:3\:10}$ & 5.04712 & 1.16e-14 & 4.30e-16 & 1.75e-16 & 2.24e-15 & 2.83e-16 & 4.05e-15 & 1.57e-14 & 3.31e-16 \\
1-0 \: $23_{\:0\:23} - 22_{\:1\:22}$ & 5.05889 & 1.01e-15 & 2.62e-15 & 1.50e-15 & 5.51e-15 & 1.69e-15 & 8.17e-15 & 7.57e-16 & 1.83e-15 \\
1-0 \: $15_{\:4\:11} - 14_{\:5\:10}$ & 5.05986 & 4.34e-15 & 6.73e-16 & 1.48e-15 & 5.77e-15 & 5.27e-15 & 1.32e-14 & 1.83e-15 & 3.95e-15 \\
1-0 \: $16_{\:3\:14} - 15_{\:2\:13}$ & 5.06240 & 7.30e-15 & 3.57e-16 & 4.56e-15 & 3.22e-16 & 1.22e-14 & 5.19e-16 & 7.79e-15 & 2.66e-16 \\
1-0 \: $16_{\:2\:14} - 15_{\:3\:13}$ & 5.06429 & 4.76e-15 & 3.28e-16 & 2.35e-15 & 3.27e-16 & 5.21e-15 & 5.16e-16 & 5.05e-15 & 2.30e-16 \\
2-1 \: $7_{\:5\:2} - 6_{\:4\:3}$ & 5.07548 & 3.92e-15 & 4.14e-16 & 2.36e-15 & 2.63e-15 & 8.00e-15 & 7.34e-16 & 7.24e-15 & 3.30e-16 \\
1-0 \: $12_{\:4\:9} - 11_{\:3\:8}$ & 5.10815 & 1.30e-14 & 8.01e-16 & 6.29e-15 & 4.80e-16 & 1.99e-14 & 8.55e-16 & 1.36e-14 & 9.14e-16 \\
2-1 \: $16_{\:3\:14} - 15_{\:2\:13}$ & 5.12485 & 3.86e-15 & 4.57e-16 & 2.77e-15 & 3.81e-15 & 8.63e-15 & 9.75e-15 & 8.69e-15 & 3.71e-16 \\
1-0 \: $11_{\:4\:8} - 10_{\:3\:7}$ & 5.13018 & 1.09e-14 & 1.03e-15 & 1.73e-16 & 3.85e-15 & 1.56e-14 & 1.19e-15 & 2.00e-14 & 1.25e-15 \\
1-0 \: $14_{\:3\:12} - 13_{\:2\:11}$ & 5.16010 & 4.24e-15 & 4.52e-15 & 4.44e-15 & 5.26e-16 & 5.80e-15 & 8.63e-16 & 2.33e-15 & 4.31e-15 \\
1-0 \: $20_{\:1\:20} - 19_{\:0\:19}$ & 5.16471 & 5.50e-15 & 4.97e-16 & 5.75e-15 & 4.88e-16 & 1.35e-14 & 7.86e-16 & 4.97e-15 & 3.97e-16 \\
1-0 \: $14_{\:2\:12} - 13_{\:3\:11}$ & 5.16709 & 3.44e-15 & 3.74e-15 & 2.74e-15 & 2.88e-15 & 6.55e-15 & 8.35e-16 & 2.31e-15 & 2.97e-15 \\
2-1 \: $15_{\:2\:13} - 14_{\:3\:12}$ & 5.18489 & 2.61e-15 & 4.52e-15 & 4.02e-15 & 6.36e-16 & 7.14e-15 & 1.02e-15 & 6.56e-15 & 4.55e-16 \\
1-0 \: $13_{\:3\:10} - 12_{\:4\:9}$ & 5.18710 & 1.04e-14 & 9.59e-16 & 1.71e-16 & 4.35e-15 & 5.13e-15 & 8.60e-15 & 1.17e-14 & 1.01e-15 \\
2-1 \: $5_{\:5\:0} - 4_{\:4\:1}$ & 5.20586 & 9.56e-15 & 1.02e-14 & 2.28e-15 & 8.82e-15 & 9.61e-16 & 1.84e-14 & 7.04e-15 & 1.51e-14 \\
1-0 \: $15_{\:1\:14} - 14_{\:2\:13}$ & 5.22140 & 1.46e-14 & 1.77e-15 & 1.41e-14 & 1.12e-15 & 3.32e-14 & 1.94e-15 & 1.33e-14 & 1.38e-14 \\
1-0 \: $13_{\:2\:11} - 12_{\:3\:10}$ & 5.22307 & 9.86e-15 & 1.29e-15 & 6.96e-15 & 1.08e-15 & 2.52e-14 & 1.94e-15 & 9.57e-15 & 1.51e-15 \\
2-1 \: $14_{\:3\:12} - 13_{\:2\:11}$ & 5.22645 & 5.03e-15 & 5.88e-15 & 2.28e-15 & 5.26e-15 & 8.52e-15 & 1.05e-14 & 2.83e-15 & 5.11e-15 \\
\enddata
\tablecomments{line flux measurements have same definition as in Table \ref{tab: fluxes}. A complete compilation of line fluxes and upper limits for all spectra included in this paper is available upon request to the first author.}
\end{deluxetable*}

\bibliography{waterHR_paper}{}

\begin{thebibliography}{}
\expandafter\ifx\csname natexlab\endcsname\relax\def\natexlab#1{#1}\fi
\providecommand{\url}[1]{\href{#1}{#1}}

\bibitem[{{Acke} {et~al.}(2008){Acke}, {Verhoelst}, {van den Ancker}, {Deroo},
  {Waelkens}, {Chesneau}, {Tatulli}, {Benisty}, {Puga}, {Waters}, {Verhoeff},
  \& {de Koter}}]{acke08}
{Acke}, B., {Verhoelst}, T., {van den Ancker}, M.~E., {et~al.} 2008, \aap, 485,
  209

\bibitem[{{Ag{\'u}ndez} {et~al.}(2018){Ag{\'u}ndez}, {Roueff}, {Le Petit}, \&
  {Le Bourlot}}]{agundez18}
{Ag{\'u}ndez}, M., {Roueff}, E., {Le Petit}, F., \& {Le Bourlot}, J. 2018,
  \aap, 616, A19

\bibitem[{{Alecian} {et~al.}(2013){Alecian}, {Wade}, {Catala}, {Grunhut},
  {Landstreet}, {Bagnulo}, {B{\"o}hm}, {Folsom}, {Marsden}, \&
  {Waite}}]{alecian13}
{Alecian}, E., {Wade}, G.~A., {Catala}, C., {et~al.} 2013, \mnras, 429, 1001

\bibitem[{{Anderson} {et~al.}(2021){Anderson}, {Blake}, {Cleeves}, {Bergin},
  {Zhang}, {Schwarz}, {Salyk}, \& {Bosman}}]{anderson21}
{Anderson}, D.~E., {Blake}, G.~A., {Cleeves}, L.~I., {et~al.} 2021, \apj, 909,
  55

\bibitem[{{Andrews} {et~al.}(2018){Andrews}, {Huang}, {P{\'e}rez}, {Isella},
  {Dullemond}, {Kurtovic}, {Guzm{\'a}n}, {Carpenter}, {Wilner}, {Zhang}, {Zhu},
  {Birnstiel}, {Bai}, {Benisty}, {Hughes}, {{\"O}berg}, \& {Ricci}}]{dsharp}
{Andrews}, S.~M., {Huang}, J., {P{\'e}rez}, L.~M., {et~al.} 2018, \apjl, 869,
  L41

\bibitem[{{Antonellini} {et~al.}(2015){Antonellini}, {Kamp},
  {Riviere-Marichalar}, {Meijerink}, {Woitke}, {Thi}, {Spaans}, {Aresu}, \&
  {Lee}}]{antonellini15}
{Antonellini}, S., {Kamp}, I., {Riviere-Marichalar}, P., {et~al.} 2015, \aap,
  582, A105

\bibitem[{{Bailer-Jones} {et~al.}(2018){Bailer-Jones}, {Rybizki}, {Fouesneau},
  {Mantelet}, \& {Andrae}}]{bj18}
{Bailer-Jones}, C.~A.~L., {Rybizki}, J., {Fouesneau}, M., {Mantelet}, G., \&
  {Andrae}, R. 2018, \aj, 156, 58

\bibitem[{{Banzatti}(2013)}]{banz13}
{Banzatti}, A. 2013, PhD thesis, Eidgenossische Technische Hochschule, Zurich,
  Switzerland

\bibitem[{{Banzatti} {et~al.}(2018){Banzatti}, {Garufi}, {Kama}, {Benisty},
  {Brittain}, {Pontoppidan}, \& {Rayner}}]{banz18}
{Banzatti}, A., {Garufi}, A., {Kama}, M., {et~al.} 2018, \aap, 609, L2

\bibitem[{{Banzatti} {et~al.}(2014){Banzatti}, {Meyer}, {Manara},
  {Pontoppidan}, \& {Testi}}]{banz14}
{Banzatti}, A., {Meyer}, M.~R., {Manara}, C.~F., {Pontoppidan}, K.~M., \&
  {Testi}, L. 2014, \apj, 780, 26

\bibitem[{{Banzatti} {et~al.}(2019){Banzatti}, {Pascucci}, {Edwards}, {Fang},
  {Gorti}, \& {Flock}}]{banz19}
{Banzatti}, A., {Pascucci}, I., {Edwards}, S., {et~al.} 2019, \apj, 870, 76

\bibitem[{{Banzatti} \& {Pontoppidan}(2015)}]{bp15}
{Banzatti}, A., \& {Pontoppidan}, K.~M. 2015, \apj, 809, 167

\bibitem[{{Banzatti} {et~al.}(2015){Banzatti}, {Pontoppidan}, {Bruderer},
  {Muzerolle}, \& {Meyer}}]{banz15}
{Banzatti}, A., {Pontoppidan}, K.~M., {Bruderer}, S., {Muzerolle}, J., \&
  {Meyer}, M.~R. 2015, \apjl, 798, L16

\bibitem[{{Banzatti} {et~al.}(2017){Banzatti}, {Pontoppidan}, {Salyk},
  {Herczeg}, {van Dishoeck}, \& {Blake}}]{banz17}
{Banzatti}, A., {Pontoppidan}, K.~M., {Salyk}, C., {et~al.} 2017, \apj, 834,
  152

\bibitem[{{Banzatti} {et~al.}(2012){Banzatti}, {Meyer}, {Bruderer}, {Geers},
  {Pascucci}, {Lahuis}, {Juh{\'a}sz}, {Henning}, \& {{\'A}brah{\'a}m}}]{banz12}
{Banzatti}, A., {Meyer}, M.~R., {Bruderer}, S., {et~al.} 2012, \apj, 745, 90

\bibitem[{{Banzatti} {et~al.}(2020){Banzatti}, {Pascucci}, {Bosman}, {Pinilla},
  {Salyk}, {Herczeg}, {Pontoppidan}, {Vazquez}, {Watkins}, {Krijt}, {Hendler},
  \& {Long}}]{banz20}
{Banzatti}, A., {Pascucci}, I., {Bosman}, A.~D., {et~al.} 2020, \apj, 903, 124

\bibitem[{{Banzatti} {et~al.}(2022){Banzatti}, {Abernathy}, {Brittain},
  {Bosman}, {Pontoppidan}, {Boogert}, {Jensen}, {Carr}, {Najita}, {Grant},
  {Sigler}, {Sanchez}, {Kern}, \& {Rayner}}]{banz22}
{Banzatti}, A., {Abernathy}, K.~M., {Brittain}, S., {et~al.} 2022, \aj, 163,
  174

\bibitem[{{Bast} {et~al.}(2011){Bast}, {Brown}, {Herczeg}, {van Dishoeck}, \&
  {Pontoppidan}}]{bast11}
{Bast}, J.~E., {Brown}, J.~M., {Herczeg}, G.~J., {van Dishoeck}, E.~F., \&
  {Pontoppidan}, K.~M. 2011, \aap, 527, A119

\bibitem[{{Bitner} {et~al.}(2008){Bitner}, {Richter}, {Lacy}, {Herczeg},
  {Greathouse}, {Jaffe}, {Salyk}, {Blake}, {Hollenbach}, {Doppmann}, {Najita},
  \& {Currie}}]{bitner08}
{Bitner}, M.~A., {Richter}, M.~J., {Lacy}, J.~H., {et~al.} 2008, \apj, 688,
  1326

\bibitem[{{Blevins} {et~al.}(2016){Blevins}, {Pontoppidan}, {Banzatti},
  {Zhang}, {Najita}, {Carr}, {Salyk}, \& {Blake}}]{blevins16}
{Blevins}, S.~M., {Pontoppidan}, K.~M., {Banzatti}, A., {et~al.} 2016, \apj,
  818, 22

\bibitem[{{Bosman} {et~al.}(2022){Bosman}, {Bergin}, {Calahan}, \&
  {Duval}}]{bosman22}
{Bosman}, A.~D., {Bergin}, E.~A., {Calahan}, J., \& {Duval}, S.~E. 2022, \apjl,
  930, L26

\bibitem[{{Bosman} {et~al.}(2017){Bosman}, {Bruderer}, \& {van
  Dishoeck}}]{bosman17}
{Bosman}, A.~D., {Bruderer}, S., \& {van Dishoeck}, E.~F. 2017, \aap, 601, A36

\bibitem[{{Brittain} {et~al.}(2016){Brittain}, {Najita}, {Carr},
  {{\'A}d{\'a}mkovics}, \& {Reynolds}}]{brittain16}
{Brittain}, S.~D., {Najita}, J.~R., {Carr}, J.~S., {{\'A}d{\'a}mkovics}, M., \&
  {Reynolds}, N. 2016, \apj, 830, 112

\bibitem[{{Brittain} {et~al.}(2013){Brittain}, {Najita}, {Carr}, {Liskowsky},
  {Troutman}, \& {Doppmann}}]{brittain13}
{Brittain}, S.~D., {Najita}, J.~R., {Carr}, J.~S., {et~al.} 2013, \apj, 767,
  159

\bibitem[{{Brittain} {et~al.}(2003){Brittain}, {Rettig}, {Simon}, {Kulesa},
  {DiSanti}, \& {Dello Russo}}]{brittain03}
{Brittain}, S.~D., {Rettig}, T.~W., {Simon}, T., {et~al.} 2003, \apj, 588, 535

\bibitem[{{Brittain} {et~al.}(2007){Brittain}, {Simon}, {Najita}, \&
  {Rettig}}]{brittain07}
{Brittain}, S.~D., {Simon}, T., {Najita}, J.~R., \& {Rettig}, T.~W. 2007, \apj,
  659, 685

\bibitem[{{Brown} {et~al.}(2012){Brown}, {Herczeg}, {Pontoppidan}, \& {van
  Dishoeck}}]{brown12}
{Brown}, J.~M., {Herczeg}, G.~J., {Pontoppidan}, K.~M., \& {van Dishoeck},
  E.~F. 2012, \apj, 744, 116

\bibitem[{{Brown} {et~al.}(2013){Brown}, {Pontoppidan}, {van Dishoeck},
  {Herczeg}, {Blake}, \& {Smette}}]{brown13}
{Brown}, J.~M., {Pontoppidan}, K.~M., {van Dishoeck}, E.~F., {et~al.} 2013,
  \apj, 770, 94

\bibitem[{{Brown} {et~al.}(2007){Brown}, {Blake}, {Dullemond}, {Mer{\'\i}n},
  {Augereau}, {Boogert}, {Evans}, {Geers}, {Lahuis}, {Kessler-Silacci},
  {Pontoppidan}, \& {van Dishoeck}}]{brown07}
{Brown}, J.~M., {Blake}, G.~A., {Dullemond}, C.~P., {et~al.} 2007, \apjl, 664,
  L107

\bibitem[{{Carmona} {et~al.}(2008){Carmona}, {van den Ancker}, {Henning},
  {Pavlyuchenkov}, {Dullemond}, {Goto}, {Thi}, {Bouwman}, \&
  {Waters}}]{carmona08}
{Carmona}, A., {van den Ancker}, M.~E., {Henning}, T., {et~al.} 2008, \aap,
  477, 839

\bibitem[{{Carr} \& {Najita}(2008)}]{cn08}
{Carr}, J.~S., \& {Najita}, J.~R. 2008, Science, 319, 1504

\bibitem[{{Carr} \& {Najita}(2011)}]{cn11}
---. 2011, \apj, 733, 102

\bibitem[{{Carr} {et~al.}(2004){Carr}, {Tokunaga}, \& {Najita}}]{carr04}
{Carr}, J.~S., {Tokunaga}, A.~T., \& {Najita}, J. 2004, \apj, 603, 213

\bibitem[{{Ciesla} \& {Cuzzi}(2006)}]{cieslacuzzi06}
{Ciesla}, F.~J., \& {Cuzzi}, J.~N. 2006, \icarus, 181, 178

\bibitem[{{Cushing} {et~al.}(2004){Cushing}, {Vacca}, \& {Rayner}}]{spextool}
{Cushing}, M.~C., {Vacca}, W.~D., \& {Rayner}, J.~T. 2004, \pasp, 116, 362

\bibitem[{{Cutri} {et~al.}(2021){Cutri}, {Wright}, {Conrow}, {Fowler},
  {Eisenhardt}, {Grillmair}, {Kirkpatrick}, {Masci}, {McCallon}, {Wheelock},
  {Fajardo-Acosta}, {Yan}, {Benford}, {Harbut}, {Jarrett}, {Lake}, {Leisawitz},
  {Ressler}, {Stanford}, {Tsai}, {Liu}, {Helou}, {Mainzer}, {Gettngs},
  {Gonzalez}, {Hoffman}, {Marsh}, {Padgett}, {Skrutskie}, {Beck}, {Papin}, \&
  {Wittman}}]{allwise}
{Cutri}, R.~M., {Wright}, E.~L., {Conrow}, T., {et~al.} 2021, VizieR Online
  Data Catalog, II/328

\bibitem[{{Doppmann} {et~al.}(2011){Doppmann}, {Najita}, {Carr}, \&
  {Graham}}]{doppmann11}
{Doppmann}, G.~W., {Najita}, J.~R., {Carr}, J.~S., \& {Graham}, J.~R. 2011,
  \apj, 738, 112

\bibitem[{{Du} \& {Bergin}(2014)}]{du14}
{Du}, F., \& {Bergin}, E.~A. 2014, \apj, 792, 2

\bibitem[{{Du} {et~al.}(2017){Du}, {Bergin}, {Hogerheijde}, {van Dishoeck},
  {Blake}, {Bruderer}, {Cleeves}, {Dominik}, {Fedele}, {Lis}, {Melnick},
  {Neufeld}, {Pearson}, \& {Y{\i}ld{\i}z}}]{du17}
{Du}, F., {Bergin}, E.~A., {Hogerheijde}, M., {et~al.} 2017, \apj, 842, 98

\bibitem[{{Dubernet} {et~al.}(2009){Dubernet}, {Daniel}, {Grosjean}, \&
  {Lin}}]{dubernet09}
{Dubernet}, M.~L., {Daniel}, F., {Grosjean}, A., \& {Lin}, C.~Y. 2009, \aap,
  497, 911

\bibitem[{{Dullemond} \& {Dominik}(2004)}]{radmc}
{Dullemond}, C.~P., \& {Dominik}, C. 2004, \aap, 417, 159

\bibitem[{{Dunkin} {et~al.}(1997){Dunkin}, {Barlow}, \& {Ryan}}]{dunkin97}
{Dunkin}, S.~K., {Barlow}, M.~J., \& {Ryan}, S.~G. 1997, \mnras, 286, 604

\bibitem[{{Fang} {et~al.}(2018){Fang}, {Pascucci}, {Edwards}, {Gorti},
  {Banzatti}, {Flock}, {Hartigan}, {Herczeg}, \& {Dupree}}]{fang18}
{Fang}, M., {Pascucci}, I., {Edwards}, S., {et~al.} 2018, \apj, 868, 28

\bibitem[{{Faure} \& {Josselin}(2008)}]{faure08}
{Faure}, A., \& {Josselin}, E. 2008, \aap, 492, 257

\bibitem[{{Fedele} {et~al.}(2012){Fedele}, {Bruderer}, {van Dishoeck},
  {Herczeg}, {Evans}, {Bouwman}, {Henning}, \& {Green}}]{fedele12}
{Fedele}, D., {Bruderer}, S., {van Dishoeck}, E.~F., {et~al.} 2012, \aap, 544,
  L9

\bibitem[{{Fedele} {et~al.}(2011){Fedele}, {Pascucci}, {Brittain}, {Kamp},
  {Woitke}, {Williams}, {Dent}, \& {Thi}}]{fedele11}
{Fedele}, D., {Pascucci}, I., {Brittain}, S., {et~al.} 2011, \apj, 732, 106

\bibitem[{{Finzi} {et~al.}(1977){Finzi}, {Hovis}, {Panfilov}, {Hess}, \&
  {Moore}}]{finzi77}
{Finzi}, J., {Hovis}, F.~E., {Panfilov}, V.~N., {Hess}, P., \& {Moore}, C.~B.
  1977, \jcp, 67, 4053

\bibitem[{{Furlan} {et~al.}(2009){Furlan}, {Watson}, {McClure}, {Manoj},
  {Espaillat}, {D'Alessio}, {Calvet}, {Kim}, {Sargent}, {Forrest}, \&
  {Hartmann}}]{furlan09}
{Furlan}, E., {Watson}, D.~M., {McClure}, M.~K., {et~al.} 2009, \apj, 703, 1964

\bibitem[{{Glassgold} {et~al.}(2009){Glassgold}, {Meijerink}, \&
  {Najita}}]{glassgold09}
{Glassgold}, A.~E., {Meijerink}, R., \& {Najita}, J.~R. 2009, \apj, 701, 142

\bibitem[{{Goldsmith} \& {Langer}(1999)}]{GL99}
{Goldsmith}, P.~F., \& {Langer}, W.~D. 1999, \apj, 517, 209

\bibitem[{{Gordon} {et~al.}(2022){Gordon}, {Rothman}, {Hargreaves}, {Hashemi},
  {Karlovets}, {Skinner}, {Conway}, {Hill}, {Kochanov}, {Tan}, {Wcis{\l}o},
  {Finenko}, {Nelson}, {Bernath}, {Birk}, {Boudon}, {Campargue}, {Chance},
  {Coustenis}, {Drouin}, {Flaud}, {Gamache}, {Hodges}, {Jacquemart}, {Mlawer},
  {Nikitin}, {Perevalov}, {Rotger}, {Tennyson}, {Toon}, {Tran}, {Tyuterev},
  {Adkins}, {Baker}, {Barbe}, {Can{\`e}}, {Cs{\'a}sz{\'a}r}, {Dudaryonok},
  {Egorov}, {Fleisher}, {Fleurbaey}, {Foltynowicz}, {Furtenbacher}, {Harrison},
  {Hartmann}, {Horneman}, {Huang}, {Karman}, {Karns}, {Kassi}, {Kleiner},
  {Kofman}, {Kwabia-Tchana}, {Lavrentieva}, {Lee}, {Long}, {Lukashevskaya},
  {Lyulin}, {Makhnev}, {Matt}, {Massie}, {Melosso}, {Mikhailenko}, {Mondelain},
  {M{\"u}ller}, {Naumenko}, {Perrin}, {Polyansky}, {Raddaoui}, {Raston},
  {Reed}, {Rey}, {Richard}, {T{\'o}bi{\'a}s}, {Sadiek}, {Schwenke},
  {Starikova}, {Sung}, {Tamassia}, {Tashkun}, {Vander Auwera}, {Vasilenko},
  {Vigasin}, {Villanueva}, {Vispoel}, {Wagner}, {Yachmenev}, \&
  {Yurchenko}}]{hitran20}
{Gordon}, I.~E., {Rothman}, L.~S., {Hargreaves}, R.~J., {et~al.} 2022, \jqsrt,
  277, 107949

\bibitem[{{Guzm{\'a}n-D{\'\i}az} {et~al.}(2021){Guzm{\'a}n-D{\'\i}az},
  {Mendigut{\'\i}a}, {Montesinos}, {Oudmaijer}, {Vioque}, {Rodrigo}, {Solano},
  {Meeus}, \& {Marcos-Arenal}}]{guzmandiaz21}
{Guzm{\'a}n-D{\'\i}az}, J., {Mendigut{\'\i}a}, I., {Montesinos}, B., {et~al.}
  2021, \aap, 650, A182

\bibitem[{{Hartmann} {et~al.}(1986){Hartmann}, {Hewett}, {Stahler}, \&
  {Mathieu}}]{hartmann86}
{Hartmann}, L., {Hewett}, R., {Stahler}, S., \& {Mathieu}, R.~D. 1986, \apj,
  309, 275

\bibitem[{{Hein Bertelsen} {et~al.}(2016{\natexlab{a}}){Hein Bertelsen},
  {Kamp}, {van der Plas}, {van den Ancker}, {Waters}, {Thi}, \&
  {Woitke}}]{heinbert16}
{Hein Bertelsen}, R.~P., {Kamp}, I., {van der Plas}, G., {et~al.}
  2016{\natexlab{a}}, \aap, 590, A98

\bibitem[{{Hein Bertelsen} {et~al.}(2016{\natexlab{b}}){Hein Bertelsen},
  {Kamp}, {van der Plas}, {van den Ancker}, {Waters}, {Thi}, \&
  {Woitke}}]{heinbert16_var}
---. 2016{\natexlab{b}}, \mnras, 458, 1466

\bibitem[{{Herczeg} {et~al.}(2011){Herczeg}, {Brown}, {van Dishoeck}, \&
  {Pontoppidan}}]{herczeg11}
{Herczeg}, G.~J., {Brown}, J.~M., {van Dishoeck}, E.~F., \& {Pontoppidan},
  K.~M. 2011, \aap, 533, A112

\bibitem[{{Herczeg} \& {Hillenbrand}(2014)}]{herczeg14}
{Herczeg}, G.~J., \& {Hillenbrand}, L.~A. 2014, \apj, 786, 97

\bibitem[{{Herczeg} {et~al.}(2012){Herczeg}, {Karska}, {Bruderer},
  {Kristensen}, {van Dishoeck}, {J{\o}rgensen}, {Visser}, {Wampfler}, {Bergin},
  {Y{\i}ld{\i}z}, {Pontoppidan}, \& {Gracia-Carpio}}]{herczeg12}
{Herczeg}, G.~J., {Karska}, A., {Bruderer}, S., {et~al.} 2012, \aap, 540, A84

\bibitem[{{Huang} {et~al.}(2018){Huang}, {Andrews}, {Dullemond}, {Isella},
  {P{\'e}rez}, {Guzm{\'a}n}, {{\"O}berg}, {Zhu}, {Zhang}, {Bai}, {Benisty},
  {Birnstiel}, {Carpenter}, {Hughes}, {Ricci}, {Weaver}, \& {Wilner}}]{huang18}
{Huang}, J., {Andrews}, S.~M., {Dullemond}, C.~P., {et~al.} 2018, \apjl, 869,
  L42

\bibitem[{Hunter(2007)}]{matplotlib}
Hunter, J.~D. 2007, Computing in Science \& Engineering, 9, 90

\bibitem[{{Kaeufl} {et~al.}(2004){Kaeufl}, {Ballester}, {Biereichel},
  {Delabre}, {Donaldson}, {Dorn}, {Fedrigo}, {Finger}, {Fischer}, {Franza},
  {Gojak}, {Huster}, {Jung}, {Lizon}, {Mehrgan}, {Meyer}, {Moorwood}, {Pirard},
  {Paufique}, {Pozna}, {Siebenmorgen}, {Silber}, {Stegmeier}, \&
  {Wegerer}}]{crires}
{Kaeufl}, H.-U., {Ballester}, P., {Biereichel}, P., {et~al.} 2004, in Society
  of Photo-Optical Instrumentation Engineers (SPIE) Conference Series, Vol.
  5492, Ground-based Instrumentation for Astronomy, ed. A.~F.~M. {Moorwood} \&
  M.~{Iye}, 1218--1227

\bibitem[{{Kamp} {et~al.}(2013){Kamp}, {Thi}, {Meeus}, {Woitke}, {Pinte},
  {Meijerink}, {Spaans}, {Pascucci}, {Aresu}, \& {Dent}}]{kamp13}
{Kamp}, I., {Thi}, W.~F., {Meeus}, G., {et~al.} 2013, \aap, 559, A24

\bibitem[{{Kamp} {et~al.}(2021){Kamp}, {Honda}, {Nomura}, {Audard}, {Fedele},
  {Waters}, {Aikawa}, {Banzatti}, {Bowey}, {Bradford}, {Dominik}, {Furuya},
  {Habart}, {Ishihara}, {Johnstone}, {Kennedy}, {Kim}, {Kral}, {Lai},
  {Larsson}, {McClure}, {Miotello}, {Momose}, {Nakagawa}, {Naylor}, {Nisini},
  {Notsu}, {Onaka}, {Pantin}, {Podio}, {Riviere Marichalar}, {Rocha},
  {Roelfsema}, {Shimonishi}, {Tang}, {Takami}, {Tazaki}, {Wolf}, {Wyatt}, \&
  {Ysard}}]{kamp21}
{Kamp}, I., {Honda}, M., {Nomura}, H., {et~al.} 2021, \pasa, 38, e055

\bibitem[{{K{\"a}ufl} {et~al.}(2015){K{\"a}ufl}, {Kerber}, {Asmus}, {Baksai},
  {Di Lieto}, {Duhoux}, {Heikamp}, {Hummel}, {Ives}, {Jakob}, {Kirchbauer},
  {Mehrgan}, {Momany}, {Pantin}, {Pozna}, {Riquelme}, {Sandrock},
  {Siebenmorgen}, {Smette}, {Stegmeier}, {Taylor}, {Tristram}, {Valdes}, {van
  den Ancker}, {Weilenmann}, \& {Wolff}}]{visirUP}
{K{\"a}ufl}, H.~U., {Kerber}, F., {Asmus}, D., {et~al.} 2015, The Messenger,
  159, 15

\bibitem[{{Krijt} {et~al.}(2022){Krijt}, {Kama}, {McClure}, {Teske}, {Bergin},
  {Shorttle}, {Walsh}, \& {Raymond}}]{krijt22}
{Krijt}, S., {Kama}, M., {McClure}, M., {et~al.} 2022, arXiv e-prints,
  arXiv:2203.10056

\bibitem[{{Kurtovic} {et~al.}(2018){Kurtovic}, {P{\'e}rez}, {Benisty}, {Zhu},
  {Zhang}, {Huang}, {Andrews}, {Dullemond}, {Isella}, {Bai}, {Carpenter},
  {Guzm{\'a}n}, {Ricci}, \& {Wilner}}]{kurtovic18}
{Kurtovic}, N.~T., {P{\'e}rez}, L.~M., {Benisty}, M., {et~al.} 2018, \apjl,
  869, L44

\bibitem[{{Labdon} {et~al.}(2019){Labdon}, {Kraus}, {Davies}, {Kreplin},
  {Kluska}, {Harries}, {Monnier}, {ten Brummelaar}, {Baron}, {Millan-Gabet},
  {Kloppenborg}, {Eisner}, {Sturmann}, \& {Sturmann}}]{labdon19}
{Labdon}, A., {Kraus}, S., {Davies}, C.~L., {et~al.} 2019, \aap, 627, A36

\bibitem[{{Labiano} {et~al.}(2021){Labiano}, {Argyriou},
  {{\'A}lvarez-M{\'a}rquez}, {Glasse}, {Glauser}, {Patapis}, {Law}, {Brandl},
  {Justtanont}, {Lahuis}, {Mart{\'\i}nez-Galarza}, {Mueller}, {Noriega-Crespo},
  {Royer}, {Shaughnessy}, \& {Vandenbussche}}]{miri_res}
{Labiano}, A., {Argyriou}, I., {{\'A}lvarez-M{\'a}rquez}, J., {et~al.} 2021,
  \aap, 656, A57

\bibitem[{{Lacy} {et~al.}(2002){Lacy}, {Richter}, {Greathouse}, {Jaffe}, \&
  {Zhu}}]{texes}
{Lacy}, J.~H., {Richter}, M.~J., {Greathouse}, T.~K., {Jaffe}, D.~T., \& {Zhu},
  Q. 2002, \pasp, 114, 153

\bibitem[{{Lagage} {et~al.}(2004){Lagage}, {Pel}, {Authier}, {Belorgey},
  {Claret}, {Doucet}, {Dubreuil}, {Durand}, {Elswijk}, {Girardot}, {K{\"a}ufl},
  {Kroes}, {Lortholary}, {Lussignol}, {Marchesi}, {Pantin}, {Peletier},
  {Pirard}, {Pragt}, {Rio}, {Schoenmaker}, {Siebenmorgen}, {Silber}, {Smette},
  {Sterzik}, \& {Veyssiere}}]{visir}
{Lagage}, P.~O., {Pel}, J.~W., {Authier}, M., {et~al.} 2004, The Messenger,
  117, 12

\bibitem[{{Larsson} {et~al.}(2002){Larsson}, {Liseau}, \&
  {Men'shchikov}}]{larsson02}
{Larsson}, B., {Liseau}, R., \& {Men'shchikov}, A.~B. 2002, \aap, 386, 1055

\bibitem[{{Liu} {et~al.}(2011){Liu}, {Zhang}, {Wu}, {Qin}, \& {Miller}}]{liu11}
{Liu}, T., {Zhang}, H., {Wu}, Y., {Qin}, S.-L., \& {Miller}, M. 2011, \apj,
  734, 22

\bibitem[{{Liu} {et~al.}(2019){Liu}, {Pascucci}, \& {Henning}}]{liu19}
{Liu}, Y., {Pascucci}, I., \& {Henning}, T. 2019, \aap, 623, A106

\bibitem[{{Long} {et~al.}(2018){Long}, {Pinilla}, {Herczeg}, {Harsono},
  {Dipierro}, {Pascucci}, {Hendler}, {Tazzari}, {Ragusa}, {Salyk}, {Edwards},
  {Lodato}, {van de Plas}, {Johnstone}, {Liu}, {Boehler}, {Cabrit}, {Manara},
  {Menard}, {Mulders}, {Nisini}, {Fischer}, {Rigliaco}, {Banzatti}, {Avenhaus},
  \& {Gully-Santiago}}]{long18}
{Long}, F., {Pinilla}, P., {Herczeg}, G.~J., {et~al.} 2018, \apj, 869, 17

\bibitem[{{Long} {et~al.}(2021){Long}, {Andrews}, {Vega}, {Wilner}, {Chandler},
  {Ragusa}, {Teague}, {P{\'e}rez}, {Calvet}, {Carpenter}, {Henning}, {Kwon},
  {Linz}, \& {Ricci}}]{long21}
{Long}, F., {Andrews}, S.~M., {Vega}, J., {et~al.} 2021, \apj, 915, 131

\bibitem[{{Manara} {et~al.}(2019){Manara}, {Tazzari}, {Long}, {Herczeg},
  {Lodato}, {Rota}, {Cazzoletti}, {van der Plas}, {Pinilla}, {Dipierro},
  {Edwards}, {Harsono}, {Johnstone}, {Liu}, {Menard}, {Nisini}, {Ragusa},
  {Boehler}, \& {Cabrit}}]{manara19}
{Manara}, C.~F., {Tazzari}, M., {Long}, F., {et~al.} 2019, \aap, 628, A95

\bibitem[{{Mandell} {et~al.}(2012){Mandell}, {Bast}, {van Dishoeck}, {Blake},
  {Salyk}, {Mumma}, \& {Villanueva}}]{mandell12}
{Mandell}, A.~M., {Bast}, J., {van Dishoeck}, E.~F., {et~al.} 2012, \apj, 747,
  92

\bibitem[{{Markwardt}(2009)}]{mpfit}
{Markwardt}, C.~B. 2009, in Astronomical Society of the Pacific Conference
  Series, Vol. 411, Astronomical Data Analysis Software and Systems XVIII, ed.
  D.~A. {Bohlender}, D.~{Durand}, \& P.~{Dowler}, 251

\bibitem[{{Meijerink} {et~al.}(2009){Meijerink}, {Pontoppidan}, {Blake},
  {Poelman}, \& {Dullemond}}]{meijerink09}
{Meijerink}, R., {Pontoppidan}, K.~M., {Blake}, G.~A., {Poelman}, D.~R., \&
  {Dullemond}, C.~P. 2009, \apj, 704, 1471

\bibitem[{{Mundt} \& {Eisl{\"o}ffel}(1998)}]{mundt98}
{Mundt}, R., \& {Eisl{\"o}ffel}, J. 1998, \aj, 116, 860

\bibitem[{{Najita} {et~al.}(2003){Najita}, {Carr}, \& {Mathieu}}]{najita03}
{Najita}, J., {Carr}, J.~S., \& {Mathieu}, R.~D. 2003, \apj, 589, 931

\bibitem[{{Najita} {et~al.}(2011){Najita}, {{\'A}d{\'a}mkovics}, \&
  {Glassgold}}]{najita11}
{Najita}, J.~R., {{\'A}d{\'a}mkovics}, M., \& {Glassgold}, A.~E. 2011, \apj,
  743, 147

\bibitem[{{Najita} {et~al.}(2013){Najita}, {Carr}, {Pontoppidan}, {Salyk}, {van
  Dishoeck}, \& {Blake}}]{najita13}
{Najita}, J.~R., {Carr}, J.~S., {Pontoppidan}, K.~M., {et~al.} 2013, \apj, 766,
  134

\bibitem[{{Najita} {et~al.}(2018){Najita}, {Carr}, {Salyk}, {Lacy}, {Richter},
  \& {DeWitt}}]{najita18}
{Najita}, J.~R., {Carr}, J.~S., {Salyk}, C., {et~al.} 2018, \apj, 862, 122

\bibitem[{{Notsu} {et~al.}(2016){Notsu}, {Nomura}, {Ishimoto}, {Walsh},
  {Honda}, {Hirota}, \& {Millar}}]{notsu16}
{Notsu}, S., {Nomura}, H., {Ishimoto}, D., {et~al.} 2016, \apj, 827, 113

\bibitem[{{Pani{\'c}} {et~al.}(2021){Pani{\'c}}, {Haworth}, {Petr-Gotzens},
  {Miley}, {van den Ancker}, {Vioque}, {Siess}, {Parker}, {Clarke}, {Kamp},
  {Kennedy}, {Oudmaijer}, {Pascucci}, {Richards}, {Ratzka}, \& {Qi}}]{panic21}
{Pani{\'c}}, O., {Haworth}, T.~J., {Petr-Gotzens}, M.~G., {et~al.} 2021,
  \mnras, 501, 4317

\bibitem[{{Panoglou} {et~al.}(2012){Panoglou}, {Cabrit}, {Pineau Des
  For{\^e}ts}, {Garcia}, {Ferreira}, \& {Casse}}]{panoglou12}
{Panoglou}, D., {Cabrit}, S., {Pineau Des For{\^e}ts}, G., {et~al.} 2012, \aap,
  538, A2

\bibitem[{{Pascucci} {et~al.}(2013){Pascucci}, {Herczeg}, {Carr}, \&
  {Bruderer}}]{pascucci13}
{Pascucci}, I., {Herczeg}, G., {Carr}, J.~S., \& {Bruderer}, S. 2013, \apj,
  779, 178

\bibitem[{{Pascucci} \& {Sterzik}(2009)}]{pascucci09}
{Pascucci}, I., \& {Sterzik}, M. 2009, \apj, 702, 724

\bibitem[{{Pascucci} {et~al.}(2020){Pascucci}, {Banzatti}, {Gorti}, {Fang},
  {Pontoppidan}, {Alexander}, {Ballabio}, {Edwards}, {Salyk}, {Sacco},
  {Flaccomio}, {Blake}, {Carmona}, {Hall}, {Kamp}, {K{\"a}ufl}, {Meeus},
  {Meyer}, {Pauly}, {Steendam}, \& {Sterzik}}]{pascucci20}
{Pascucci}, I., {Banzatti}, A., {Gorti}, U., {et~al.} 2020, \apj, 903, 78

\bibitem[{{Perez Chavez} {et~al.}(2021){Perez Chavez}, {Banzatti}, {Wheeler},
  \& {Pontoppidan}}]{perez21_spexodisks}
{Perez Chavez}, J.~A., {Banzatti}, A., {Wheeler}, C., \& {Pontoppidan}, K.
  2021, in American Astronomical Society Meeting Abstracts, Vol.~53, American
  Astronomical Society Meeting Abstracts, 115.05

\bibitem[{{Pinilla} {et~al.}(2018){Pinilla}, {Tazzari}, {Pascucci}, {Youdin},
  {Garufi}, {Manara}, {Testi}, {van der Plas}, {Barenfeld}, {Canovas}, {Cox},
  {Hendler}, {P{\'e}rez}, \& {van der Marel}}]{pinilla18}
{Pinilla}, P., {Tazzari}, M., {Pascucci}, I., {et~al.} 2018, \apj, 859, 32

\bibitem[{{Pontoppidan} {et~al.}(2011){Pontoppidan}, {Blake}, \&
  {Smette}}]{pont11}
{Pontoppidan}, K.~M., {Blake}, G.~A., \& {Smette}, A. 2011, \apj, 733, 84

\bibitem[{{Pontoppidan} {et~al.}(2009){Pontoppidan}, {Meijerink}, {Dullemond},
  \& {Blake}}]{pont09}
{Pontoppidan}, K.~M., {Meijerink}, R., {Dullemond}, C.~P., \& {Blake}, G.~A.
  2009, \apj, 704, 1482

\bibitem[{{Pontoppidan} {et~al.}(2010{\natexlab{a}}){Pontoppidan}, {Salyk},
  {Blake}, \& {K{\"a}ufl}}]{pont10}
{Pontoppidan}, K.~M., {Salyk}, C., {Blake}, G.~A., \& {K{\"a}ufl}, H.~U.
  2010{\natexlab{a}}, \apjl, 722, L173

\bibitem[{{Pontoppidan} {et~al.}(2010{\natexlab{b}}){Pontoppidan}, {Salyk},
  {Blake}, \& {K{\"a}ufl}}]{pont10b}
---. 2010{\natexlab{b}}, \apjl, 722, L173

\bibitem[{{Pontoppidan} {et~al.}(2018){Pontoppidan}, {Bergin}, {Melnick},
  {Bradford}, {Staguhn}, {Leisawitz}, {Meixner}, {Fortney}, {Salyk}, {Blake},
  {Zhang}, {Banzatti}, {Kataria}, {Meshkat}, {de Val-Borro}, {Stevenson}, \&
  {Fraine}}]{pont18}
{Pontoppidan}, K.~M., {Bergin}, E.~A., {Melnick}, G., {et~al.} 2018, arXiv
  e-prints, arXiv:1804.00743

\bibitem[{{Prato} {et~al.}(2003){Prato}, {Greene}, \& {Simon}}]{prato03}
{Prato}, L., {Greene}, T.~P., \& {Simon}, M. 2003, \apj, 584, 853

\bibitem[{{Rab} {et~al.}(2022){Rab}, {Weber}, {Grassi}, {Ercolano}, {Picogna},
  {Caselli}, {Thi}, {Kamp}, \& {Woitke}}]{rab22}
{Rab}, C., {Weber}, M., {Grassi}, T., {et~al.} 2022, arXiv e-prints,
  arXiv:2210.15486

\bibitem[{{Rayner} {et~al.}(2016){Rayner}, {Tokunaga}, {Jaffe}, {Bonnet},
  {Ching}, {Connelley}, {Kokubun}, {Lockhart}, \& {Warmbier}}]{ishell}
{Rayner}, J., {Tokunaga}, A., {Jaffe}, D., {et~al.} 2016, in Society of
  Photo-Optical Instrumentation Engineers (SPIE) Conference Series, Vol. 9908,
  Ground-based and Airborne Instrumentation for Astronomy VI, ed. C.~J.
  {Evans}, L.~{Simard}, \& H.~{Takami}, 990884

\bibitem[{Rayner {et~al.}(2022)Rayner, Tokunaga, Jaffe, Bond, Bonnet, Ching,
  Connelley, Cushing, Kokubun, Lockhart, Vacca, \& Warmbier}]{ishell22}
Rayner, J., Tokunaga, A., Jaffe, D., {et~al.} 2022, Publications of the
  Astronomical Society of the Pacific, 134, 015002.
\newblock \url{https://doi.org/10.1088/1538-3873/ac3cb4}

\bibitem[{{Rieke} {et~al.}(2015){Rieke}, {Wright}, {B{\"o}ker}, {Bouwman},
  {Colina}, {Glasse}, {Gordon}, {Greene}, {G{\"u}del}, {Henning}, {Justtanont},
  {Lagage}, {Meixner}, {N{\o}rgaard-Nielsen}, {Ray}, {Ressler}, {van Dishoeck},
  \& {Waelkens}}]{miri}
{Rieke}, G.~H., {Wright}, G.~S., {B{\"o}ker}, T., {et~al.} 2015, \pasp, 127,
  584

\bibitem[{{Riviere-Marichalar} {et~al.}(2012){Riviere-Marichalar},
  {M{\'e}nard}, {Thi}, {Kamp}, {Montesinos}, {Meeus}, {Woitke}, {Howard},
  {Sandell}, {Podio}, {Dent}, {Mendigut{\'\i}a}, {Pinte}, {White}, \&
  {Barrado}}]{rivmarichalar12}
{Riviere-Marichalar}, P., {M{\'e}nard}, F., {Thi}, W.~F., {et~al.} 2012, \aap,
  538, L3

\bibitem[{{Ros} \& {Johansen}(2013)}]{ros13}
{Ros}, K., \& {Johansen}, A. 2013, \aap, 552, A137

\bibitem[{{Salyk} {et~al.}(2009){Salyk}, {Blake}, {Boogert}, \&
  {Brown}}]{salyk09}
{Salyk}, C., {Blake}, G.~A., {Boogert}, A.~C.~A., \& {Brown}, J.~M. 2009, \apj,
  699, 330

\bibitem[{{Salyk} {et~al.}(2011{\natexlab{a}}){Salyk}, {Blake}, {Boogert}, \&
  {Brown}}]{salyk11}
---. 2011{\natexlab{a}}, \apj, 743, 112

\bibitem[{{Salyk} {et~al.}(2013){Salyk}, {Herczeg}, {Brown}, {Blake},
  {Pontoppidan}, \& {van Dishoeck}}]{salyk13}
{Salyk}, C., {Herczeg}, G.~J., {Brown}, J.~M., {et~al.} 2013, \apj, 769, 21

\bibitem[{{Salyk} {et~al.}(2019){Salyk}, {Lacy}, {Richter}, {Zhang},
  {Pontoppidan}, {Carr}, {Najita}, \& {Blake}}]{salyk19}
{Salyk}, C., {Lacy}, J., {Richter}, M., {et~al.} 2019, \apj, 874, 24

\bibitem[{{Salyk} {et~al.}(2008){Salyk}, {Pontoppidan}, {Blake}, {Lahuis}, {van
  Dishoeck}, \& {Evans}}]{salyk08}
{Salyk}, C., {Pontoppidan}, K.~M., {Blake}, G.~A., {et~al.} 2008, \apjl, 676,
  L49

\bibitem[{{Salyk} {et~al.}(2011{\natexlab{b}}){Salyk}, {Pontoppidan}, {Blake},
  {Najita}, \& {Carr}}]{salyk11_spitz}
{Salyk}, C., {Pontoppidan}, K.~M., {Blake}, G.~A., {Najita}, J.~R., \& {Carr},
  J.~S. 2011{\natexlab{b}}, \apj, 731, 130

\bibitem[{{Salyk} {et~al.}(2022){Salyk}, {Pontoppidan}, {Banzatti},
  {K{\"a}ufl}, {Hall}, {Pascucci}, {Carmona}, {Blake}, {Alexander}, \&
  {Kamp}}]{salyk22}
{Salyk}, C., {Pontoppidan}, K.~M., {Banzatti}, A., {et~al.} 2022, \aj, 164, 136

\bibitem[{{Sargent} {et~al.}(2014){Sargent}, {Forrest}, {Watson}, {D'Alessio},
  {Calvet}, {Furlan}, {Kim}, {Green}, {Pontoppidan}, {Richter}, \&
  {Tayrien}}]{sargent14}
{Sargent}, B.~A., {Forrest}, W., {Watson}, D.~M., {et~al.} 2014, \apj, 792, 83

\bibitem[{{Sch{\"o}ier} {et~al.}(2005){Sch{\"o}ier}, {van der Tak}, {van
  Dishoeck}, \& {Black}}]{lamda}
{Sch{\"o}ier}, F.~L., {van der Tak}, F.~F.~S., {van Dishoeck}, E.~F., \&
  {Black}, J.~H. 2005, \aap, 432, 369

\bibitem[{{Simon} {et~al.}(2016){Simon}, {Pascucci}, {Edwards}, {Feng},
  {Gorti}, {Hollenbach}, {Rigliaco}, \& {Keane}}]{simon16}
{Simon}, M.~N., {Pascucci}, I., {Edwards}, S., {et~al.} 2016, \apj, 831, 169

\bibitem[{{Sullivan} {et~al.}(2019{\natexlab{a}}){Sullivan}, {Prato},
  {Edwards}, {Avilez}, \& {Schaefer}}]{sullivan19}
{Sullivan}, K., {Prato}, L., {Edwards}, S., {Avilez}, I., \& {Schaefer}, G.~H.
  2019{\natexlab{a}}, \apj, 884, 28

\bibitem[{{Sullivan} {et~al.}(2019{\natexlab{b}}){Sullivan}, {Wilking},
  {Greene}, {Lisalda}, {Gibb}, \& {Ejeta}}]{sullivan19_rv}
{Sullivan}, T., {Wilking}, B.~A., {Greene}, T.~P., {et~al.} 2019{\natexlab{b}},
  \aj, 158, 41

\bibitem[{{Tabone} {et~al.}(2020){Tabone}, {Godard}, {Pineau des For{\^e}ts},
  {Cabrit}, \& {van Dishoeck}}]{tabone20}
{Tabone}, B., {Godard}, B., {Pineau des For{\^e}ts}, G., {Cabrit}, S., \& {van
  Dishoeck}, E.~F. 2020, \aap, 636, A60

\bibitem[{{Tennyson} {et~al.}(2001){Tennyson}, {Zobov}, {Williamson},
  {Polyansky}, \& {Bernath}}]{tennyson01}
{Tennyson}, J., {Zobov}, N.~F., {Williamson}, R., {Polyansky}, O.~L., \&
  {Bernath}, P.~F. 2001, Journal of Physical and Chemical Reference Data, 30,
  735

\bibitem[{{Thi} {et~al.}(2013){Thi}, {Kamp}, {Woitke}, {van der Plas},
  {Bertelsen}, \& {Wiesenfeld}}]{thi13}
{Thi}, W.~F., {Kamp}, I., {Woitke}, P., {et~al.} 2013, \aap, 551, A49

\bibitem[{{Tripathi} {et~al.}(2017){Tripathi}, {Andrews}, {Birnstiel}, \&
  {Wilner}}]{tripathi17}
{Tripathi}, A., {Andrews}, S.~M., {Birnstiel}, T., \& {Wilner}, D.~J. 2017,
  \apj, 845, 44

\bibitem[{{Ubeira Gabellini} {et~al.}(2019){Ubeira Gabellini}, {Miotello},
  {Facchini}, {Ragusa}, {Lodato}, {Testi}, {Benisty}, {Bruderer}, {T.
  Kurtovic}, {Andrews}, {Carpenter}, {Corder}, {Dipierro}, {Ercolano},
  {Fedele}, {Guidi}, {Henning}, {Isella}, {Kwon}, {Linz}, {McClure}, {Perez},
  {Ricci}, {Rosotti}, {Tazzari}, \& {Wilner}}]{margi19}
{Ubeira Gabellini}, M.~G., {Miotello}, A., {Facchini}, S., {et~al.} 2019,
  \mnras, 486, 4638

\bibitem[{{van der Marel} {et~al.}(2016){van der Marel}, {van Dishoeck},
  {Bruderer}, {Andrews}, {Pontoppidan}, {Herczeg}, {van Kempen}, \&
  {Miotello}}]{vdmarel16}
{van der Marel}, N., {van Dishoeck}, E.~F., {Bruderer}, S., {et~al.} 2016,
  \aap, 585, A58

\bibitem[{{van der Plas} {et~al.}(2015){van der Plas}, {van den Ancker},
  {Waters}, \& {Dominik}}]{vdplas15}
{van der Plas}, G., {van den Ancker}, M.~E., {Waters}, L.~B.~F.~M., \&
  {Dominik}, C. 2015, \aap, 574, A75

\bibitem[{{van der Tak} {et~al.}(2007){van der Tak}, {Black}, {Sch{\"o}ier},
  {Jansen}, \& {van Dishoeck}}]{vdt07}
{van der Tak}, F.~F.~S., {Black}, J.~H., {Sch{\"o}ier}, F.~L., {Jansen}, D.~J.,
  \& {van Dishoeck}, E.~F. 2007, \aap, 468, 627

\bibitem[{{van Dishoeck} {et~al.}(2014){van Dishoeck}, {Bergin}, {Lis}, \&
  {Lunine}}]{vandishoeck14}
{van Dishoeck}, E.~F., {Bergin}, E.~A., {Lis}, D.~C., \& {Lunine}, J.~I. 2014,
  in Protostars and Planets VI, ed. H.~{Beuther}, R.~S. {Klessen}, C.~P.
  {Dullemond}, \& T.~{Henning}, 835

\bibitem[{{Walsh} {et~al.}(2015){Walsh}, {Nomura}, \& {van Dishoeck}}]{walsh15}
{Walsh}, C., {Nomura}, H., \& {van Dishoeck}, E. 2015, \aap, 582, A88

\bibitem[{{Wang} {et~al.}(2019){Wang}, {Bai}, \& {Goodman}}]{wang19}
{Wang}, L., {Bai}, X.-N., \& {Goodman}, J. 2019, \apj, 874, 90

\bibitem[{Waskom(2021)}]{seaborn}
Waskom, M.~L. 2021, Journal of Open Source Software, 6, 3021.
\newblock \url{https://doi.org/10.21105/joss.03021}

\bibitem[{{Wichittanakom} {et~al.}(2020){Wichittanakom}, {Oudmaijer},
  {Fairlamb}, {Mendigut{\'\i}a}, {Vioque}, \& {Ababakr}}]{wichittanakom20}
{Wichittanakom}, C., {Oudmaijer}, R.~D., {Fairlamb}, J.~R., {et~al.} 2020,
  \mnras, 493, 234

\bibitem[{{Woitke} {et~al.}(2018){Woitke}, {Min}, {Thi}, {Roberts}, {Carmona},
  {Kamp}, {M{\'e}nard}, \& {Pinte}}]{woitke18}
{Woitke}, P., {Min}, M., {Thi}, W.~F., {et~al.} 2018, \aap, 618, A57

\bibitem[{{Woitke} {et~al.}(2016){Woitke}, {Min}, {Pinte}, {Thi}, {Kamp},
  {Rab}, {Anthonioz}, {Antonellini}, {Baldovin-Saavedra}, {Carmona}, {Dominik},
  {Dionatos}, {Greaves}, {G{\"u}del}, {Ilee}, {Liebhart}, {M{\'e}nard},
  {Rigon}, {Waters}, {Aresu}, {Meijerink}, \& {Spaans}}]{woitke16}
{Woitke}, P., {Min}, M., {Pinte}, C., {et~al.} 2016, \aap, 586, A103

\bibitem[{{W{\"o}lfer} {et~al.}(2021){W{\"o}lfer}, {Facchini}, {Kurtovic},
  {Teague}, {van Dishoeck}, {Benisty}, {Ercolano}, {Lodato}, {Miotello},
  {Rosotti}, {Testi}, \& {Ubeira Gabellini}}]{wolfer21}
{W{\"o}lfer}, L., {Facchini}, S., {Kurtovic}, N.~T., {et~al.} 2021, \aap, 648,
  A19

\bibitem[{{Wu} {et~al.}(2004){Wu}, {Wei}, {Zhao}, {Shi}, {Yu}, {Qin}, \&
  {Huang}}]{wu04}
{Wu}, Y., {Wei}, Y., {Zhao}, M., {et~al.} 2004, \aap, 426, 503

\bibitem[{{Yang} {et~al.}(2022){Yang}, {Green}, {Pontoppidan}, {Bergner},
  {Cleeves}, {Evans}, {Garrod}, {Jin}, {Kim}, {Kim}, {Lee}, {Sakai},
  {Shingledecker}, {Shope}, {Tobin}, \& {van Dishoeck}}]{yang22}
{Yang}, Y.-L., {Green}, J.~D., {Pontoppidan}, K.~M., {et~al.} 2022, arXiv
  e-prints, arXiv:2208.10673

\bibitem[{{Yvart} {et~al.}(2016){Yvart}, {Cabrit}, {Pineau des For{\^e}ts}, \&
  {Ferreira}}]{yvart16}
{Yvart}, W., {Cabrit}, S., {Pineau des For{\^e}ts}, G., \& {Ferreira}, J. 2016,
  \aap, 585, A74

\end{thebibliography}
\bibliographystyle{aasjournal}

\end{document}